\documentclass[a4paper,fleqn,usenatbib,useAMS,onecolumn]{mnras}

\usepackage{graphicx}	
\usepackage{amsmath}	
\usepackage{amssymb}	
\usepackage{multicol}        
\usepackage{bm}		
\usepackage{pdflscape}	
\usepackage{dcolumn}   
\usepackage{grffile}
\usepackage{color}
\usepackage{grffile}

\newcommand\Eqn[1]     {Eq.\,(\ref{#1})}
\newcommand\Eqns[2]    {Eqs.\,(\ref{#1}) and~(\ref{#2})}

\newcommand\Fig[1]     {Fig.\,{\ref{#1}}}

\newcommand\nn         {\nonumber}

\def\mnras{{Mon.~ Not.~ R.~ Astron.~ Soc.~}}

\def\prd{{Phys.~ Rev.~ D.~}}

\def\apj{{Astrophys.~ J.~}}
\def\apjs{{Astrophys.~ J.~ Suppl.~}}
\def\apjl{{Astrophys.~ J.~ Lett.~}}

\def\nat{{Nature (London)~}}

\def\nar{{New~Astro.~Rev.}}

\def\mnras{{MNRAS}}

\def\prd{{PRD}}

\def\apj{{ApJ}}
\def\apjs{{ApJS}}
\def\apjl{{ApJL}}
\def\aap{{A\&A}}
\def\nat{{Nature}}

\def\physrep{{Phys.~ Rep.~}}

\newcommand{\be}{\begin{equation}}
\newcommand{\ee}{\end{equation}}
\newcommand{\ba}{\begin{eqnarray}}
\newcommand{\ea}{\end{eqnarray}}

\def\pp1{{\prime}}
\def\pp2{{\prime\prime}}

\def\2D{{\rm 2D}}

\def\Vmu{{V_{\mu}}}

\def\Vu{{V_{\mu}}}

\def\SN{{\mathcal S}/{\mathcal N}}

\def\g{{\rm g}}
\def\h{{\rm h}}
\def\c{{\rm c}}
\def\s{{\rm s}}

\def\bx{{\bf x}}
\def\br{{\bf r}}

\def\bk{{\bf k}}
\def\bq{{\bf q}}

\def\1Loop{{\rm 1Loop}}

\def\Mpc{\, h^{-1}{\rm Mpc}}

\def\dr{d^3r}
\def\dx{d^3x}
\def\dk{d^3k}
\def\dq{d^3q}

\def\nbar{\bar{n}}
\def\nbarg{\bar{n}_{\rm g}}

\def\dirac{\delta^{\rm D}}
\def\h{{\rm h}}

\def\la{\mathrel{\mathpalette\fun <}}
\def\ga{\mathrel{\mathpalette\fun >}}
\def\fun#1#2{\lower3.6pt\vbox{\baselineskip0pt\lineskip.9pt
        \ialign{$\mathsurround=0pt#1\hfill##\hfil$\crcr#2\crcr\sim\crcr}}}

\def\Nh{{N_{\rm h}}}
\def\Ng{{N_{\rm g}}}

\def\Veff{V_{\rm eff}}
\def\Fg{{\mathcal F}_{\rm g}}
\def\Fgt{\tilde{\mathcal F}_{\rm g}}
\def\Gb{\overline{\mathcal G}}

\def\Lmin{L_{\rm min}}

\def\SN{{\mathcal S}/{\mathcal N}}
\def\NS{{\mathcal N}/{\mathcal S}}

\title[Optimal estimation of the galaxy power spectrum]
{Towards optimal estimation of the galaxy power spectru}

\author[{\it Smith \& Marian}]
{Robert E.~Smith$^{1,2}$\thanks{r.e.smith@sussex.ac.uk} and Laura Marian$^1$\thanks{l.marian@sussex.ac.uk}\\
$^1$ Department of Physics and Astronomy, University of Sussex, Brighton BN1 9QH, UK\\
$^2$ Max-Planck-Institut f\"ur Astrophysik, Karl-Schwarzschild-Str.1, Postfach 1523, 85740 Garching, Germany
}

\begin{document}

\label{firstpage}
\pagerange{\pageref{firstpage}--\pageref{lastpage}}


\maketitle


\begin{abstract}
The galaxy power spectrum encodes a wealth of information about
cosmology and the matter fluctuations. Its unbiased and optimal
estimation is therefore of great importance.  In this paper we
generalise the framework of Feldman et al. (1994) to take into account
the fact that galaxies are not simply a Poisson sampling of the
underlying dark matter distribution. Besides finite survey-volume
effects and flux-limits, our optimal estimation scheme incorporates
several of the key tenets of galaxy formation: galaxies form and
reside exclusively in dark matter haloes; a given dark matter halo may
host several galaxies of various luminosities; galaxies inherit part
of their large-scale bias from their host halo.  Under these broad
assumptions, we prove that the optimal weights \emph{do not}
explicitly depend on galaxy luminosity, other than through defining
the maximum survey volume and effective galaxy density at a given
position.  Instead, they depend on the bias associated with the host
halo; the first and second factorial moments of the halo occupation
distribution; a selection function, which gives the fraction of
galaxies that can be observed in a halo of mass $M$ at position $\br$
in the survey; and an effective number density of galaxies.  If one
wishes to reconstruct the matter power spectrum, then, provided the
model is correct, this scheme provides the only unbiased estimator.
The practical challenges with implementing this approach are also
discussed.
\end{abstract}




\section{introduction}
The power spectrum of matter fluctuations, or equivalently the
two-point correlation function, is a fundamental tool for constraining
the cosmological parameters. It contains detailed information about
the large-scale geometrical structure of space-time, the constituents
of energy-density and their evolution with redshift, and also provides
us with information about the primordial scalar fluctuation
spectrum. However, we do not directly observe the matter density
field, instead we observe galaxy angular positions and measure radial
velocities, or redshifts, from spectra. Given a galaxy redshift
survey, two things are crucial: how to obtain an unbiased estimate of
the information in the matter fluctuations; and obtaining an estimate
that has the highest signal-to-noise possible, i.e. an optimal
measurement.

The first point may be rephrased as the need to understand the
relation between galaxy and matter fluctuations -- more commonly
referred to as galaxy bias.  The second point, that of optimality, in
fact also relies on our understanding of bias, since only through
knowing how the galaxies are embedded in the mass distribution can one
devise efficient survey strategies; for example, if all galaxies
formed in pairs then one would only require information about one
galaxy from each pair to obtain all of the useful cosmological
information.

The development of galaxy correlation functions as a tool for
constraining the cosmological model was first realized by Peebles and
collaborators in a series of pioneering papers in the 1970s
\citep{Peebles1973,HauserPeebles1973,PeeblesHauser1974,Peebles1974,PeeblesGroth1975,Peebles1975,SeldnerPeebles1977,GrothPeebles1977,FryPeebles1978,SeldnerPeebles1978,SeldnerPeebles1979,FryPeebles1980}.
Subsequent studies built on this, and estimators were developed that
took better account of fluctuations in the mean number-density of
galaxies
\citep{DavisPeebles1983,LandySzalay1993,Hamilton1993b,Bernstein1994}. These
new estimators were only optimal in the case that the clustering was
very weak and when galaxies represented a Poisson sampling of the
underlying matter fluctuations.

The development of techniques for the direct estimation of the galaxy
power spectrum began in earnest in the early 1990s
\citep{BaumgartFry1991,PeacockNicholson1991,Fisheretal1993}. This
culminated in the seminal work of \citep[][hereafter
  FKP]{Feldmanetal1994}. In their seminal approach, galaxies were
assumed to be a Poisson sampling of the mass density field. They
showed that provided one subtracted an appropriate shot-noise term,
and deconvolved for the survey window function, one could obtain an
unbiased estimate of the matter power spectrum. Subsequent analysis
focused on obtaining quadratic and decorrelated band estimates
\citep{VogeleySzalay1996,Hamilton1997a,Hamilton1997b,Hamilton2000,HamiltonTegmark2000}.

In the last two decades our understanding of galaxy formation has made
rapid progress and the current best models strongly suggest that
galaxies are not related to the underlying dark matter in the simple
way that was envisioned in FKP.
\citep{WhiteRees1978,WhiteFrenk1991,Kauffmannetal1999,Bensonetal2000,Springeletal2005}.
Furthermore, improved observational studies have subsequently
discovered that galaxy clustering is in fact dependent on a number of
properties of the galaxy distribution: e.g. luminosity
\citep{Parketal1994,Norbergetal2001,Norbergetal2002a,Zehavietal2002,Zehavietal2005,Swansonetal2008,Zehavietal2011},
colour
\citep{Brownetal2000,Zehavietal2002short,Zehavietal2005short,Swansonetal2008,Zehavietal2011short},
morphology \citep{DavisGeller1976,Guzzoetal1997,Norbergetal2002a},
stellar mass \citep{Lietal2006} etc.

\citet[][hereafter PVP]{Percivaletal2004a} attempted to correct the
FKP framework to take into account the effects of luminosity dependent
bias. To this end, PVP assumed that the probability of finding a
galaxy of a given luminosity in a certain patch of space would be a
Poisson variate, whose mean was proportional to the local density of
dark matter multiplied by a luminosity-dependent bias factor.  Their
work demonstrated two important facts: firstly that an optimal
weighting scheme depended sensitively on the assumptions about the
bias and secondly, if their assumptions about the bias were correct,
the FKP method was a biased estimator of the matter power spectrum.

In this paper we argue that the approach of PVP, whilst qualitatively
reasonable, is in fact still at odds with our current understanding of
galaxy formation and therefore unlikely to be the {\em true} optimal
estimator. The key ideas from galaxy formation and evolution that we
would like to build into our estimator are: galaxies only form in dark
matter haloes \citep{WhiteRees1978}; haloes can host a number of
galaxies of various luminosities; the large-scale bias associated with
a given galaxy, is largely inherited from the bias of the host dark
matter halo. In a recent paper \citep[][hereafter
  SM14]{SmithMarian2014}, we generalised the FKP formalism to account
for the clustering of galaxy clusters -- which turned out to have a
similar mathematical structure to the PVP scheme. We now undertake to
generalise the FKP formalism to take into account these ideas from
galaxy formation. As we will show, these effects will lead us to a new
optimal estimator and method for reconstructing the matter power
spectrum.

Before moving on, it is worth noting that current state-of-the-art
galaxy redshift surveys, such as the Baryon Oscillation Spectroscopic
Survey \citep[][hereafter
  BOSS]{Andersonetal2012,Andersonetal2014a,Andersonetal2014b}, Galaxy
And Mass Assembly \citep[][hereafter GAMA]{Blakeetal2013}, and
WiggleZ \citep{Blakeetal2011}, have all used the FKP power spectrum
estimation procedure. Future surveys, such as DESI \citep{DESI2013},
Euclid \citep{EUCLID2011} and SKA \citep{SKA2004}, will have
significantly larger volumes and so unbiased and optimised data
anlaysis will be crucial if we are to obtain the tightest
constraints on the cosmological parameters.

The paper is broken down as follows: In \S\ref{sec:survey} we describe
generic properties of a galaxy redshift survey and present a new
theoretical quantity, the halo-galaxy double-delta expansion. We
explore its statistical properties.  In \S\ref{sec:theory1} we show
how one may obtain unbiased estimates of the matter correlation
function and power spectrum. In \S\ref{sec:fluctuations} we derive the
covariance matrix of the fluctuations in the galaxy power spectrum. In
\S\ref{sec:optimal} we derive the optimal weights. In
\S\ref{sec:Fisher} we present a new expression for the Fisher
information matrix for optimally-weighted galaxy power spectra. In
\S\ref{sec:practical} we enumerate the steps for a practical
implementation of this approach. Finally, in \S\ref{sec:conclusions}
we summarize our findings and draw our conclusions.


\section{Survey specifications and the $\Fg$--field}\label{sec:survey}


\subsection{Preliminaries: a generic galaxy redshift survey}


Let us begin by defining our fiducial galaxy survey: suppose that we
have observed $N^{\rm tot}_\g$ galaxies and to the $i$th galaxy we
assign a luminosity $L_i$, redshift $z_i$ and angular position on the
sky $\bm\Omega_i=\bm\Omega(\theta_i,\phi_i)$. If we specify the
background FLRW spacetime, then we may convert the redshift into a
comoving radial geodesic distance $\chi_i=\chi(z_i)$. A galaxy's
comoving position vector may now be expressed as
$\br_i=\br(\chi_i,\bm\Omega_i)$.

The survey mask function depends on both the position and luminosity
of galaxies, given an adopted flux limit. In this work we shall take
the angular and radial parts of the survey mask function to be
separable, though this assumption does not change our results:
\be \Theta(\br|L) = \Theta(\bm\Omega) \Theta(\chi|L) \ .
\ee
Note that if the flux-limit is not uniform across the survey then the
radial function $\Theta(\chi,L)$ would still be a function of the
angular position vector $\bm\Omega$, and the survey mask cannot be
separated as in the equation above. The angular part of the mask may
be written as:
\be \Theta(\bm\Omega) = 
\left\{
\begin{array}{cc}
1 \ ; & [{\bm\Omega} \in \{{\bm\Omega}_{\mu}\} ] \\
0 \ ; & [\rm otherwise] 
\end{array} \ ,
\right.
\ee
where $\{\bm\Omega_{\mu}\}$ is the set of angular positions that lie
inside the survey area. The radial mask function may be written:
\be \Theta(\chi|L) = \left\{
\begin{array}{cc}
1 \ ; & [{\chi} \leq \chi_{\rm max}(L) ] \\
0 \ ; & [\rm otherwise] 
\end{array} \ ,
\right.
\ee
where $\chi_{\rm max}(L)$ is the maximum distance out to which a
galaxy of luminosity $L$ could have been detected.

The survey volume for galaxies with luminosity $L$ is simply the
integral of the mask function over all space:
\be \Vmu(L) = \int  \Theta(\br|L)
dV,\ee
where $dV$ is the comoving volume element at position vector $\br$
(for a flat universe $dV=d^3r=\chi^2d\Omega d\chi$) . In what follows
it will be also useful to note that the relation $\chi_{\rm max}(L)$
may be inverted to obtain the minimum galaxy luminosity that could
have been detected at radial position $\chi(z)$ in the survey.  We
shall write this as:
\be 
\left[\Lmin(\br)/ h^{-2} L_{\odot}\right] =
10^{-\frac{2}{5}\left(m_{\rm lim} -25 -M_{\odot}\right)}\left[d_{\rm L}(\br)/\Mpc\right]^{-2}\ ,
\ee
where $m_{\rm lim}$ is the apparent magnitude limit of the survey,
$M_{\odot}$ is the absolute magnitude of the sun, $h$ is the
dimensionless Hubble parameter and $d_{\rm L}$ is the luminosity
distance (for a flat universe $d_{\rm L}(z)=(1+z)\chi(z)$).  Thus for
any general function ${\mathcal B(\chi, L)}$, we have the useful
integral relations:
\be 
\int_0^{\infty} dL \int_{0}^{\infty}d\chi \Theta(\chi|L) \mathcal B(\chi, L) 
= \int_0^{\infty} dL \int_{0}^{\chi_{\rm max}(L)}  d\chi \mathcal B(\chi, L) 
= \int_{0}^{\infty}d\chi  \int_{\Lmin(\chi)}^{\infty} dL  \mathcal B(\chi, L) \ . \label{eq:ThetaInt}
\ee


\subsection{The halo-galaxy double-delta expansion}

Our understanding of galaxy formation tells us that galaxies form
exclusively in dark matter haloes, and that each dark matter halo may
host several galaxies of various luminosities. It therefore follows
that the large-scale bias associated with any given galaxy is directly
proportional to the bias of the host halo.  We shall mathematically
encode these ideas in our density field as follows: our $N^{\rm
  tot}_\g$ galaxies are distributed inside $N_h$ dark matter
haloes. Thus the $i$th dark matter halo of mass $M_i$ and centre of
mass position $\bx_i$ will host a number of galaxies that depends on
its mass, $N_\g(M_i)$. The $j$th galaxy will have a position vector
$\br_j$ relative to the centre of the halo and a luminosity $L_j$.

In order to study the statistical properties of the galaxy field in
this scenario we need to simultaneously account for both the spatial
distribution of the haloes, as well as the galaxies inside them. We
therefore introduce a new function, dubbed the `galaxy-halo
double-delta expansion'. This is a Dirac delta function expansion over
the halo positions and masses, as well as over the positions and
luminosities of the galaxies inside the haloes. It is written:
\be n_{\g}(\br,L,\bx,M)=
\sum_{i=1}^{N_h}\delta^{\rm D}(\bx-\bx_i)\delta^{\rm D}(M-M_i) 
\sum_{j=1}^{N_\g(M_i)}\delta^{\rm D}(\br-\br_j-\bx_i)\delta^{\rm D}(L-L_j) \ ,
\label{eq:density} 
\ee
where $N_{\rm h}$ is the total number of host dark matter haloes in
the survey volume and $N_{\rm g}(M_i)$ is the total number of galaxies
in the $i$th halo. In the above function, the order of variables is
important: $\br$ refers to the spatial vector in the galaxy field, $L$
the luminosity, $\bx$ the spatial vector in the halo field, and $M$
the halo mass. Note that the units of the above function are inverse
squared volume, inverse mass, and inverse luminosity
\footnote{We note that this equation is the more rigorous starting
  point for all Halo Model calculations of the galaxy field. However,
  so far as we are aware it has not been written down before. This in
  part owes to the fact that the galaxy-clustering expressions could
  be deduced by analogy with the mass clustering. However, for the
  case of the optimal weights in a realistic survey that approach is
  not feasible.}.

Next, in analogy with SM14, we define a field $\Fg$, which is related
to the over-density of galaxies. Our survey will be finite and will
contain masked regions and an apparent magnitude limit $m_{\rm
  lim}$. Hence, the overdensity field of galaxies with magnitudes
above some threshold luminosity may be written using \Eqn{eq:density}
in the following way:
\be \Fg(\br) = 
\int^{\infty}_{0} dL  
\int \dx \int_0^{\infty} dM \Theta(\br|L) \frac{w(\br,L,\bx,M)}{\sqrt{A}} \left[n_{\g}(\br,L,\bx,M)-\alpha
  n_{\s}(\br,L,\bx,M)\right] \label{eq:galden}  \ ,
\ee
where $w(\br,L,\bx,M)$ is a weighting function that we will wish to
determine in an optimal way, and $A$ is a normalisation parameter that
will be chosen later. 

The function $n_{\s}(\br,L,\bx,M)$ is the random galaxy-halo
double-delta expansion. This immediately leads to an important
question: what constitutes a random catalogue? Conventionally, in the
FKP approach one would distribute the mock galaxies randomly within
the survey volume -- preserving the number counts as a function of
redshift. However, since we know (or have assumed in this model) that
galaxies form only inside dark matter haloes, we do not want to remove
this property. Instead it is the dark matter haloes which should be
randomly distributed, and not the galaxies. Therefore, the function
$n_{\s}(\br,L,\bx,M)$ represents the distribution of galaxies in a
mock sample {\em whose dark matter haloes possess no intrinsic spatial
  correlations}, and have a number density that is $1/\alpha$ of the
true galaxy-halo double-delta field. Note that for this random
distribution, while the halo centres are not correlated, the galaxies
still follow the density distribution inside each dark matter halo. In
addition, the haloes possess a mass spectrum and the galaxy
luminosities are conditioned on the halo mass.

Both quantities defined by \Eqns{eq:density}{eq:galden} are of central
importance and will be extensively used in this paper. It is therefore
worthwhile for us to take some time to understand their meaning and
how one should employ them to infer the statistical properties of the
galaxy density field. This we do in the following section.


\subsection{Calculation of the expectation of the galaxy density field}


As a demonstration of how one can use the halo-galaxy double-delta
expansion and take statistical averages we calculate the expectation
of $\Fg$.  We use \Eqn{eq:galden} to break $\left<\Fg(\br)\right>$
into two parts:
\be \left<\Fg(\br)\right> = 
 \int^{\infty}_{0} dL  
\Theta(\br|L) \left[\left<{\mathcal N}_{\g}(\br,L) \right>-
\alpha \left<{\mathcal N}_{\s}(\br,L)\right>\right] \ ,
\label{eq:expecFg1}
\ee
where we introduced the weighted mean number density of
galaxies per unit luminosity, at the spatial position
$\br(\chi,\bm\Omega)$:
\ba \left< {\mathcal N}_{\g}(\br,L) \right> & \equiv & 
\int \dx \int_{0}^{\infty} dM \frac{w(\br,L,\bx,M)}{\sqrt{A}}  
\left< n_{\g}(\br,L,\bx,M) \right>  
\nn \\
& = & 
\int \dx \int_{0}^{\infty} dM \frac{w(\br,L,\bx,M)}{\sqrt{A}}   \left<
\sum_{i=1}^{N_h}\delta^{\rm D}(\bx-\bx_i)\delta^{\rm D}(M-M_i) 
\sum_{j=1}^{N_\g(M_i)}\delta^{\rm D}(\br-\br_j-\bx_i)\delta^{\rm D}(L-L_j) \right> ,
\label{eq:ng}
\ea
with a similar expression for $\left<{\mathcal
  N}_{\s}(\br,L)\right>$. The function in \Eqn{eq:ng} is related to
the galaxy luminosity function.

To proceed further we now need to understand what taking the
`expectation value' actually means. Following \citet{Smith2012}, this
operation can be broken down into three steps. First, the fluctuations
in the underlying dark matter field are sampled -- we shall denote
this averaging through a sub-script $s$. Second, given the dark matter
field, the haloes may be obtained as a sampling of the density field
-- we shall denote this operation through sub-script $h$. Third, given
a set of dark matter haloes, and sufficient knowledge of the
properties of the halo, galaxies may then be sampled into each halo --
we shall denote this operation through sub-script $g$. Hence,
\Eqn{eq:ng} can be rewritten,
\ba
\left< {\mathcal N}_{\g}(\br,L) \right> =  
\int \dx \int_{0}^{\infty} dM \frac{w(\br,L,\bx,M)}{\sqrt{A}}\left<
\sum_{i=1}^{N_h}\delta^{\rm D}(\bx-\bx_i)\delta^{\rm D}(M-M_i) 
\left<\sum_{j=1}^{N_\g(M_i)}\delta^{\rm D}(\br-\br_j-\bx_i)\delta^{\rm D}(L-L_j) \right>_\g 
\right>_{\rm s,h} 
\label{eq:mean1}
\ea
Let us now compute the average over the galaxy sampling for the $i$th
dark matter halo:
\ba 
\left<\sum_{j=1}^{N_{\g}(M_i)}\delta^{\rm D}(\br-\br_j-\bx_i)\delta^{\rm D}(L-L_j) \right>_\g
& \equiv & \sum_{N_{\g}=0}^{\infty} P(N_{\g}|\lambda(M_i))
\int \prod_{k=1}^{N_{\g}}\left\{\dr_k dL_k\right\}
p(\br_1,\dots,\br_{N_\g},L_1,\dots, L_{N_\g}|M_i,\bx_i)
\nn \\
& & \hspace{-0.47cm}\times  \left[\dirac(\br-\br_1-\bx_i)\dirac(L-L_1)+\dots
+\dirac(\br-\br_{N_\g}-\bx_i)\dirac(L-L_{N_\g})\right] \ ,
\label{eq:den1}
\ea
where in the above we have introduced the following quantities:
$P(N_{\g}|\lambda(M_i))$ is the discrete probability that there are
$N_\g$ galaxies in the $i$th dark matter halo and this we assume
depends on some function of the dark matter halo mass $M_i$;
$p(\br_1,\dots,\br_{N_{\g}},L_1,\dots, L_{N_\g}|M_i,\bx_i)$ is the
joint probability density function for finding the $N_{\rm g}$
galaxies being located at positions $\{\br_1,\dots,\br_{N_\g}\}$
relative to the halo centre $\bx_i$, and with luminosities
$\{L_1,\dots,L_{N_\g}\}$, conditioned on $M_i$ and $\bx_i$. We have
assumed that the properties and distribution of the galaxies in the
$i$th halo are independent of all other external haloes. If the
probability for finding a galaxy at a given position inside a halo is
determined by the density profile of the matter in the halo, and if
the probability that the galaxy has a luminosity $L$ depends only on
the halo mass, then this joint probability can be written in the
following manner:
\be p(\br_1,\dots,\br_{N_{\g}},L_1,\dots, L_{N_\g}|M_i,\bx_i) = 
\prod_{k=1}^{N_{\g}} \left\{p(\br_k|M_i,\bx_i) p(L_k|M_i)\right\} 
= \prod_{k=1}^{N_{\g}} \left\{ U(\br_k|M_i,\bx_i) 
\Phi(L_k|M_i) \right\} \label{eq:probrl} \ ,
\ee
where in the above equation we have used the density profile of
galaxies in the halo, normalised by the total number of galaxies in
that halo, $U$, to define
\be p(\br|M,\bx) \equiv U(\br|M,\bx)\equiv \rho_{\g}(\br|M, \bx)/N_{\g}(M) \ . \ee
We have also used
\be p(L_k|M_i) \equiv \Phi(L_k|M_i) \ ,\ee
as the probability density that a galaxy hosted by a halo of mass
$M$, has a luminosity $L$ \footnote{Note that this is closely related
  to the conditional luminosity function introduced by
  \citet[c.f.][]{Yangetal2003}, which in our notation would be
  $\Phi_{\rm Yang\,et\,al}(L|M)=N^{(1)}_{\g}(M)\Phi(L|M)$.}. In
writing \Eqn{eq:probrl} we have assumed that, for a given galaxy, its
spatial location inside the dark matter halo is independent of its
luminosity. As will be shown later, this assumption will not be
crucial for the derivation of the optimal weights.

On integrating over the Dirac delta functions in \Eqn{eq:den1} we find
\ba
\left<\sum_{j=1}^{N_{\g}(M_i)}\delta^{\rm D}(\br-\br_j-\bx_i)\delta^{\rm D}(L-L_j) \right>_\g
& = & \sum_{N_{\g}=0}^{\infty} P(N_{\g}|\lambda(M_i))N_g 
U(\br-\bx_i|M_i) \Phi(L|M_i)  \nn \\
& = & N^{(1)}_\g(M_i) U(\br-\bx_i|M_i) \Phi(L|M_i)  \ , 
\ea
where we have suppressed the dependence of $U$ on the halo centre. The
second equality follows from the definition of the first factorial
moment of the galaxy distribution:
\be 
N^{(1)}_\g(M_i)\equiv \sum_{N_{\g}=0}^{\infty} P(N_{\g}|\lambda(M_i))N_g \ .
\ee 

Returning to our main calculation, on substituting the last two equations
into \Eqn{eq:mean1}, we now obtain
\ba \left< {\mathcal N}_{\g}(\br,L) \right> & = & 
\int \dx \int_{0}^{\infty} dM \frac{w(\br,L,\bx,M)}{\sqrt{A}} \left<
\sum_{i=1}^{N_h}\delta^{\rm D}(\bx-\bx_i)\delta^{\rm D}(M-M_i) 
N^{(1)}_\g(M_i) U(\br-\bx_i|M_i) \Phi(L|M_i) 
\right>_{\rm s,h} 
\nn \\
& = & 
\int \dx \int_{0}^{\infty} dM \frac{w(\br,L,\bx,M)}{\sqrt{A}} 
\int\dx_1\dots\dx_{N_\h} dM_1\dots dM_{N_\h}
p(\bx_1,\dots\bx_{N_\h},M_1,\dots,M_{N_\h}) 
\nn \\
& & \times \sum_{i=1}^{N_h}
\delta^{\rm D}(\bx-\bx_i)\delta^{\rm D}(M-M_i) 
N^{(1)}_\g(M_i) U(\br-\bx_i|M_i) \Phi(L|M_i)  \ .
\label{eq:den3}
\ea
In the above, we followed SM14 to introduce
$p(\bx_1,\dots,\bx_{N_{\h}},M_1,\dots,M_{N_\h})$ as the joint
probability density for the $N_\h$ dark matter halo centres being
located at positions $\{\bx_1,\dots,\bx_{N_\h}\}$, and with masses
$\{M_1,\dots,M_{N_\h}\}$. On integrating over the Dirac delta
functions, the mean number density of galaxies becomes,
\ba \left< {\mathcal N}_{\g}(\br,L) \right>& = & \int \dx \int_{0}^{\infty} dM \frac{w(\br,L,\bx,M)}{\sqrt{A}}
N_{\h} p(\bx,M) N^{(1)}_\g(M)
U(\br-\bx|M) \Phi(L|M) 
 \label{eq:mean2} \ .
\ea
The joint distribution function for obtaining a halo of mass $M$ at
position $\bx$ can be written as the product of two independent
one-point probability density functions \citep{SmithWatts2005}:
\be p(\bx,M)=p(M)p(\bx)=\frac{\nbar(M)}{\nbar_\h}\times\frac{1}{\Vu} 
= \frac{\nbar(M)}{N_{\h}}\ ,\ee
where $\nbar(M)$ is the mean mass function of dark matter haloes,
which tells us the number density of haloes of mass $M$, per unit
mass, and $\nbar_\h=N_{\h}/\Vu$ is the mean number density of haloes.
On substituting this expression into \Eqn{eq:mean2} we find that the
mean density of galaxies, per unit luminosity, at spatial location
$\br$ may be written:
\be 
\left< {\mathcal N}_{\g}(\br,L) \right> = \frac{1}{\sqrt{A}}\phi_w(\br,L) \ , \label{eq:mean3} 
\ee
where we have defined 
\be \phi_w(\br,L) \equiv \int_{0}^{\infty} dM \nbar(M)
N^{(1)}_\g(M)\Phi(L|M) \int \dx w(\br,L,\bx,M)
U(\br-\bx|M) \ .
\ee
If we were to set the weight function to unity, the above expression
would be the galaxy luminosity function \citep{Yangetal2003}:
\be \phi(L)\equiv \int_{0}^{\infty} dM \nbar(M)  N^{(1)}_\g(M)\Phi(L|M) \ .\ee

Turning to the second expectation value in \Eqn{eq:expecFg1}, we note
that the only difference between $\left<{\mathcal N}_{\g}(\br,L) \right>$ and
$\left<{\mathcal N}_{\s}(\br,L)\right>$ is the artificially increased space-density
of clusters and the absence of any intrinsic clustering.  Hence, we
also have,
\be 
\alpha\left< {\mathcal N}_{\s}(\br,L) \right> = \frac{1}{\sqrt{A}}\phi_w(\br,L) \label{eq:Ns}
\ .\ee
Returning to \Eqn{eq:expecFg1} and inserting \Eqns{eq:mean3}{eq:Ns} we
arrive at the result:
\be \left<\Fg(\br)\right>=0 \ .\ee
Hence, the $\Fg$--field, like the over-density field of matter, is
truly a mean-zero field.

Note that we have neglected to take into account the statistical
properties of obtaining the $\Nh$ clusters in the survey volume. In what
follows we shall assume that the survey volumes are sufficiently large
that this may be essentially treated as a deterministic
quantity. However, it can be taken into account
\citep[e.g. see][]{ShethLemson1999,SmithWatts2005,Smith2009}.


\section{Clustering Estimators}\label{sec:theory1}

We now move on to the more interesting problem of using the
halo-galaxy double-delta expansion to compute the clustering
properties of the galaxy distribution. We begin first with the
correlation function, and then through Fourier transforms look at the
power spectrum. This task will be somewhat laborious, however it will
enable us to develop and establish a number of important concepts and
results.


\subsection{The two-point correlation function of galaxies}

The two-point correlation function of the field $\Fg$ can be computed 
using our double-delta expansion through:
\ba 
\left<\Fg(\br_1)\Fg(\br_2)\right> & = & \frac{1}{A}\int dL_1 dL_2 d^3x_1 d^3x_2 
dM_1 dM_2 \Theta(\br_1|L_1) \Theta(\br_2|L_2) w(\br_1, L_1, \bx_1, M_1) w(\br_2, L_2, \bx_2, M_2)\nn\\
& &  \times \left[\frac{}{} \left<n_\g(\br_1, L_1, \bx_1, M_1) 
n_\g(\br_2,L_2,\bx_2, M_2)\right> 
-\alpha\left<n_\g(\br_1, L_1, \bx_1, M_1) n_{\s}(\br_2, L_2,\bx_2, M_2)\right> \right. 
\nn \\ & &  \left. \frac{}{} 
-\alpha\left<n_{\s}(\br_1, L_1, \bx_1, M_1) n_\g(\br_2, L_2, \bx_2, M_2)\right> 
+\alpha^2\left<n_{\s}(\br_1, L_1, \bx_1, M_1) n_{\s}(\br_2, L_2, \bx_2, M_2)\right>
\right] \label{eq:FF} \ .
\ea
The expectation terms in the square bracket on the right-hand side of
this equation can be evaluated in a similar manner as was done for the
case of the mean density.  In Appendix~\ref{app:corr}, we provide a
detailed derivation of the terms $\left<n_{\g 1}n_{\g 2}\right>$,
$\left<n_{\s 1}n_{\g 2}\right>$, $\left<n_{\g 1}n_{\s 2}\right>$ and
$\left<n_{\s 1}n_{\s 2}\right>$.  On substituting
Eqns~(\ref{eq:ngng3A}), (\ref{eq:ngng3B}), (\ref{eq:ngns}) and
(\ref{eq:nsns}) into \Eqn{eq:FF} and on integrating over the delta
functions, we find that the correlation function may be written as the
sum of three terms:
\ba 
\left<\Fg(\br_1)\Fg(\br_2)\right> & = &  \prod_{i=1}^2 \left\{ 
\int \dx_i dM_i \nbar(M_i) b(M_i)  N^{(1)}_{\g}(M_i) 
{\mathcal W}^U_{(1)}(\br_i,\bx_i,M_i) \right\} \xi(|\bx_1-\bx_2|) \nn \\
& &  +(1+\alpha) \int d^3x \, dM \nbar(M) N_{\g}^{(2)}(M)
{\mathcal W}^U_{(1)}(\br_1,\bx, M){\mathcal W}^U_{(1)}(\br_2,\bx, M) \nn \\
& & +(1+\alpha) \int d^3x \,dM  \nbar(M) N^{(1)}_{\rm g}(M) 
{\mathcal W}^U_{(2)}(\br_1,\bx, M) \dirac(\br_1-\br_2)\label{eq:xihm2} \ ,
\ea
where $b(M)$ is the large-scale linear bias of dark matter haloes,
$\xi(\bx)$ is the dark matter correlation function, and
$N_{\g}^{(2)}(M)$ is the second factorial moment of the galaxy numbers
(for more details on these quantities see Appendix~\ref{app:corr}). In
the above expression we have also defined the quantity:
\be 
   {\mathcal W}^U_{(l)}(\br,\bx, M)\equiv U^l(\br-\bx|M){\mathcal W}_{(l)}(\br,\bx,M) \ ,
\label{eq:weight_def}
\ee 
with
\be {\mathcal W}_{(l)}(\br,\bx,M) \equiv \frac{1}{A^{l/2}}\int dL 
\Phi(L|M) \Theta(\br|L)  w^l(\br,L,\bx,M) 
\label{eq:W} \ .\ee

Based on the above analysis, we see that an obvious estimator for the
$\Fg$ correlation function is,
\be 
\hat{\xi}_{\Fg}(\br)\equiv \int \dr' \Fg(\br')\Fg(\br+\br') 
\hspace{0.3cm} ; \hspace{0.3cm}(\br\ne 0) \ .
\ee
The expectation of the estimator is:
\ba
\left<\hat{\xi}_{\Fg}(\br)\right> & = &
\int \prod_{i=1}^{2} \left\{\dx_i \,dM_i \nbar(M_i)b(M_i) N^{(1)}_{\g}(M_i)\right\}
\xi(|\bx_1-\bx_2|) \int \dr' {\mathcal W}^U_{(1)}(\br',\bx_1, M_1)
{\mathcal W}^U_{(1)}(\br+\br',\bx_2, M_2) \nn \\
& &  +(1+\alpha) \int \dx\, dM \nbar(M) N_{\g}^{(2)}(M) \int \dr'\,  
{\mathcal W}^U_{(1)}(\br',\bx, M) \,{\mathcal W}^U_{(1)}(\br+\br',\bx, M)
\hspace{0.3cm} ; \hspace{0.3cm}(\br\ne 0)
.\label{eq:xihm2Expec} 
\ea
In general $\hat{\xi}_{\Fg}$ is a biased estimator for the matter
correlation function $\xi$. Thus, in order to make a robust comparison
between theory and observations, one must either compute the theory
predictions as in \Eqn{eq:xihm2} or generate Monte-Carlo mock samples
and use the same estimator to compare theory and observation.


\subsection{An estimator for the matter correlation function in the large-scale limit}
\label{ssec:xiLS}


In the large-scale limit, the clustering of galaxies can be used to
obtain an unbiased estimate of the matter correlation function.  To
see this note that, since the dark matter haloes in simulations are
cuspy, on large scales the mass- or galaxy-number-normalised density
profiles of galaxies behave approximately as Dirac delta functions. We
shall therefore take
\be 
U^{\rm LS}(\br|M) \rightarrow \dirac(\br)  \ ,
\ee
and on implementing this in \Eqn{eq:xihm2Expec}, we find after
integrating over the Dirac delta functions:
\ba
\left<\hat{\xi}_{\Fg}(\br)\right> & \approx &
\xi(\br) \int \prod_{i=1}^{2} \left\{ dM_i \nbar(M_i)b(M_i) N^{(1)}_{\g}(M_i)\right\}
\int \dr' \,{\mathcal W}_{(1)}(\br',\br', M_1)\,{\mathcal W}_{(1)}(\br+\br',\br+\br', M_2)
\ ; \hspace{0.1cm} (\br\ne0)
\label{eq:xihm3} 
\ea
In this limit there are a number of interesting things that happen:
firstly, the weight function has now become independent of the halo
positions, i.e. we no longer differentiate between galaxy and halo
positions, taking them to be the same. Hence, we may now write:
\ba 
w(\br,L,\bx,M) \rightarrow w^{\rm LS}(\br,L,M)  \ \ ; \ \ 
{\mathcal W}_{(l)}(\br,\bx,M) \rightarrow {\mathcal W}^{\rm LS}_{(l)}(\br,M)  \ .\ea
Secondly, the dark matter correlation function has separated out and
so we may easily invert \Eqn{eq:xihm3} to obtain an unbiased estimate
for the dark matter clustering. The estimator is:
\be
\hat{\xi}(\br) \approx \frac{\hat{\xi}_{\Fg}(\br)}{\Sigma_0(\br)} \hspace{0.3cm} ; 
\hspace{0.3cm} \Sigma_0(\br) \equiv 
\int \dr' \,{\mathcal G}^{(1)}_{(1,1)}(\br')\,{\mathcal G}^{(1)}_{(1,1)}(\br+\br') 
\ \  \ ; \ \ \ (\br\ne0) \label{eq:xiest}
\ee
where we have defined the new set of weighted window functions:
\be 
{\mathcal G}^{(n)}_{(l,m)}(\br) \equiv 
\int dM \nbar(M) b^{m}(M) N^{(n)}_{\g}(M) \left[{\mathcal W}^{\rm LS}_{(l)}(\br,M)\right]^n
  \ . \label{eq:G}
\ee
The function $\Sigma_0(\br)$ represents the correlation function of
the averaged survey window functions. Our correlation function
estimator \Eqn{eq:xiest}, is therefore a generalization of that of
\citep{LandySzalay1993}.


\subsection{The galaxy power spectrum}


We now turn to the Fourier space dual of the correlation function --
the galaxy power spectrum. To obtain this let us begin by defining our
3D Fourier transform convention for a function $B$ and its inverse as:
\ba
\tilde{B}(\bk) \equiv  \int \dr B(\br){\rm e}^{i\,\bk\cdot\br}
\hspace{0.5cm} 
\Leftrightarrow
\hspace{0.5cm} 
 B(\br) =  \int \frac{\dk}{(2\pi)^3} \tilde{B}(\bk)  {\rm e}^{-i\, \bk\cdot\br}  \ .
\nn\ea
We distinguish real- and Fourier-space quantities that share the same
symbol through use of the tilde notation. We also define the power
spectrum $P_{B}(k)$ of any infinite statistically
homogeneous random field $\tilde{B}(\bk)$ to be:
\be \left<\tilde{B}(\bk)\tilde{B}(\bk')\right>
\equiv (2\pi)^3\delta^{\rm D}(\bk+\bk')P_{B}(\bk) \ . \nn \ee
Note, if the field $B$ were statistically isotropic, the power
spectrum would simply be a function of the scalar $k$. In addition
the power spectrum and two-point correlation function of the field $B$
form a Fourier pair:
\be \xi_{B}(|\bx-\bx'|) = \int
\frac{\dq}{(2\pi)^3}\int \frac{\dq'}{(2\pi)^3} 
P_{B}(\bq)(2\pi)^3\dirac(\bq+\bq'){\rm e}^{-i\bq\cdot\bx}{\rm
  e}^{-i\bq'\cdot\bx'} \ . \nn \ee
With these definitions in hand we may now transform \Eqn{eq:xihm2},
and on considering the case where $\bk_2=-\bk$, we find that the
expectation of the square of the amplitude of the Fourier modes of
$\tilde{\Fg}$ is given by:
\ba 
\left<|\tilde\Fg(\bk)|^2\right> & = & \int \frac{\dq}{(2\pi)^3} P(\bq)
\prod_{i=1}^2 \left\{\int dM_i \,\nbar(M_i) b(M_i)  N^{(1)}_{\g}(M_i)\right\}
\tilde{\mathcal W}^U_{(1)}(\bk,-\bq, M_1) \tilde{\mathcal W}^U_{(1)}(-\bk, \bq, M_2) \nn \\
& & +(1+\alpha)\int \frac{\dq}{(2\pi)^3} dM \,\nbar(M) N_{\g}^{(2)}(M)
\tilde{\mathcal W}^U_{(1)}(\bk,-\bq, M)\tilde{\mathcal W}^U_{(1)}(-\bk, \bq, M) \nn \\
& & +(1+\alpha) \int \dr \,\dx \,dM \, \nbar(M) N^{(1)}_{\rm g}(M) 
\tilde{\mathcal W}^U_{(2)}(\br, \bx, M) 
\label{eq:Phm2} \ ,
\ea
where $P(\bq)$ is the matter power spectrum and we have set the
Fourier transform of the effective survey window function to be:
\be \tilde{\mathcal W}^U_{(l)}(\bk, \bq, M)\equiv 
 \int \dr \dx \,{\mathcal W}^U_{(l)}(\br, \bx, M){\rm e}^{i\bk\cdot\br} {\rm e}^{i\bq\cdot\bx} \ .
\ee
As in the case of the correlation function of the field $\Fg$, its
power spectrum does not provide a direct estimate of the matter power
spectrum. 


\subsection{The power spectrum in the large-scale limit}

Let us now consider the power spectrum of $\Fgt$ in the large-scale
limit. As discussed in \S\ref{ssec:xiLS} we expect the density
profiles to behave like Dirac delta functions in real space, in
Fourier space the density profiles on large scales simply obey:
$\tilde{U}(\bk|M)\xrightarrow{k\rightarrow0} 1$. Under this condition
\Eqn{eq:Phm2} simplifies to:
\be
\left<|\Fgt(\bk)|^2\right> \approx \int \frac{\dq}{(2\pi)^3} P(\bq) \left|\tilde
{\mathcal G}^{(1)}_{(1,1)}(\bk-\bq)\right|^2 +  P_{\rm shot} \ ,
\label{eq:PhmLS}
\ee
where the second term on the right-hand side is a $k$-independent
effective shot-noise term,
\be
P_{\rm shot} \equiv  (1+\alpha)\left[\tilde{\mathcal G}^{(2)}_{(1,0)}({\bf 0})+
\tilde{\mathcal G}^{(1)}_{(2,0)}({\bf 0})\right] \ .
\label{eq:Pshot}
\ee
In the limit that the survey volume is large, the window functions
$\tilde{\mathcal G}^{(n)}_{(l,m)}(\bk)$ will be very narrowly peaked
around $\bk={\bf 0}$. Provided the matter power spectrum is a
smoothly-varying function of scale, the window functions
$\tilde{\mathcal G}^{(n)}_{(l,m)}(\bk)$ will behave in a way that is
similar to the Dirac delta function. Hence, Eq.~(\ref{eq:PhmLS})
becomes:
\be
\left<|\Fgt(\bk)|^2\right> \approx  P(\bk) \int  \frac{\dq}{(2\pi)^3}  
\left|\tilde{\mathcal G}^{(1)}_{(1,1)}(\bk-\bq)\right|^2 + P_{\rm shot} \ .
\ee

Let us focus on the integral factor on the right-hand-side of the
above expression. If we now perform the transformation of variables
$\bq\rightarrow\bk-\bq$ and use Parseval's theorem, we find:
\be
\int \frac{\dq}{(2\pi)^3} \left|\tilde{\mathcal G}^{(1)}_{(1,1)}(\bk-\bq)\right|^2 
= \int \frac{\dq}{(2\pi)^3} \left|\tilde{\mathcal G}^{(1)}_{(1,1)}(\bq)\right|^2 
= \int \dr \left|{\mathcal G}^{(1)}_{(1,1)}(\br)\right|^2\ .
\nn \ee
Upon back-substitution of \Eqn{eq:G} into the above expression, we
obtain:
\be 
 \int \dr \left|{\mathcal G}^{(1)}_{(1,1)}(\br)\right|^2
 =  \frac{1}{A}
 \int \dr
\left[ \int  dM \nbar(M) N^{(1)}_{\g}(M) b(M) 
 \int  dL  \Phi(L|M) \Theta(\br|L) w(\br,L,M)\right]^2 \ .
\nn \ee 
Note that we have not yet specified the parameter $A$, which we now
take to be:
\be
A \equiv  \int \dr
\left[ \int  dM \nbar(M) N^{(1)}_{\g}(M) b(M)  \int  dL  \Phi(L|M) \Theta(\br|L) w(\br,L,M)\right]^2 \ .
\label{eq:norm_def}
\ee
Note, we will now drop the super-script LS notation for $w$ and
${\mathcal W}_{(l)}$, since for the remainder of this study we shall
be working only in the large-scale limit. Thus, our estimator for the
matter power spectrum can be written simply:
\be
\hat{P}(\bk)  =  |\Fgt(\bk)|^2 - P_{\rm shot} \ . \label{eq:Pk}
\ee
If our modelling of the galaxy distribution is correct, then the above
estimator constitutes the only unbiased estimator of the matter power
spectrum. Before proceeding further, note that in the above expression
we have obtained the power spectrum per mode. In fact, we are more
interested in its band-power estimate. Hence, our final estimator in
the large-scale and large-volume limit is:
\be
 \overline{P}(k_i) = 
\frac{1}{V_i}\int_{V_i} \dk \hat{P}(\bk) = \frac{1}{V_i}\int_{V_i} 
\dk |\tilde\Fg(\bk)|^2 - P_{\rm shot}\label{eq:PggEstLS} \ ,
\ee
where in the above we have summed over all modes in a $k$-space shell
of thickness $\Delta k$ and volume
\be 
V_{i} \equiv \int_{V_i} \dk 
= 4\pi \int^{k_i+\Delta k/2}_{k_i-\Delta k/2} k^2 dk 
= 4\pi k_i^2 \Delta k
\left[1+\frac{1}{12}\left(\frac{\Delta k}{k_i}\right)^2\right] \ .
\ee
%

\section{Statistical fluctuations in the galaxy power spectrum}
\label{sec:fluctuations}

In order to obtain the optimal estimator we need to know how the
signal-to-noise (hereafter $\mathcal \SN$) varies when we vary the
shape of our weight function $w$. Thus, we need to understand the
noise properties of our power spectrum estimator, i.e. compute its
covariance matrix. The covariance matrix of two band-power estimates
is given by:
\ba 
{\rm Cov}\!\left[\overline{P}(k_i),\overline{P}(k_j)\right] 
& \equiv & 
\left< \overline{P}(k_i)\overline{P}(k_j)\right> -
\left<\overline{P}(k_i)\right>\left<\overline{P}(k_j)\right> 
=\frac{1}{V_{i}}\int_{V_i} \dk_1 \frac{1}{V_j}\int_{V_j} \dk_2\, 
{\rm Cov}\!\left[\hat{P}(\bk_1),\hat{P}(\bk_2)\right]  \ , \label{eq:covdef}
\nn \ea
where the last factor on the right-hand side of the above expression
is the covariance of the power in two separate Fourier modes. For the
case of large survey volumes and in the large-scale limit the matter
power spectrum is given by \Eqn{eq:PggEstLS}. Hence,
\ba 
{\rm Cov}\!\left[\hat{P}(\bk_1),\hat{P}(\bk_2)\right] & \approx &  
{\rm Cov}\!\left[|\Fgt(\bk_1)|^2,|\Fgt(\bk_2)|^2\right]
=\left<|\Fgt(\bk_1)|^2|\Fgt(\bk_2)|^2\right>-
\left<|\Fgt(\bk_1)|^2\right> \left<|\Fgt(\bk_2)|^2 \right> \label{eq:CovFF}  \ .
\ea
The approximation in the equation above follows from the discussion in
Appendix~\ref{app:covm}.

In Appendix~\ref{app:covGen}, we derive a general expression for the
covariance matrix of $|\Fgt(\bk_1)|^2$ and $|\Fgt(\bk_2)|^2$, with all
$n$-point connected spectra and shot-noise terms included -- this is
obtained by combining \Eqns{eq:covF0}{eq:covF} with \Eqn{eq:four1}.
Under the assumption of a Gaussian matter density field, our general
expression simplifies to \Eqn{eq:GaussCov}. Furthermore, we also show
in Appendix~\ref{app:B3} that in the large-scale limit
\Eqn{eq:GaussCov} can be written as:
\ba
{\rm Cov}\!\left[|\Fgt(\bk_1)|^2,|\Fgt(\bk_2)|^2\right] &\hspace{-0.1cm} =\hspace{-0.1cm} &
\left|\int \frac{\dq}{(2\pi)^3} P(\bq) \tilde{\mathcal G}^{(1)}_{(1,1)}(\bk_1+\bq) 
\tilde{\mathcal G}^{(1)}_{(1,1)}(\bk_2-\bq) + (1+\alpha)
\left[\tilde{\mathcal G}^{(1)}_{(2,0)}(\bk_1+\bk_2) + \tilde{\mathcal G}^{(2)}_{(1,0)}(\bk_1+\bk_2)\right] 
\right|^2 \nn \\
&  \hspace{-2.5cm} + & \hspace{-1.2cm}
\left|\int \frac{\dq}{(2\pi)^3} P(\bq) \tilde{\mathcal G}^{(1)}_{(1,1)}(\bk_1+\bq) 
\tilde{\mathcal G}^{(1)}_{(1,1)}(-\bk_2-\bq) + (1+\alpha)
\left[\tilde{\mathcal G}^{(1)}_{(2,0)}(\bk_1-\bk_2) + \tilde{\mathcal G}^{(2)}_{(1,0)}(\bk_1-\bk_2)
\right] \right|^2 \ .
\label{eq:CovLS}
\ea
In the limit where the survey volume is large, the functions
$\tilde{\mathcal G}^{(n)}_{(l,m)}$ are very narrowly peaked around
$k=0$. Furthermore, if the power spectrum does not rapidly vary over
the scale of the effective window function, then we may treat it as a
constant in \Eqn{eq:CovLS}. Thus,
\ba
{\rm Cov}\!\left[|\Fgt(\bk_1)|^2,|\Fgt(\bk_2)|^2\right]
& \approx &
\left| P(\bk_1) \tilde{\mathcal Q}^{(1,1)}_{(1,1|1,1)}(\bk_1+\bk_2)
+ (1+\alpha)\left[\tilde{\mathcal Q}^{(2)}_{(1|0)}(\bk_1+\bk_2) +
\tilde{\mathcal Q}^{(1)}_{(2|0)}(\bk_1+\bk_2)\right]\right|^2
\nn \\
& & 
+\left|P(\bk_1)\tilde{\mathcal Q}^{(1,1)}_{(1,1|1,1)}(\bk_1-\bk_2)
+(1+\alpha)\left[\tilde{\mathcal Q}^{(2)}_{(1|0)}(\bk_1-\bk_2) 
+\tilde{\mathcal Q}^{(1)}_{(2|0)}(\bk_1-\bk_2)\right]\right|^2
\label{eq:CovLS2} \ ,
\ea
where we have introduced the functions:
\be {\mathcal Q}_{(l_1,l_2|m_1,m_2)}^{(n_1,n_2)}(\br)
\equiv {\mathcal G}_{(l_1,m_1)}^{(n_1)}(\br) {\mathcal G}_{(l_2,m_2)}^{(n_2)}(\br) \ ,
\ee
and made use of the convolution theorem to write their Fourier transforms:
\be
{\mathcal Q}_{(l_1,l_2|m_1,m_2)}^{(n_1,n_2)}(\bk) = \int \frac{\dq}{(2\pi)^3}
{\mathcal G}_{(l_1,m_1)}^{(n_1)}(\bq) {\mathcal G}_{(l_2,m_2)}^{(n_2)}(\bk-\bq) \ .
\nn \ee
Note that we also used the trivial identity ${\mathcal
  Q}^{(n)}_{(l|m)}={\mathcal G}^{(n)}_{(l,m)}$.  

Returning to \Eqn{eq:covdef}, we find that after substitution of
\Eqn{eq:CovLS2} into \Eqn{eq:CovFF}, the bin-averaged estimates of the
power spectrum can be written:
\be
{\rm Cov}\!\left[\overline{P}(k_i),\overline{P}(k_j)\right] =
2 \int_{V_{i}} \frac{\dk_1}{V_i} 
\int_{V_{j}}  \frac{\dk_2}{V_j} 
\left| P(\bk_1) \tilde{\mathcal Q}^{(1,1)}_{(1,1|1,1)}(\bk_1\hspace{-0.1cm}+\hspace{-0.1cm}\bk_2)
+ (1+\alpha)\left[
\tilde{\mathcal Q}^{(2)}_{(1|0)}(\bk_1+  \bk_2)
+\tilde{\mathcal Q}^{(1)}_{(2|0)}(\bk_1+\bk_2)
\right]\right|^2.
\label{eq:CovLS3}
\ee
\Eqn{eq:CovLS3} follows from the integrals over $\bk_2$ in
\Eqn{eq:CovLS2} being invariant under the transformation
$\bk_2\rightarrow-\bk_2$.  Furthermore, if the $k$-space shells are
narrow compared to the scale over which the power spectrum varies,
then the shell-averaged power spectrum can be pulled out of the
integrals. In Appendix~\ref{app:qresult} we detail the computation of
\Eqn{eq:CovLS3} and show that the covariance can be reexpressed as:
\be
{\rm Cov}\!\left[\overline{P}(k_i),\overline{P}(k_j)\right] = 
\frac{2(2\pi)^3}{V_{i}}{\overline P}^2(k_i)\delta^{K}_{i,j}
\int \dr \left\{ \left[{\mathcal G}^{(1)}_{(1,1)}(\br)\right]^2 + 
\frac{(1+\alpha)}{\overline{P}(k_i)} \left[ 
{\mathcal G}^{(2)}_{(1,0)}(\br) + {\mathcal G}^{(1)}_{(2,0)}(\br)\right]
\right\}^2 \ ,
\label{eq:cov_final}
\ee
with the functions ${\mathcal G}^{(n)}_{(l,m)}(\br)$ defined by
\Eqn{eq:G}.
\section{Optimal estimator}
\label{sec:optimal}
Our aim is to find the optimal weighting scheme that will maximize the
$\SN$ ratio on a given band-power estimate of the galaxy power
spectrum.
\subsection{The optimal weight equation}
To begin, note that maximizing the $\SN$ ratio is equivalent to
minimizing its inverse, the noise-to-signal ratio $\NS$. The square of
the latter can be expressed as:
\ba 
F[w(\br,L,M)]\equiv\frac{\sigma^2_{P}(k_i)}{ \overline{P}^2(k_i)} 
& = &
\frac{2(2\pi)^3}{V_{i}}
 \int \dr \left\{ \left[{\mathcal G}^{(1)}_{(1,1)}(\br)\right]^2 + 
\frac{(1+\alpha)}{\overline{P}(k_i)} \left[ 
{\mathcal G}^{(2)}_{(1,0)}(\br) + {\mathcal G}^{(1)}_{(2,0)}(\br)\right]
\right\}^2 
\label{eq:Func1} \ .
\ea
In the above expression, we have written the squared noise-to-signal
$F[w]$ as a functional of the weights $w(\br, L, M)$. The standard way
for finding the optimal weights is to perform the variation of the
functional $F$ with respect to the weights
$w(\br,L,M)$. Operationally, the functional variation of $F[w]$ is
carried out by comparing $F[w]$ with the functional obtained for
weight functions that possess a small path variation
$w(\br,L,M)\rightarrow w(\br,L,M)+\delta w(\br,L,M)$. This variation
can be defined:
\be
\delta F[w] \equiv F[w(\br,L,M) + \delta w(\br,L,M)] - F[w(\br,L,M)] =
\int \dr\,dL\,dM \left\{
\frac{\delta F}{\delta w(\br,L,M)}\right\} \delta w(\br,L,M) \ .
\ee
Extremisation means that the functional derivative is stationary
for small variations around the optimal weights:
\be \frac{\delta F}{\delta w(\br,L,M)}=0\ .\ee
Recall that the definition of the weights in \Eqn{eq:W} includes the
normalization constant $A$ specified by \Eqn{eq:norm_def}. Since the
normalization constant $A$ is itself a function of the weights, it
follows that $F[w]$ is in fact a ratio of two weight-dependent
functionals:
\be
F[w] \equiv  \frac{\mathcal{N}[w]}{\mathcal{D}[w]} \ , \label{eq:Func2}
\ee
with the definitions:
\ba
\mathcal{N}[w] & \equiv &  \int \dr \left\{ \left[\Gb^{(1)}_{(1,1)}(\br)\right]^2 + 
c \left[\Gb^{(2)}_{(1,0)}(\br) + \Gb^{(1)}_{(2,0)}(\br)\right]\right\}^2 \ ; 
\label{eq:N_def} \\
{\mathcal D}[w] & \equiv & A^2[w] = 
\left[ \int\dr \left[\Gb^{(1)}_{(1,1)}(\br)\right]^2 \right]^2 \ .
\label{eq:D_def} 
\ea
In the above, we introduced the scaled effective window functions:
\be
\overline{\mathcal G}^{(n)}_{(l,m)}(\br)=A^{n l/2} {\mathcal G}^{(n)}_{(l,m)}(\br) \ ,
\label{eq:Gscaled}
\ee
as well as the constant $c\equiv (1+\alpha)/\overline{P}(k_i)$, which
helps keep the equations as compact as possible. We also dropped the
overall constant $2(2\pi)^3/V_{i}$ from the functional
$\mathcal{N}[w]$, since it plays no role in the minimization process.
 
Minimizing $F[w]$ is equivalent to solving the functional problem:
\be
\frac{1}{\mathcal{D}[w]}\left(\delta\mathcal{N}[w]-\frac{\mathcal{N}[w]}
{\mathcal{D}[w]}\delta\mathcal{D}[w]\right)=0 \hspace{0.3cm}\Longleftrightarrow 
\hspace{0.3cm} \delta\mathcal{N}[w] - F[w] \delta \mathcal{D}[w] = 0.
\label{eq:optw1}
\ee
Therefore, to find the optimal weights satisfying \Eqn{eq:optw1}, we
first need to compute the variations of $\mathcal{N}$ and
$\mathcal{D}$ with a perturbation $\delta w$. This calculation is
outlined in Appendix~\ref{app:functionals}. Putting together
Eqns.~(\ref{eq:optw1}),~(\ref{eq:deriv_N}),~(\ref{eq:deriv_D}), we
arrive at the following general equation for the optimal weights:
\be
\left\{[\Gb^{(1)}_{(1,1)}(\br)]^2 
+ c \left[\Gb^{(2)}_{(1,0)}(\br) 
+\Gb^{(1)}_{(2,0)}(\br)\right]\right\}
\left\{\Gb^{(1)}_{(1,1)}(\br) b(M) + c\left[
w(\br, L, M) + \overline{\mathcal W}_1(\br, M) 
\beta(M) N_{\g}^{(1)}(M) \right]\right\}=
\Gb^{(1)}_{(1,1)}(\br)\,b(M) \ . 
\label{eq:optw2}
\ee
In the above, $\overline{\mathcal W}_1$ was introduced by
\Eqn{eq:Wbar}, and the function $\beta(M)$ specifies the relation
between the first and second factorial moments of galaxies in a halo
of mass $M$, as discussed in \cite{CooraySheth2002}:
\be
N_{\g}^{(2)}(M)=\beta(M)\left[N_{\g}^{(1)}(M)\right]^2 \ .
\label{eq:beta} \ee
Note that for a Poisson distribution, $\beta=1$, although we do not
make this assumption here.

On inspection of \Eqn{eq:optw2} we notice that, with the exception of
the weights $w(\br,L,M)$, none of the terms carries any explicit
dependence on the luminosity of the galaxies. We therefore conclude
that \emph{the optimal weights are independent of luminosity}. Hence,
without any loss of generality, we may now redefine the weights to be:
\be w(\br,L,M) \Rightarrow w(\br,M) \label{eq:Lweight} \ .\ee
One immediate consequence of this is that the functions
$\overline{\mathcal W}_{(l)}(\br,M)$ can now be written in the much
simplified form:
\be \overline{\mathcal W}_{(l)}(\br,M) = w^l(\br,M)
\int dL \Phi(L|M) \Theta(\br|L) = w^l(\br,M) {\mathcal S}(\br, M)  \ .
\nn \ee
In the above we have introduced the function:
\be
\mathcal{S}(\br, M) \equiv  \Theta(\bm\Omega)  \int^{\infty}_{0} dL \Theta(\chi|L) \Phi(L|M) =
  \Theta(\bm\Omega)  \int^{\infty}_{\Lmin(\chi)} dL \Phi(L|M) \ ,
\label{eq:S}  
\ee
with the second equality following from \Eqn{eq:ThetaInt}. ${\mathcal
  S}(\br, M)$ is the number of galaxies in a halo of mass $M$
observable at comoving distance $\br$ relative to the total number of
galaxies in that halo. The range of $S$ is the interval $[0, 1]$, and
it has the following limiting behaviour: for $M\ge M_{\rm min}$ we
have $\lim_{\chi\rightarrow0}{\mathcal S}(\br,M)=1$ and
$\lim_{\chi\rightarrow\infty}{\mathcal S}(\br,M)=0$; and for $M<M_{\rm
  min}$ we have ${\mathcal S}(\br,M)=0$, where $M_{\rm min}$ is the
minimum halo mass required for a dark matter halo to be able to host a
galaxy. Note that for a volume-limited survey $\mathcal{S}={\rm
  constant}$.

A further consequence of \Eqn{eq:Lweight} is that the effective survey
window functions given by \Eqn{eq:Gscaled} reduce to:
\be \overline{\mathcal G}^{(n)}_{(l,m)}(\br) \equiv \int dM \nbar(M)
b^{m}(M) N^{(n)}_{\g}(M) \left[w^l(\br,M) {\mathcal S}(\br, M)\right]^n \ .
\label{eq:Gscaled2} 
\ee
Implementing these considerations in \Eqn{eq:optw2}, we
arrive at the equation governing the optimal weights:
\be
\left\{\left[\Gb^{(1)}_{(1,1)}(\br)\right]^2 + c \left[\Gb^{(2)}_{(1,0)}(\br) + \Gb^{(1)}_{(2,0)}(\br)\right]\right\}
\left\{\Gb^{(1)}_{(1,1)}(\br) + c\frac{w(\br, M)}{ b(M)} 
\left[1 +  \beta(M) N_{\g}^{(1)}(M)   \mathcal{S}(\br, M)\right]\right\} = \Gb^{(1)}_{(1,1)}(\br) \ .
\label{eq:optw3}
\ee
%
\subsection{The optimal weights}

We now seek a general solution for the weight equation \Eqn{eq:optw3}.
To begin, we notice that the only part of the weight equation that
carries any mass dependence is the second bracket on the
left-hand side of \Eqn{eq:optw3}. If we set the radial vector to a
constant $\br=\br_0$, then the optimal weights at fixed position
inside the angular mask must have the mass dependence:
\be w(\br_0,M)  \propto  \frac{b(M)}
    {\frac{}{} \hspace{-0.1cm} 1 + \beta(M) N_{\g}^{(1)}(M)
        {\mathcal S}(\br_0,M)}
\label{eq:optw_mass} \ .
\ee
The weights are therefore proportional to the bias of the dark matter
halo in which the galaxy is hosted and inversely proportional to the
factor $[1+\beta(M) N_{\g}^{(1)}(M) {\mathcal S}(\br_0,M)]$.  Since
this last term depends on the galaxy selection function ${\mathcal
  S}$, the weight function is not separable in position and mass, as
was found by SM14 for the case of optimal weighting of a sample of
galaxy clusters. Nevertheless, without any loss of generality, we can
factor out this part of the weight function from the general weight
solution:
\be
w(\br, M)=\tilde{w}(\br) \left[\frac{b(M)}{\frac{}{} \hspace{-0.1cm}1 
+ \beta(M) N_{\g}^{(1)}(M) {\mathcal S}(\br, M)}\right]
\label{eq:optw_gen1} \ ,
\ee
where $\tilde{w}(\br)$ is a function of position only that needs to be
determined.  It is clear from the above equation that the term on the
right-hand side encompasses the whole mass dependence of the optimal
weights. If we now reinsert this expression into \Eqn{eq:optw3} we see
that the weight equation reduces to:
\be
\left\{\left[\Gb^{(1)}_{(1,1)}(\br)\right]^2 + c \left[\Gb^{(2)}_{(1,0)}(\br) + \Gb^{(1)}_{(2,0)}(\br)\right]\right\}
\left\{\Gb^{(1)}_{(1,1)}(\br) + c\tilde{w}(\br)\right\} = \Gb^{(1)}_{(1,1)}(\br) \ .
\label{eq:optw4}
\ee
In order to proceed further we need to recompute the effective survey
window functions $\Gb$ functions from \Eqn{eq:Gscaled2} with the new
weight function \Eqn{eq:optw_gen1}. It is straightforward to show
that:
\ba 
\Gb^{(1)}_{(1, 1)}(\br) & = & \tilde{w}(\br) \nbar_{\rm eff}(\br)\, , \mbox{where}
\hspace{0.2cm} n_{\rm eff}(\br)\equiv \int dM \nbar(M) b^2(M) 
\left[\frac{ N_{\g}^{(1)}(M)S(\br, M)} {1+\beta(M) N_{\g}^{(1)}(M) S(\br, M)}\right]\ ;  \\ 
\Gb^{(2)}_{(1, 0)}(\br) & = & \tilde{w}^2(\br) \int dM \nbar(M) b^2(M) \beta(M) \left[\frac
{N_{\g}^{(1)}(M) S(\br, M)}{1+\beta(M) N_{\g}^{(1)}(M) S(\br, M)}\right]^2  \ ; \\
\Gb^{(1)}_{(2, 0)}(\br) & = & \tilde{w}^2(\br) \int dM \nbar(M) b^2(M)  \frac
{N_{\g}^{(1)}(M)S(\br, M)}{\left[1+\beta(M) N_{\g}^{(1)}(M) S(\br, M)\right]^2}  \ .
\label{eq:Ggen} 
\ea
From the above equations we also notice the useful relation:
\be
\Gb^{(2)}_{(1, 0)}(\br) + \Gb^{(1)}_{(2, 0)}(\br) = \tilde{w}(\br) 
\Gb^{(1)}_{(1, 1)}(\br) = \tilde{w}^2(\br) \nbar_{\rm eff}(\br) \ .
\nn \ee
Replacing all these ingredients back into \Eqn{eq:optw4}, after a
little algebra we find that $\tilde{w}$ has the solution:
\be
\tilde{w}(\br)=1/\left[c + \nbar_{\rm eff}(\br)\right] \ .
\label{eq:optw_r} 
\ee

On putting together \Eqns{eq:optw_gen1}{eq:optw_r}, back substituting
the constant $c=(1+\alpha)/\overline{P}_i$, we arrive at the general
solution for the optimal weights:
\be
w(\br, M)=\frac{b(M)}{\left[\frac{}{} \hspace{-0.1cm} 1 + \beta(M) 
N_{\g}^{(1)}(M) {\mathcal S}(\br, M)\right]}\frac{1}
{\left[(1+\alpha) + \nbar_{\rm eff}(\br) \overline{P}_i \right]} \ .
\label{eq:optw_gen2}
\ee
This expression is the central result of this paper.
Before inspecting how these weights behave for a specific case, a
number of interesting points may be noted. First, if we were able to
identify galaxies in a survey whose host halo masses were drawn from
some narrow range, then the first factor in \Eqn{eq:optw_gen2} would
be constant and so the weights would revert back to a scheme that
structurally resembles FKP, although with a different effective number
density. Second, we note that there is no natural limit where the
above weighting scheme follows that derived by PVP. This dissimilarity
emphasises how important an effect modifications of the underlying
model assumptions can be on the matter power spectrum estimation and
optimisation.


\begin{figure}
\centering{
  \includegraphics[width=8.6cm,clip=]{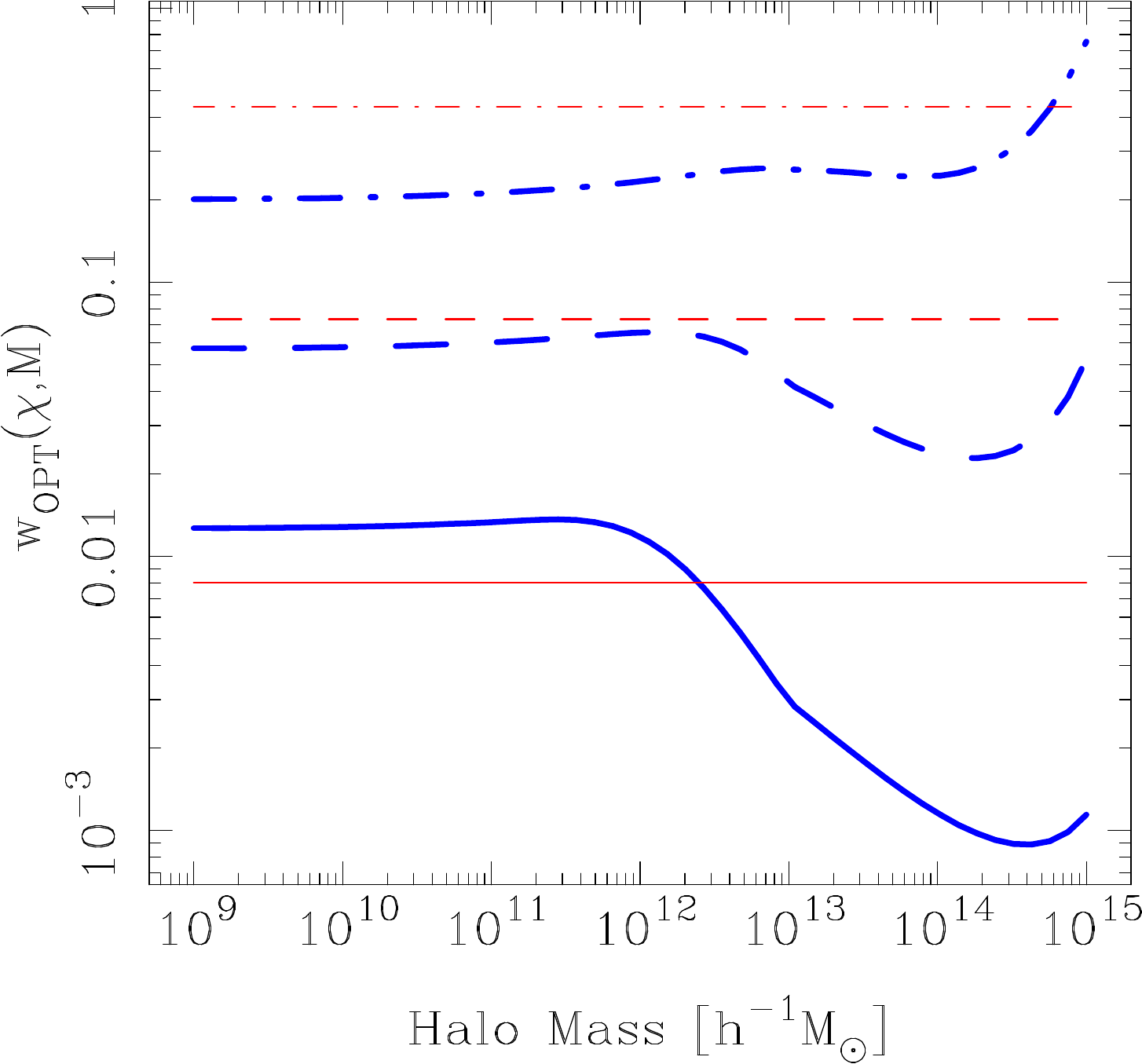}\hspace{0.2cm}
  \includegraphics[width=8.6cm,clip=]{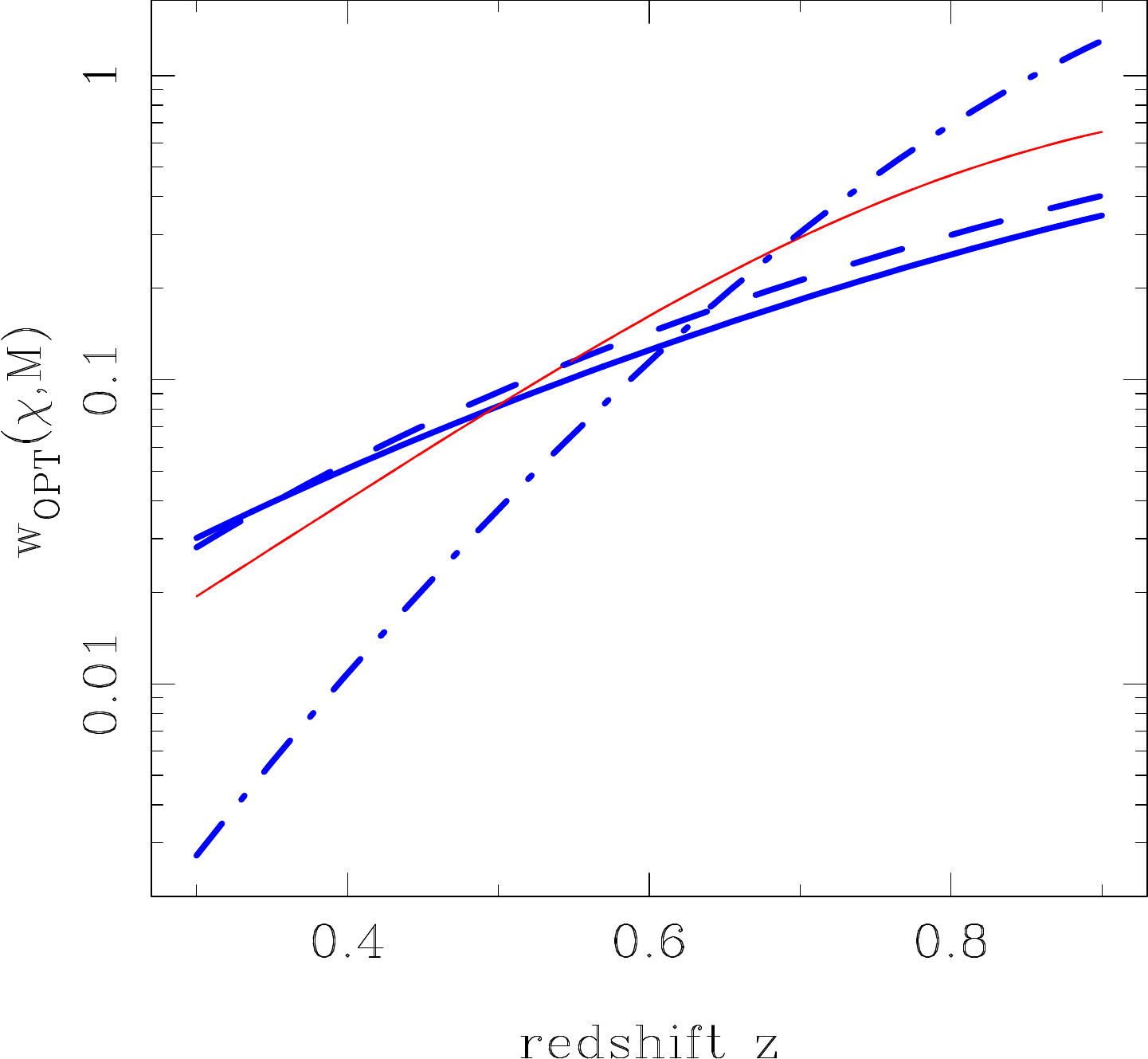}}
\caption{\small{{\bf Left panel}: evolution of the optimal weights in
    a fiducial flux-limited survey as a function of halo mass. The
    thick blue and thin red lines represent the optimal weights and
    the FKP weights, respectively. The solid, dashed and dot-dashed
    line styles denote the results for increasing $\chi$,
    respectively. {\bf Right panel}: evolution of the optimal weights
    as a function of redshift for several halo masses. Thick blue and
    thin red lines denote optimal and FKP weights. The solid, dashed
    and dot-dashed lines show the results for galaxy-, group- and
    cluster-scale halo masses, respectively. We have taken the
    flux-limit to be $m_{\rm lim}=22$.} \label{fig:weights}}
\end{figure}


Figure~\ref{fig:weights} demonstrates how the optimal weights vary as
a function of the galaxies host halo mass and redshift. At low
redshifts, the galaxy selection ${\mathcal S}\rightarrow 1$. For
galaxies that are hosted by low-mass haloes, $N^{(1)}_{\g}(M)<1$ and
so $w\propto b(M)$. On the other hand, for the high mass clusters
$N^{(1)}_{\g}(M)\gg1$, and \mbox{$w(\br,M)\propto
  b(M)/N^{(1)}_{\g}(M)$}. Hence we would expect the galaxies in
low-mass haloes to be weighted more strongly than those in high-mass
haloes, since the bias is rather a slowly evolving function of halo
mass.  At higher redshift, we would expect that this trend reverses,
since $\{{\mathcal S},\nbar_{\rm eff}\}\rightarrow 0$ and so the
weights effectively follow the bias of the host haloes. These trends
are exactly what is seen in the figure. For reference,
\Fig{fig:weights} also compares the optimal weights with the original
FKP weight function, given by: $w_{\rm
  FKP}(\br)\propto\left[1+\nbar(\br)P(k)\right]^{-1}$
\footnote{Note
  that in order to evaluate the weight functions we took $m_{\rm
    lim}=22$ and adopted the CLF model of \citet{Yangetal2003} to
  compute ${\mathcal S}(\br,M)$ and $N^{(1)}_{\rm g}(M)$.  For the
  function $\beta(M)$ we employed the model presented in
  \citet{CooraySheth2002} derived from semi-analytic galaxies.}.
The upturn at large masses in the left panel, is driven by the mass
dependence of the ratio $b(M)/N(M)$. For large masses, $b(M)$ is a
steep function of mass $b(M)\propto M^{1.5}$ \citep{SeljakWarren2004},
whereas for most halo occupation distribution models, $N(M)\propto
M^{1}$ \citep{Zehavietal2011}. Hence, leading to an upturn for large
masses.

\subsection{Time evolution of the optimal weights}

Before moving on, we briefly discuss the redshift dependence of the
optimal weights in \Eqn{eq:optw_gen2}. So far, we have considered that
$\nbar(M)$, $b(M)$, $\Phi(L|M)$, $N^{(1)}(M)$, $\beta(M)$ and
$\xi(\br)$ are all independent of time (here we will parameterise time
evolution through the comoving distance $\chi$). This is approximately
correct if the survey volume is sufficiently small so that these
functions do not evolve appreciably over the survey. In general,
however, they are time dependent.  Therefore we would have:
$\nbar(M)\rightarrow \nbar(M,\chi)$, $b(M)\rightarrow b(M,\chi)$,
$\Phi(L|M)\rightarrow \Phi(L|M,\chi)$, $N^{(1)}(M)\rightarrow
N^{(1)}(M,\chi) $, $\beta(M)\rightarrow \beta(M,\chi)$ and
$\xi(\br_1-\br_2)\rightarrow
\xi(\br_1-\br_2,\chi_1,\chi_2)=D(\chi_1)D(\chi_2)\xi(\br_1-\br_2)$.

In the last equality we have assumed that the correlation function
obeys linear theory, hence the resulting growth factors. Working
under this assumption, we redefine the $\Gb$ functions to absorb the
growth factors:
\be \overline{\mathcal G}^{(n)}_{(l,m)}(\br) \equiv  \int dM \nbar(M,\chi)
\left[D(\chi)b(M,\chi)\right]^{m} N^{(n)}_{\g}(M,\chi) \left[w^l(\br,M) {\mathcal S}(\br,M)\right]^n \ 
\label{eq:Gscaled3} \ .
\ee
Formally this is equivalent to redefining the halo bias parameter:
$b(M)\rightarrow D(\chi) b(M,\chi)$, and we prefer this latter
approach. Thus, \Eqn{eq:optw_gen2} becomes:
\be w(\br, M)=\frac{D(\chi)b(M,\chi)}{\left[\frac{}{} \hspace{-0.1cm} 1 +
    \beta(M,\chi) N_{\g}^{(1)}(M,\chi) {\mathcal S}(\br,
    M)\right]}\frac{1} {\left[(1+\alpha) +
    \overline{P}_i \,\nbar_{\rm eff}(\chi)\right]} \ ,
\label{eq:optw_gen3}
\ee
where the new effective number density is:
\be \nbar_{\rm eff}(\br)\equiv \int dM \nbar(M,\chi)
D^2(\chi)b^2(M,\chi) \left[\frac{ N_{\g}^{(1)}(M,\chi)S(\br, M)}
  {1+\beta(M,\chi) N_{\g}^{(1)}(M,\chi) S(\br, M)}\right]\ .  \ee
%

\section{Information content of galaxy clustering}\label{sec:Fisher}

The ability of a set of band-power estimates of the galaxy power
spectrum to constrain the cosmological model can, theoretically, be
determined through construction of the Fisher information
matrix. Under the assumption that the density field is Gaussianly
distributed, one finds that the power spectrum for a given Fourier
mode is exponentially distributed about the mean power, and that the
band-power estimate is $\chi^2$ distributed
\citep{Takahashietal2011}. Owing to the central limit theorem, in the
limit of a large number of Fourier modes per $k$-space shell, the
power spectrum estimates thus approach the Gaussian
distribution. Under the assumption that the power spectrum estimator
is Gaussianly distributed, it can be shown that the Fisher matrix has
the form \citep{Tegmarketal1997,Tegmark1997} (but see
\citet{Abramo2012}):
\def\Cov{\mathbf{C}}
\be {\mathcal F}_{\alpha\beta}  =  
\frac{1}{2} {\rm Tr}\left[\Cov^{-1}\Cov_{,\alpha}\Cov^{-1}\Cov_{,\beta}\right]
+\sum_{i,j} \frac{\partial \overline{P}_i}{\partial \alpha}C^{-1}_{ij}
\frac{\partial \overline{P}_j}{\partial \beta} 
 \approx  \sum_{i,j} 
\frac{\partial \overline{P}_i}{\partial \alpha}C^{-1}_{ij}
\frac{\partial \overline{P}_j}{\partial \beta} \ ,
\nn \ee
where the approximate equality follows from the fact that the second
term on the right-hand side of the first equality dominates over the
first term, since it scales directly in proportion with the number of
Fourier modes, whereas the first term is independent of the number of
modes. In the above we have made use of the notation
$\partial/\partial\alpha \equiv \partial/\partial\theta_{\alpha}$ to
denote partial derivatives with respect to the cosmological parameters
$\theta_{\alpha}$. On taking the covariance matrix to be diagonal, as
is the case in \Eqn{eq:cov_final}, the above expression for the Fisher
matrix becomes:
\be {\mathcal F}_{\alpha\beta}  =  \sum_{i,j}\frac{\partial \log
  \overline{P}_i}{\partial \alpha} \overline{P}_i
\frac{\delta^{K}_{ij}}{\sigma^2_P(k_i)} \overline{P}_j \frac{\partial
  \log \overline{P}_j}{\partial \beta} 
= \sum_{i}\frac{\partial \log \overline{P}_i}{\partial \alpha}
\frac{\partial \log \overline{P}_i}{\partial \beta}
\left(\frac{{\mathcal S}}{{\mathcal N}}\right)^2\!\!(k_i) \ .  
\ee
If we now define the effective survey volume through the expression,
\be 
V_{\rm eff}(k_i)  \equiv \frac{2(2\pi)^3}{V_i} \left(\frac{{\mathcal S}}{{\mathcal N}}\right)^2\!\!(k_i) 
 \ , \label{eq:Fish3}
\ee
and take the continuum limit for the Fourier modes, we find that the
Fisher matrix can be expressed as \citep{Tegmark1997}:
\be {\mathcal F}_{\alpha\beta} =  \frac{1}{2}\int \frac{\dk}{(2\pi)^3}
\frac{\partial \log P(k)}{\partial \alpha}
\frac{\partial \log P(k)}{\partial \beta} V_{\rm eff}(k)\ . \label{eq:Fisher}
\ee
Thus in order to determine the information content of the galaxy
power-spectrum obtained using a general weight function $w$, we simply
need to calculate $\Veff[w](k)$ or equivalently $\SN[w](k)$. It is
clear from \Eqns{eq:cov_final}{eq:Func1} that a general expression for
the $\SN$ is given by:
\be \left(\frac{{\mathcal S}}{{\mathcal N}}\right)^2(k_i) =
\frac{V_i}{2 (2\pi)^3}\int \dr
  \left[\Gb^{(1)}_{(1,1)}(\br)\right]^2 
\left\{\int \dr
  \left(\left[\Gb^{(1)}_{(1,1)}(\br)\right]^2 +
  \frac{(1+\alpha)}{\overline{P}_i} \left[ \Gb^{(2)}_{(1,0)}(\br) +
    \Gb^{(1)}_{(2,0)}(\br)\right] \right)^2 \right\}^{-1} .
\label{eq:SN_gen}
\ee

\noindent In the case of the optimal weights from \Eqn{eq:optw_gen3},
a little algebra leads to the simplified result:
\be
\left(\frac{{\mathcal S}}{{\mathcal N}}\right)^2(k_i)=\frac{V_i}{2 (2\pi)^3} \int \dr \left[\frac{\overline{P}_i\,
\nbar_{\rm eff}(\br)}{(1+\alpha) + \overline{P}_i \,\nbar_{\rm eff}(\br)}\right]^2 \ .
\label{eq:SNopt}
\ee
The above expression will be useful for forecasting how well a future
galaxy redshift survey may constrain cosmological parameters, after an
optimal power spectrum analysis has been performed.

\section{Practical challenges in implementing the optimal weights}\label{sec:practical}

In order to implement the optimal weighting scheme, one requires
knowledge of: the halo mass function $\nbar(M)$; the halo bias
function $b(M)$; the conditional probability density $\Phi(L|M)$; the
first and second factorial moments of the halo occupation distribution
as parameterised by $N^{(1)}(M)$ and $\beta(M)$; and a way to
associate each galaxy in the survey to a host halo. A possible route
for achieving this is as follows:

\begin{itemize}
\item Pure halo-dependent quantities: $\nbar(M)$ and $b(M)$. These
  functions can be determined directly from numerical simulations;
  there also exist a number of accurate semi-analytic fitting
  functions in the literature \citep[for recent examples
    see][]{Tinkeretal2008,Crocceetal2010,Watsonetal2013}.  However, in
  order to employ these one needs to specify the underlying
  cosmological model -- we do not consider this too troublesome, since
  it is also required to turn redshifts into distances.

\item Galaxy formation dependent functions: $\Phi(L|M)$,
  $N_\g^{(1)}(M)$, $N_\g^{(2)}(M)$. These require a model of galaxy
  formation or additional measurements.  On adopting a
  state-of-the-art SAM, these functions can be measured directly
  \citep{Bensonetal2000,CooraySheth2002}. They may also be obtained
  from the data through the CLF approach
  \citep{Yangetal2003,vandenBoschetal2013}.

\item Associating galaxies to groups: this step could be performed
  through application of standard friends-of-friends group finding
  algorithms or more sophisticated colour-magnitude-redshift grouping
  methods \citep{Ekeetal2004,Koesteretal2007,Rykoffetal2014}.

\item Determine group halo mass: through the use of good quality
  mock catalogues, such as can be facilitated through a SAM, one may
  apply the same grouping algorithms as were used on the real data to
  the mock data, thus finding the mapping between each group and the
  most likely halo mass \citep{Ekeetal2004}.

\item Implement the optimal weighting scheme and measure $P(k)$.

\end{itemize}

Owing to the fact that the steps enumerated above can not be performed
without error, it is likely that this will introduce additional
scatter that we have not accounted for in our optimal estimator. We
expect that this scatter will not bias the measurements, but will most
likely lead to a reduction in signal-to-noise. We shall leave it as a
task for future work to explore how well this method can be
implemented in detail.


\section{Conclusions}\label{sec:conclusions}

In this paper we have developed the theory for the unbiased and
optimal estimation of the matter power spectrum from the galaxy power
spectrum. Our approach generalises the original approach of FKP, by
taking into account central ideas from the theory of galaxy formation:
galaxies form and reside exclusively in dark matter haloes; a given
dark matter halo may host many galaxies of various luminosities;
galaxies inherit part of their large-scale bias from their host halo.

In \S\ref{sec:survey} we described the generic properties of a galaxy
redshift survey and presented a new theoretical quantity: the
galaxy-halo double delta expansion. We demonstrated how one may use
this expansion of the halo and galaxy fields to answer basic
statistical questions concerning the galaxy distribution. In
particular we gave a derivation of the galaxy luminosity function in
this framework.

In \S\ref{sec:theory1} we presented estimators for the galaxy
correlation function and power spectrum. It was proved that, in the
large-scale and large-survey-volume limits, these were unbiased
estimates of the dark matter correlation function and power
spectrum. We demonstrated that, similar to FKP, in our scheme the
matter power spectrum could be obtained by subtracting an effective
shot-noise component followed by the deconvolution of the power
spectrum associated with an effective survey window function.

In \S\ref{sec:fluctuations} we derived general expressions for the
covariance matrix of the weighted galaxy power spectrum, including all
non-Gaussian terms arising from the nonlinear evolution of matter
fluctuations, discreteness effects and finite survey geometry
effects. These results generalise the earlier results of
\citep{MeiksinWhite1999,Scoccimarroetal1999,Smith2009}. In the limits
of large-scales, large survey volumes, and Gaussian fluctuations, the
covariance matrix was found to be diagonal.
 
In \S\ref{sec:optimal} we found an equation that governs the optimal
weights to be applied to galaxies. We found a general solution of the
weight equation. Interestingly, the solution did not carry any
explicit dependence on galaxy luminosity. Instead the weights were
found to be simply a function of two variables: the spatial position
within the survey and the mass of the dark matter halo hosting the
galaxies.

In \S\ref{sec:Fisher} we presented a new expression for the Fisher
information matrix, for a weighted galaxy power spectrum measurement.
We also presented a formula for the signal-to-noise obtained for the
optimal weights.

Finally, in \S\ref{sec:practical} we outlined the practical steps that
would need to be followed if one were to carry out the optimal power
spectrum analysis.

In a companion work \citep{SmithMarian2015b}, we explore the
signal-to-noise and cosmological information gains achievable through
the optimal weighting scheme. In a future work, we will also explore
how well one may implement such a scheme with real data.


\section*{Acknowledgments} 

We thank Simon White for useful discussions.  RES acknowledges part
support from ERC Advanced grant 246797 GALFORMOD.  LM thanks MPA for
its kind hospitality while part of this work was being performed.



\bibliographystyle{mnras}

\begin{thebibliography}{}
\makeatletter
\relax
\def\mn@urlcharsother{\let\do\@makeother \do\$\do\&\do\#\do\^\do\_\do\%\do\~}
\def\mn@doi{\begingroup\mn@urlcharsother \@ifnextchar [ {\mn@doi@}
  {\mn@doi@[]}}
\def\mn@doi@[#1]#2{\def\@tempa{#1}\ifx\@tempa\@empty \href
  {http://dx.doi.org/#2} {doi:#2}\else \href {http://dx.doi.org/#2} {#1}\fi
  \endgroup}
\def\mn@eprint#1#2{\mn@eprint@#1:#2::\@nil}
\def\mn@eprint@arXiv#1{\href {http://arxiv.org/abs/#1} {{\tt arXiv:#1}}}
\def\mn@eprint@dblp#1{\href {http://dblp.uni-trier.de/rec/bibtex/#1.xml}
  {dblp:#1}}
\def\mn@eprint@#1:#2:#3:#4\@nil{\def\@tempa {#1}\def\@tempb {#2}\def\@tempc
  {#3}\ifx \@tempc \@empty \let \@tempc \@tempb \let \@tempb \@tempa \fi \ifx
  \@tempb \@empty \def\@tempb {arXiv}\fi \@ifundefined
  {mn@eprint@\@tempb}{\@tempb:\@tempc}{\expandafter \expandafter \csname
  mn@eprint@\@tempb\endcsname \expandafter{\@tempc}}}

\bibitem[\protect\citeauthoryear{{Abramo}}{{Abramo}}{2012}]{Abramo2012}
{Abramo} L.~R.,  2012, \mn@doi [\mnras] {10.1111/j.1365-2966.2011.20166.x},
  \href {http://adsabs.harvard.edu/abs/2012MNRAS.420.2042A} {420, 2042}

\bibitem[\protect\citeauthoryear{{Anderson} et~al.,}{{Anderson}
  et~al.}{2012}]{Andersonetal2012}
{Anderson} L.,  et~al., 2012, \mn@doi [\mnras]
  {10.1111/j.1365-2966.2012.22066.x}, \href
  {http://adsabs.harvard.edu/abs/2012MNRAS.427.3435A} {427, 3435}

\bibitem[\protect\citeauthoryear{{Anderson} et~al.,}{{Anderson}
  et~al.}{2014a}]{Andersonetal2014b}
{Anderson} L.,  et~al., 2014a, \mn@doi [\mnras] {10.1093/mnras/stt2206}, \href
  {http://adsabs.harvard.edu/abs/2014MNRAS.439...83A} {439, 83}

\bibitem[\protect\citeauthoryear{{Anderson} et~al.,}{{Anderson}
  et~al.}{2014b}]{Andersonetal2014a}
{Anderson} L.,  et~al., 2014b, \mn@doi [\mnras] {10.1093/mnras/stu523}, \href
  {http://adsabs.harvard.edu/abs/2014MNRAS.441...24A} {441, 24}

\bibitem[\protect\citeauthoryear{{Baumgart} \& {Fry}}{{Baumgart} \&
  {Fry}}{1991}]{BaumgartFry1991}
{Baumgart} D.~J.,  {Fry} J.~N.,  1991, \mn@doi [\apj] {10.1086/170166}, \href
  {http://adsabs.harvard.edu/abs/1991ApJ...375...25B} {375, 25}

\bibitem[\protect\citeauthoryear{{Benson}, {Cole}, {Frenk}, {Baugh}  \&
  {Lacey}}{{Benson} et~al.}{2000}]{Bensonetal2000}
{Benson} A.~J.,  {Cole} S.,  {Frenk} C.~S.,  {Baugh} C.~M.,   {Lacey} C.~G.,
  2000, \mn@doi [\mnras] {10.1046/j.1365-8711.2000.03101.x}, \href
  {http://adsabs.harvard.edu/abs/2000MNRAS.311..793B} {311, 793}

\bibitem[\protect\citeauthoryear{{Bernstein}}{{Bernstein}}{1994}]{Bernstein1994}
{Bernstein} G.~M.,  1994, \mn@doi [\apj] {10.1086/173915}, \href
  {http://esoads.eso.org/abs/1994ApJ...424..569B} {424, 569}

\bibitem[\protect\citeauthoryear{{Blake}, {Abdalla}, {Bridle}  \&
  {Rawlings}}{{Blake} et~al.}{2004}]{SKA2004}
{Blake} C.~A.,  {Abdalla} F.~B.,  {Bridle} S.~L.,   {Rawlings} S.,  2004,
  \mn@doi [\nar] {10.1016/j.newar.2004.09.045}, \href
  {http://adsabs.harvard.edu/abs/2004NewAR..48.1063B} {48, 1063}

\bibitem[\protect\citeauthoryear{{Blake} et~al.,}{{Blake}
  et~al.}{2011}]{Blakeetal2011}
{Blake} C.,  et~al., 2011, \mn@doi [\mnras] {10.1111/j.1365-2966.2011.18903.x},
  \href {http://adsabs.harvard.edu/abs/2011MNRAS.415.2876B} {415, 2876}

\bibitem[\protect\citeauthoryear{{Blake} et~al.,}{{Blake}
  et~al.}{2013}]{Blakeetal2013}
{Blake} C.,  et~al., 2013, \mn@doi [\mnras] {10.1093/mnras/stt1791}, \href
  {http://adsabs.harvard.edu/abs/2013MNRAS.436.3089B} {436, 3089}

\bibitem[\protect\citeauthoryear{{Brown}, {Webster}  \& {Boyle}}{{Brown}
  et~al.}{2000}]{Brownetal2000}
{Brown} M.~J.~I.,  {Webster} R.~L.,   {Boyle} B.~J.,  2000, \mn@doi [\mnras]
  {10.1046/j.1365-8711.2000.03688.x}, \href
  {http://adsabs.harvard.edu/abs/2000MNRAS.317..782B} {317, 782}

\bibitem[\protect\citeauthoryear{{Cooray} \& {Sheth}}{{Cooray} \&
  {Sheth}}{2002}]{CooraySheth2002}
{Cooray} A.,  {Sheth} R.,  2002, \physrep, \href
  {http://esoads.eso.org/abs/2002PhR...372....1C} {372, 1}

\bibitem[\protect\citeauthoryear{{Crocce}, {Fosalba}, {Castander}  \&
  {Gazta{\~n}aga}}{{Crocce} et~al.}{2010}]{Crocceetal2010}
{Crocce} M.,  {Fosalba} P.,  {Castander} F.~J.,   {Gazta{\~n}aga} E.,  2010,
  \mn@doi [\mnras] {10.1111/j.1365-2966.2009.16194.x}, \href
  {http://esoads.eso.org/abs/2010MNRAS.403.1353C} {403, 1353}

\bibitem[\protect\citeauthoryear{{Davis} \& {Geller}}{{Davis} \&
  {Geller}}{1976}]{DavisGeller1976}
{Davis} M.,  {Geller} M.~J.,  1976, \mn@doi [\apj] {10.1086/154575}, \href
  {http://adsabs.harvard.edu/abs/1976ApJ...208...13D} {208, 13}

\bibitem[\protect\citeauthoryear{{Davis} \& {Peebles}}{{Davis} \&
  {Peebles}}{1983}]{DavisPeebles1983}
{Davis} M.,  {Peebles} P.~J.~E.,  1983, \mn@doi [\apj] {10.1086/160884}, \href
  {http://esoads.eso.org/abs/1983ApJ...267..465D} {267, 465}

\bibitem[\protect\citeauthoryear{{Eke} et~al.,}{{Eke}
  et~al.}{2004}]{Ekeetal2004}
{Eke} V.~R.,  et~al., 2004, \mn@doi [\mnras]
  {10.1111/j.1365-2966.2004.07408.x}, \href
  {http://adsabs.harvard.edu/abs/2004MNRAS.348..866E} {348, 866}

\bibitem[\protect\citeauthoryear{{Feldman}, {Kaiser}  \& {Peacock}}{{Feldman}
  et~al.}{1994}]{Feldmanetal1994}
{Feldman} H.~A.,  {Kaiser} N.,   {Peacock} J.~A.,  1994, \mn@doi [\apj]
  {10.1086/174036}, \href {http://esoads.eso.org/abs/1994ApJ...426...23F} {426,
  23}

\bibitem[\protect\citeauthoryear{{Fisher}, {Davis}, {Strauss}, {Yahil}  \&
  {Huchra}}{{Fisher} et~al.}{1993}]{Fisheretal1993}
{Fisher} K.~B.,  {Davis} M.,  {Strauss} M.~A.,  {Yahil} A.,   {Huchra} J.~P.,
  1993, \mn@doi [\apj] {10.1086/172110}, \href
  {http://adsabs.harvard.edu/abs/1993ApJ...402...42F} {402, 42}

\bibitem[\protect\citeauthoryear{{Fry} \& {Gaztanaga}}{{Fry} \&
  {Gaztanaga}}{1993}]{FryGaztanaga1993}
{Fry} J.~N.,  {Gaztanaga} E.,  1993, \mn@doi [\apj] {10.1086/173015}, \href
  {http://esoads.eso.org/abs/1993ApJ...413..447F} {413, 447}

\bibitem[\protect\citeauthoryear{{Fry} \& {Peebles}}{{Fry} \&
  {Peebles}}{1978}]{FryPeebles1978}
{Fry} J.~N.,  {Peebles} P.~J.~E.,  1978, \mn@doi [\apj] {10.1086/156001}, \href
  {http://adsabs.harvard.edu/abs/1978ApJ...221...19F} {221, 19}

\bibitem[\protect\citeauthoryear{{Fry} \& {Peebles}}{{Fry} \&
  {Peebles}}{1980}]{FryPeebles1980}
{Fry} J.~N.,  {Peebles} P.~J.~E.,  1980, \mn@doi [\apj] {10.1086/158037}, \href
  {http://adsabs.harvard.edu/abs/1980ApJ...238..785F} {238, 785}

\bibitem[\protect\citeauthoryear{{Groth} \& {Peebles}}{{Groth} \&
  {Peebles}}{1977}]{GrothPeebles1977}
{Groth} E.~J.,  {Peebles} P.~J.~E.,  1977, \mn@doi [\apj] {10.1086/155588},
  \href {http://adsabs.harvard.edu/abs/1977ApJ...217..385G} {217, 385}

\bibitem[\protect\citeauthoryear{{Guzzo}, {Strauss}, {Fisher}, {Giovanelli}  \&
  {Haynes}}{{Guzzo} et~al.}{1997}]{Guzzoetal1997}
{Guzzo} L.,  {Strauss} M.~A.,  {Fisher} K.~B.,  {Giovanelli} R.,   {Haynes}
  M.~P.,  1997, \apj, \href {http://adsabs.harvard.edu/abs/1997ApJ...489...37G}
  {489, 37}

\bibitem[\protect\citeauthoryear{{Hamilton}}{{Hamilton}}{1993}]{Hamilton1993b}
{Hamilton} A.~J.~S.,  1993, \mn@doi [\apj] {10.1086/173288}, \href
  {http://adsabs.harvard.edu/abs/1993ApJ...417...19H} {417, 19}

\bibitem[\protect\citeauthoryear{{Hamilton}}{{Hamilton}}{1997a}]{Hamilton1997a}
{Hamilton} A.~J.~S.,  1997a, \mnras, \href
  {http://adsabs.harvard.edu/abs/1997MNRAS.289..285H} {289, 285}

\bibitem[\protect\citeauthoryear{{Hamilton}}{{Hamilton}}{1997b}]{Hamilton1997b}
{Hamilton} A.~J.~S.,  1997b, \mnras, \href
  {http://adsabs.harvard.edu/abs/1997MNRAS.289..295H} {289, 295}

\bibitem[\protect\citeauthoryear{{Hamilton}}{{Hamilton}}{2000}]{Hamilton2000}
{Hamilton} A.~J.~S.,  2000, \mnras, \href
  {http://esoads.eso.org/abs/2000MNRAS.312..257H} {312, 257}

\bibitem[\protect\citeauthoryear{{Hamilton} \& {Tegmark}}{{Hamilton} \&
  {Tegmark}}{2000}]{HamiltonTegmark2000}
{Hamilton} A.~J.~S.,  {Tegmark} M.,  2000, \mn@doi [\mnras]
  {10.1046/j.1365-8711.2000.03074.x}, \href
  {http://adsabs.harvard.edu/abs/2000MNRAS.312..285H} {312, 285}

\bibitem[\protect\citeauthoryear{{Hauser} \& {Peebles}}{{Hauser} \&
  {Peebles}}{1973}]{HauserPeebles1973}
{Hauser} M.~G.,  {Peebles} P.~J.~E.,  1973, \mn@doi [\apj] {10.1086/152453},
  \href {http://adsabs.harvard.edu/abs/1973ApJ...185..757H} {185, 757}

\bibitem[\protect\citeauthoryear{{Kauffmann}, {Colberg}, {Diaferio}  \&
  {White}}{{Kauffmann} et~al.}{1999}]{Kauffmannetal1999}
{Kauffmann} G.,  {Colberg} J.~M.,  {Diaferio} A.,   {White} S.~D.~M.,  1999,
  \mn@doi [\mnras] {10.1046/j.1365-8711.1999.02202.x}, \href
  {http://esoads.eso.org/abs/1999MNRAS.303..188K} {303, 188}

\bibitem[\protect\citeauthoryear{{Koester} et~al.,}{{Koester}
  et~al.}{2007}]{Koesteretal2007}
{Koester} B.~P.,  et~al., 2007, \mn@doi [\apj] {10.1086/509599}, \href
  {http://adsabs.harvard.edu/abs/2007ApJ...660..239K} {660, 239}

\bibitem[\protect\citeauthoryear{{Landy} \& {Szalay}}{{Landy} \&
  {Szalay}}{1993}]{LandySzalay1993}
{Landy} S.~D.,  {Szalay} A.~S.,  1993, \mn@doi [\apj] {10.1086/172900}, \href
  {http://adsabs.harvard.edu/abs/1993ApJ...412...64L} {412, 64}

\bibitem[\protect\citeauthoryear{{Laureijs} et~al.,}{{Laureijs}
  et~al.}{2011}]{EUCLID2011}
{Laureijs} R.,  et~al., 2011, preprint, \href
  {http://adsabs.harvard.edu/abs/2011arXiv1110.3193L} {} (\mn@eprint {arXiv}
  {1110.3193})

\bibitem[\protect\citeauthoryear{{Levi} et~al.,}{{Levi}
  et~al.}{2013}]{DESI2013}
{Levi} M.,  et~al., 2013, preprint, \href
  {http://adsabs.harvard.edu/abs/2013arXiv1308.0847L} {} (\mn@eprint {arXiv}
  {1308.0847})

\bibitem[\protect\citeauthoryear{{Li}, {Kauffmann}, {Jing}, {White},
  {B{\"o}rner}  \& {Cheng}}{{Li} et~al.}{2006}]{Lietal2006}
{Li} C.,  {Kauffmann} G.,  {Jing} Y.~P.,  {White} S.~D.~M.,  {B{\"o}rner} G.,
  {Cheng} F.~Z.,  2006, \mn@doi [\mnras] {10.1111/j.1365-2966.2006.10066.x},
  \href {http://adsabs.harvard.edu/abs/2006MNRAS.368...21L} {368, 21}

\bibitem[\protect\citeauthoryear{{Meiksin} \& {White}}{{Meiksin} \&
  {White}}{1999}]{MeiksinWhite1999}
{Meiksin} A.,  {White} M.,  1999, \mnras, \href
  {http://esoads.eso.org/abs/1999MNRAS.308.1179M} {308, 1179}

\bibitem[\protect\citeauthoryear{{Mo} \& {White}}{{Mo} \&
  {White}}{1996}]{MoWhite1996}
{Mo} H.~J.,  {White} S.~D.~M.,  1996, \mnras, \href
  {http://esoads.eso.org/abs/1996MNRAS.282..347M} {282, 347}

\bibitem[\protect\citeauthoryear{{Mo}, {Jing}  \& {White}}{{Mo}
  et~al.}{1997}]{Moetal1997}
{Mo} H.~J.,  {Jing} Y.~P.,   {White} S.~D.~M.,  1997, \mnras, \href
  {http://esoads.eso.org/abs/1997MNRAS.284..189M} {284, 189}

\bibitem[\protect\citeauthoryear{{Norberg} et~al.,}{{Norberg}
  et~al.}{2001}]{Norbergetal2001}
{Norberg} P.,  et~al., 2001, \mn@doi [\mnras]
  {10.1046/j.1365-8711.2001.04839.x}, \href
  {http://adsabs.harvard.edu/abs/2001MNRAS.328...64N} {328, 64}

\bibitem[\protect\citeauthoryear{{Norberg} et~al.,}{{Norberg}
  et~al.}{2002}]{Norbergetal2002a}
{Norberg} P.,  et~al., 2002, \mn@doi [\mnras]
  {10.1046/j.1365-8711.2002.05348.x}, \href
  {http://adsabs.harvard.edu/abs/2002MNRAS.332..827N} {332, 827}

\bibitem[\protect\citeauthoryear{{Park}, {Vogeley}, {Geller}  \&
  {Huchra}}{{Park} et~al.}{1994}]{Parketal1994}
{Park} C.,  {Vogeley} M.~S.,  {Geller} M.~J.,   {Huchra} J.~P.,  1994, \mn@doi
  [\apj] {10.1086/174508}, \href
  {http://adsabs.harvard.edu/abs/1994ApJ...431..569P} {431, 569}

\bibitem[\protect\citeauthoryear{{Peacock} \& {Nicholson}}{{Peacock} \&
  {Nicholson}}{1991}]{PeacockNicholson1991}
{Peacock} J.~A.,  {Nicholson} D.,  1991, \mnras, \href
  {http://adsabs.harvard.edu/abs/1991MNRAS.253..307P} {253, 307}

\bibitem[\protect\citeauthoryear{{Peebles}}{{Peebles}}{1973}]{Peebles1973}
{Peebles} P.~J.~E.,  1973, \mn@doi [\apj] {10.1086/152431}, \href
  {http://adsabs.harvard.edu/abs/1973ApJ...185..413P} {185, 413}

\bibitem[\protect\citeauthoryear{{Peebles}}{{Peebles}}{1974}]{Peebles1974}
{Peebles} P.~J.~E.,  1974, \mn@doi [\apjs] {10.1086/190309}, \href
  {http://adsabs.harvard.edu/abs/1974ApJS...28...37P} {28, 37}

\bibitem[\protect\citeauthoryear{{Peebles}}{{Peebles}}{1975}]{Peebles1975}
{Peebles} P.~J.~E.,  1975, \mn@doi [\apj] {10.1086/153450}, \href
  {http://adsabs.harvard.edu/abs/1975ApJ...196..647P} {196, 647}

\bibitem[\protect\citeauthoryear{{Peebles} \& {Groth}}{{Peebles} \&
  {Groth}}{1975}]{PeeblesGroth1975}
{Peebles} P.~J.~E.,  {Groth} E.~J.,  1975, \mn@doi [\apj] {10.1086/153390},
  \href {http://adsabs.harvard.edu/abs/1975ApJ...196....1P} {196, 1}

\bibitem[\protect\citeauthoryear{{Peebles} \& {Hauser}}{{Peebles} \&
  {Hauser}}{1974}]{PeeblesHauser1974}
{Peebles} P.~J.~E.,  {Hauser} M.~G.,  1974, \mn@doi [\apjs] {10.1086/190308},
  \href {http://adsabs.harvard.edu/abs/1974ApJS...28...19P} {28, 19}

\bibitem[\protect\citeauthoryear{{Percival}, {Verde}  \& {Peacock}}{{Percival}
  et~al.}{2004}]{Percivaletal2004a}
{Percival} W.~J.,  {Verde} L.,   {Peacock} J.~A.,  2004, \mn@doi [\mnras]
  {10.1111/j.1365-2966.2004.07245.x}, \href
  {http://adsabs.harvard.edu/abs/2004MNRAS.347..645P} {347, 645}

\bibitem[\protect\citeauthoryear{{Rykoff} et~al.,}{{Rykoff}
  et~al.}{2014}]{Rykoffetal2014}
{Rykoff} E.~S.,  et~al., 2014, \mn@doi [\apj] {10.1088/0004-637X/785/2/104},
  \href {http://adsabs.harvard.edu/abs/2014ApJ...785..104R} {785, 104}

\bibitem[\protect\citeauthoryear{{Scoccimarro}, {Zaldarriaga}  \&
  {Hui}}{{Scoccimarro} et~al.}{1999}]{Scoccimarroetal1999}
{Scoccimarro} R.,  {Zaldarriaga} M.,   {Hui} L.,  1999, \mn@doi [\apj]
  {10.1086/308059}, \href {http://adsabs.harvard.edu/abs/1999ApJ...527....1S}
  {527, 1}

\bibitem[\protect\citeauthoryear{{Seldner} \& {Peebles}}{{Seldner} \&
  {Peebles}}{1977}]{SeldnerPeebles1977}
{Seldner} M.,  {Peebles} P.~J.~E.,  1977, \mn@doi [\apj] {10.1086/155404},
  \href {http://adsabs.harvard.edu/abs/1977ApJ...215..703S} {215, 703}

\bibitem[\protect\citeauthoryear{{Seldner} \& {Peebles}}{{Seldner} \&
  {Peebles}}{1978}]{SeldnerPeebles1978}
{Seldner} M.,  {Peebles} P.~J.~E.,  1978, \mn@doi [\apj] {10.1086/156464},
  \href {http://adsabs.harvard.edu/abs/1978ApJ...225....7S} {225, 7}

\bibitem[\protect\citeauthoryear{{Seldner} \& {Peebles}}{{Seldner} \&
  {Peebles}}{1979}]{SeldnerPeebles1979}
{Seldner} M.,  {Peebles} P.~J.~E.,  1979, \mn@doi [\apj] {10.1086/156699},
  \href {http://adsabs.harvard.edu/abs/1979ApJ...227...30S} {227, 30}

\bibitem[\protect\citeauthoryear{{Seljak} \& {Warren}}{{Seljak} \&
  {Warren}}{2004}]{SeljakWarren2004}
{Seljak} U.,  {Warren} M.~S.,  2004, \mn@doi [\mnras]
  {10.1111/j.1365-2966.2004.08297.x}, \href
  {http://esoads.eso.org/abs/2004MNRAS.355..129S} {355, 129}

\bibitem[\protect\citeauthoryear{{Sheth} \& {Lemson}}{{Sheth} \&
  {Lemson}}{1999}]{ShethLemson1999}
{Sheth} R.~K.,  {Lemson} G.,  1999, \mn@doi [\mnras]
  {10.1046/j.1365-8711.1999.02378.x}, \href
  {http://esoads.eso.org/abs/1999MNRAS.304..767S} {304, 767}

\bibitem[\protect\citeauthoryear{{Smith}}{{Smith}}{2009}]{Smith2009}
{Smith} R.~E.,  2009, \mn@doi [\mnras] {10.1111/j.1365-2966.2009.15490.x},
  \href {http://adsabs.harvard.edu/abs/2009MNRAS.tmp.1337S} {pp 1337--+}

\bibitem[\protect\citeauthoryear{{Smith}}{{Smith}}{2012}]{Smith2012}
{Smith} R.~E.,  2012, \mn@doi [\mnras] {10.1111/j.1365-2966.2012.21745.x},
  \href {http://adsabs.harvard.edu/abs/2012MNRAS.426..531S} {426, 531}

\bibitem[\protect\citeauthoryear{{Smith} \& {Marian}}{{Smith} \&
  {Marian}}{2014}]{SmithMarian2014}
{Smith} R.~E.,  {Marian} L.,  2014, preprint, \href
  {http://adsabs.harvard.edu/abs/2014arXiv1406.1800S} {} (\mn@eprint {arXiv}
  {1406.1800})

\bibitem[\protect\citeauthoryear{{Smith} \& {Marian}}{{Smith} \&
  {Marian}}{2015}]{SmithMarian2015b}
{Smith} R.~E.,  {Marian} L.,  2015, preprint, \href
  {http://adsabs.harvard.edu/abs/2015arXiv150704365S} {} (\mn@eprint {arXiv}
  {1507.04365})

\bibitem[\protect\citeauthoryear{{Smith} \& {Watts}}{{Smith} \&
  {Watts}}{2005}]{SmithWatts2005}
{Smith} R.~E.,  {Watts} P.~I.~R.,  2005, \mn@doi [\mnras]
  {10.1111/j.1365-2966.2005.09053.x}, \href
  {http://esoads.eso.org/abs/2005MNRAS.360..203S} {360, 203}

\bibitem[\protect\citeauthoryear{{Smith}, {Scoccimarro}  \& {Sheth}}{{Smith}
  et~al.}{2007}]{Smithetal2007}
{Smith} R.~E.,  {Scoccimarro} R.,   {Sheth} R.~K.,  2007, \mn@doi [\prd]
  {10.1103/PhysRevD.75.063512}, \href
  {http://esoads.eso.org/abs/2007PhRvD..75f3512S} {75, 063512}

\bibitem[\protect\citeauthoryear{{Springel} et~al.,}{{Springel}
  et~al.}{2005}]{Springeletal2005}
{Springel} V.,  et~al., 2005, \mn@doi [\nat] {10.1038/nature03597}, \href
  {http://esoads.eso.org/abs/2005Natur.435..629S} {435, 629}

\bibitem[\protect\citeauthoryear{{Swanson}, {Tegmark}, {Blanton}  \&
  {Zehavi}}{{Swanson} et~al.}{2008}]{Swansonetal2008}
{Swanson} M.~E.~C.,  {Tegmark} M.,  {Blanton} M.,   {Zehavi} I.,  2008, \mn@doi
  [\mnras] {10.1111/j.1365-2966.2008.12948.x}, \href
  {http://esoads.eso.org/abs/2008MNRAS.385.1635S} {385, 1635}

\bibitem[\protect\citeauthoryear{{Takahashi} et~al.,}{{Takahashi}
  et~al.}{2011}]{Takahashietal2011}
{Takahashi} R.,  et~al., 2011, \mn@doi [\apj] {10.1088/0004-637X/726/1/7},
  \href {http://adsabs.harvard.edu/abs/2011ApJ...726....7T} {726, 7}

\bibitem[\protect\citeauthoryear{{Tegmark}}{{Tegmark}}{1997}]{Tegmark1997}
{Tegmark} M.,  1997, Physical Review Letters, \href
  {http://esoads.eso.org/abs/1997PhRvL..79.3806T} {79, 3806}

\bibitem[\protect\citeauthoryear{{Tegmark}, {Taylor}  \& {Heavens}}{{Tegmark}
  et~al.}{1997}]{Tegmarketal1997}
{Tegmark} M.,  {Taylor} A.~N.,   {Heavens} A.~F.,  1997, \mn@doi [\apj]
  {10.1086/303939}, \href {http://esoads.eso.org/abs/1997ApJ...480...22T} {480,
  22}

\bibitem[\protect\citeauthoryear{{Tinker}, {Kravtsov}, {Klypin}, {Abazajian},
  {Warren}, {Yepes}, {Gottl{\"o}ber}  \& {Holz}}{{Tinker}
  et~al.}{2008}]{Tinkeretal2008}
{Tinker} J.,  {Kravtsov} A.~V.,  {Klypin} A.,  {Abazajian} K.,  {Warren} M.,
  {Yepes} G.,  {Gottl{\"o}ber} S.,   {Holz} D.~E.,  2008, \mn@doi [\apj]
  {10.1086/591439}, \href {http://esoads.eso.org/abs/2008ApJ...688..709T} {688,
  709}

\bibitem[\protect\citeauthoryear{{Vogeley} \& {Szalay}}{{Vogeley} \&
  {Szalay}}{1996}]{VogeleySzalay1996}
{Vogeley} M.~S.,  {Szalay} A.~S.,  1996, \mn@doi [\apj] {10.1086/177399}, \href
  {http://adsabs.harvard.edu/abs/1996ApJ...465...34V} {465, 34}

\bibitem[\protect\citeauthoryear{{Watson}, {Iliev}, {D'Aloisio}, {Knebe},
  {Shapiro}  \& {Yepes}}{{Watson} et~al.}{2013}]{Watsonetal2013}
{Watson} W.~A.,  {Iliev} I.~T.,  {D'Aloisio} A.,  {Knebe} A.,  {Shapiro} P.~R.,
    {Yepes} G.,  2013, \mn@doi [\mnras] {10.1093/mnras/stt791}, \href
  {http://adsabs.harvard.edu/abs/2013MNRAS.433.1230W} {433, 1230}

\bibitem[\protect\citeauthoryear{{White} \& {Frenk}}{{White} \&
  {Frenk}}{1991}]{WhiteFrenk1991}
{White} S.~D.~M.,  {Frenk} C.~S.,  1991, \mn@doi [\apj] {10.1086/170483}, \href
  {http://adsabs.harvard.edu/abs/1991ApJ...379...52W} {379, 52}

\bibitem[\protect\citeauthoryear{{White} \& {Rees}}{{White} \&
  {Rees}}{1978}]{WhiteRees1978}
{White} S.~D.~M.,  {Rees} M.~J.,  1978, \mnras, \href
  {http://esoads.eso.org/abs/1978MNRAS.183..341W} {183, 341}

\bibitem[\protect\citeauthoryear{{Yang}, {Mo}  \& {van den Bosch}}{{Yang}
  et~al.}{2003}]{Yangetal2003}
{Yang} X.,  {Mo} H.~J.,   {van den Bosch} F.~C.,  2003, \mn@doi [\mnras]
  {10.1046/j.1365-8711.2003.06254.x}, \href
  {http://esoads.eso.org/abs/2003MNRAS.339.1057Y} {339, 1057}

\bibitem[\protect\citeauthoryear{{Zehavi} et~al.,}{{Zehavi}
  et~al.}{2002a}]{Zehavietal2002}
{Zehavi} I.,  et~al., 2002a, \mn@doi [\apj] {10.1086/339893}, \href
  {http://adsabs.harvard.edu/abs/2002ApJ...571..172Z} {571, 172}

\bibitem[\protect\citeauthoryear{{Zehavi}, {Blanton}, {Frieman}, {Weinberg},
  {Mo}, {Strauss}  \& {SDSS Collaboration}}{{Zehavi}
  et~al.}{2002b}]{Zehavietal2002short}
{Zehavi} I.,  {Blanton} M.~R.,  {Frieman} J.~A.,  {Weinberg} D.~H.,  {Mo}
  H.~J.,  {Strauss} M.~A.,   {SDSS Collaboration} 2002b, \mn@doi [\apj]
  {10.1086/339893}, \href {http://adsabs.harvard.edu/abs/2002ApJ...571..172Z}
  {571, 172}

\bibitem[\protect\citeauthoryear{{Zehavi} et~al.,}{{Zehavi}
  et~al.}{2005a}]{Zehavietal2005}
{Zehavi} I.,  et~al., 2005a, \mn@doi [\apj] {10.1086/431891}, \href
  {http://adsabs.harvard.edu/abs/2005ApJ...630....1Z} {630, 1}

\bibitem[\protect\citeauthoryear{{Zehavi} et~al.,}{{Zehavi}
  et~al.}{2005b}]{Zehavietal2005short}
{Zehavi} I.,  et~al., 2005b, \mn@doi [\apj] {10.1086/431891}, \href
  {http://adsabs.harvard.edu/abs/2005ApJ...630....1Z} {630, 1}

\bibitem[\protect\citeauthoryear{{Zehavi} et~al.,}{{Zehavi}
  et~al.}{2011a}]{Zehavietal2011}
{Zehavi} I.,  et~al., 2011a, \mn@doi [\apj] {10.1088/0004-637X/736/1/59}, \href
  {http://adsabs.harvard.edu/abs/2011ApJ...736...59Z} {736, 59}

\bibitem[\protect\citeauthoryear{{Zehavi}, {Zheng}, {Weinberg}, {Blanton}  \&
  {SDSS Collaboration}}{{Zehavi} et~al.}{2011b}]{Zehavietal2011short}
{Zehavi} I.,  {Zheng} Z.,  {Weinberg} D.~H.,  {Blanton} M.~R.,   {SDSS
  Collaboration} 2011b, \mn@doi [\apj] {10.1088/0004-637X/736/1/59}, \href
  {http://adsabs.harvard.edu/abs/2011ApJ...736...59Z} {736, 59}

\bibitem[\protect\citeauthoryear{{van den Bosch}, {More}, {Cacciato}, {Mo}  \&
  {Yang}}{{van den Bosch} et~al.}{2013}]{vandenBoschetal2013}
{van den Bosch} F.~C.,  {More} S.,  {Cacciato} M.,  {Mo} H.,   {Yang} X.,
  2013, \mn@doi [\mnras] {10.1093/mnras/sts006}, \href
  {http://adsabs.harvard.edu/abs/2013MNRAS.430..725V} {430, 725}

\makeatother
\end{thebibliography}
\input{OptGalaxiesWeight.MNRAS.v2.bbl}


\appendix


\section{Derivation of the two point correlations:
$\left<n_\g n_\g'\right>$, $\left<n_\g n_s'\right>$, and $\left<n_s n_s'\right>$
}\label{app:corr}


Let us begin by defining the short hand notation for the correlation:
$\left<n_\g n_\g'\right> \equiv
\left<n_\g(\br,L,\bx,M)n_\g(\br',L',\bx',M')\right>$. Following the
analysis of \S\ref{sec:survey}, this correlation may be written:
\ba
\left<n_\g n_\g'\right>  
& = & 
\left< 
\sum_{i,j=1}^{N_h}
\delta^{\rm D}(\bx-\bx_i)\delta^{\rm D}(M-M_i) 
\delta^{\rm D}(\bx'-\bx_j)\delta^{\rm D}(M'-M_j) \right.
\nn \\
& & \times \left. \left<
\sum_{k=1}^{N_\g(M_i)}\sum_{l=1}^{N_\g(M_j)}
\delta^{\rm D}(\br-\br_k-\bx_i)\delta^{\rm D}(L-L_k)
\delta^{\rm D}(\br'-\br_l-\bx_j)\delta^{\rm D}(L'-L_l)
\right>_g \right>_s \ .
\ea
If we now split the sums over $i$ and $j$ into two parts, a piece where
$i\ne j$ and a piece where $i=j$, then we find
\ba
\left<n_\g n_\g'\right>  
& = & 
\left< 
\sum_{i\ne j}^{N_h}
\delta^{\rm D}(\bx-\bx_i)\delta^{\rm D}(M-M_i) 
\delta^{\rm D}(\bx'-\bx_j)\delta^{\rm D}(M'-M_j) \right.
\nn \\
& & \times \left. \left<
\sum_{k=1}^{N_\g(M_i)}
\sum_{l=1}^{N_\g(M_j)}
\delta^{\rm D}(\br-\br_k-\bx_i)\delta^{\rm D}(L-L_k)
\delta^{\rm D}(\br'-\br_l-\bx_j)\delta^{\rm D}(L'-L_l)
\right>_g \right>_s
\nn \\
& & +
\left< \sum_{i=j}^{N_h}
\delta^{\rm D}(\bx-\bx_i)\delta^{\rm D}(M-M_i) 
\delta^{\rm D}(\bx'-\bx_i)\delta^{\rm D}(M'-M_i) \right.
\nn \\
& & \times \left. \left<
\sum_{k,l=1}^{N_\g(M_i)}
\delta^{\rm D}(\br-\br_k-\bx_i)\delta^{\rm D}(L-L_k)
\delta^{\rm D}(\br'-\br_l-\bx_i)\delta^{\rm D}(L'-L_l)
\right>_g \right>_s \ .
\ea
Consider the terms associated with the $i\ne j$ sum, since we have
assumed that the galaxy properties hosted by the $i$th halo are
independent of the galaxy properties in the $j$th halo, we may write
the average of these terms as the product of the two averages. Next,
consider the term $i=j$, and notice that we may also separate the sum
over $k$ and $l$ into two terms, a term with $k\ne l$ and a term with
$k=l$.  This leads us to write the following expression:
\ba
\left<n_\g n_\g'\right>  
& = & 
\left< 
\sum_{i\ne j}^{N_h}
\delta^{\rm D}(\bx-\bx_i)\delta^{\rm D}(M-M_i) 
\delta^{\rm D}(\bx'-\bx_j)\delta^{\rm D}(M'-M_j) \right.
\nn \\
& & \times 
\left. 
\left<
\sum_{k=1}^{N_\g(M_i)}
\delta^{\rm D}(\br-\br_k-\bx_i)\delta^{\rm D}(L-L_k)
\right>_g
\left<
\sum_{l=1}^{N_\g(M_j)}
\delta^{\rm D}(\br'-\br_l-\bx_j)\delta^{\rm D}(L'-L_l)
\right>_g 
\right>_s
\nn \\
& & 
+
\left< \sum_{i=j}^{N_h}
\delta^{\rm D}(\bx-\bx_i)\delta^{\rm D}(M-M_i) 
\delta^{\rm D}(\bx'-\bx_i)\delta^{\rm D}(M'-M_i) \right.
\nn \\
& & 
\times
\left[
\left<
\sum_{k=1}^{N_\g(M_i)}\sum_{l\ne k}^{N_\g(M_i)}
\delta^{\rm D}(\br-\br_k-\bx_i)\delta^{\rm D}(L-L_k)
\delta^{\rm D}(\br'-\br_l-\bx_i)\delta^{\rm D}(L'-L_l)
\right>_g \right. 
\nn \\
& & 
\left. \left. +
\left<
\sum_{k=l}^{N_\g(M_i)}
\delta^{\rm D}(\br-\br_k-\bx_i)\delta^{\rm D}(L-L_k)
\delta^{\rm D}(\br'-\br_k-\bx_i)\delta^{\rm D}(L'-L_k)
\right>_g 
\right]
\right>_s \ .
\ea
We are now able to compute the expectations over the galaxy
populations, and with the help of \Eqn{eq:probrl} we find:
\ba
\left<n_\g n_\g'\right>  
& = & 
\left< 
\sum_{i\ne j}^{N_h}
\delta^{\rm D}(\bx-\bx_i)\delta^{\rm D}(M-M_i) 
\delta^{\rm D}(\bx'-\bx_j)\delta^{\rm D}(M'-M_j) \right.
\nn \\
& & \times 
\left. 
\frac{}{}N^{(1)}_\g(M_i) N^{(1)}_\g(M_j)U(\br-\bx_i|M_i)U(\br'-\bx_j|M_j)
\Phi(L|M_i) \Phi(L'|M_j)\Theta(\br|L) \Theta(\br'|L')
\right>_s
\nn \\
& & 
+
\left< \sum_{i=j}^{N_h}
\delta^{\rm D}(\bx-\bx_i)\delta^{\rm D}(M-M_i) 
\delta^{\rm D}(\bx'-\bx_i)\delta^{\rm D}(M'-M_i) \right.
\nn \\
& & 
\times
\left[
\frac{}{}N_\g^{(2)}(M_i) U(\br-\bx_i|M_i)  U(\br'-\bx_i|M_i)
\Phi(L|M_i) \Phi(L'|M_i)\Theta(\br|L)\Theta(\br'|L')
\right. 
\nn \\
& & 
\left. \left. +
N_{\g}^{(1)}(M_i) \Phi(L|M_i) U(\br-\bx_i|M_i)  \Theta(\br|L) \dirac(L-L')\dirac(\br-\br')
\right]
\right>_s \label{eq:ngng1} \ ,
\ea
where in the above we have used a short-hand notation for the
factorial moments of the galaxy numbers:
\be
N_{\g}^{(l)}(M)  \equiv  \left<N_\g(N_\g-1)\dots (N_\g-l+1)|M\right> 
= \sum_{N_{\g}=0}^{\infty} P(N_{\g}|\lambda(M))N_\g(N_\g-1)\dots (N_\g-l+1)\ .
\ee
Let us now deal with the averages over the dark matter haloes and let
us write the first and second terms in \Eqn{eq:ngng1} as $\left<n_\g
n_\g'\right>_{\rm A}$ and $\left<n_\g n_\g'\right>_{\rm B}$. Considering the first
term, the expectations may be computed as in \Eqn{eq:den3}, and
we find
\ba
\left<n_\g n_\g'\right>_{\rm A}  
& = & 
\sum_{i\ne j}^{N_h}
\int \prod_{\nu=1}^{\Nh}\left\{\dx_\nu dM_\nu\right\}p(\bx_1,\dots,\bx_\Nh,M_1,\dots,M_\Nh)
\delta^{\rm D}(\bx-\bx_i)\delta^{\rm D}(M-M_i) 
\delta^{\rm D}(\bx'-\bx_j)\delta^{\rm D}(M'-M_j) 
\nn \\
& & \times 
\frac{}{}N^{(1)}_\g(M_i) N^{(1)}_\g(M_j)U(\br-\bx_i|M_i)U(\br'-\bx_j|M_j) 
\Phi(L|M_i) \Phi(L'|M_j)\Theta(\br|L) \Theta(\br'|L')
\nn \\
& = &
\Nh(\Nh-1) p(\bx,\bx',M,M') N^{(1)}_\g(M) N^{(1)}_\g(M')
U(\br-\bx|M)U(\br'-\bx'|M') \Phi(L|M) \Phi(L'|M') \Theta(\br|L) \Theta(\br'|L')
\ .\label{eq:ngng2A}
\ea
The joint probability density functions for the halo centres and
masses may be expressed in terms of products of their 1-point PDFs and
correlation functions. For the case of two-points we have:
\ba 
p(\bx_1,\bx_1,M_1,M_2) & \equiv & p(\bx_1,M_1) p(\bx_2,M_2)
\left[1+\xi^\c(\bx_1,\bx_2,M_1,M_2)\right] =
\frac{\nbar(M_1)\nbar(M_2)}{\Nh^2} \left[1+\xi^\c(\bx_1,\bx_2,M_1,M_2)\right] \ .
\ea
In addition, if we assume that the cluster density field is some local
function of the underlying dark matter density
\citep{FryGaztanaga1993,MoWhite1996,Moetal1997,Smithetal2007}, the
cross-correlation function of clusters of masses $M_1$ and $M_2$, at
leading order, can be written:
\be \xi^\c(|\bx_1-\bx_2|,M_1,M_2) = b(M_1)b(M_2)\xi(|\bx_1-\bx_2|) \ ,
\label{eq:linbias}\ee
where $\xi(r)$ is the correlation of the underlying matter
fluctuations. On using this relation in \Eqn{eq:ngng2A} we find,
\ba
\left<n_\g n_\g'\right>_{\rm A}
& = & 
\nbar(M)\nbar(M')\left[1+b(M_1)b(M_2)\xi(|\bx_1-\bx_2|)\right]
N^{(1)}_\g(M) N^{(1)}_\g(M') U(\br-\bx|M)U(\br'-\bx'|M')\nn \\
& & \times \Phi(L|M) \Phi(L'|M') \Theta(\br|L) \Theta(\br'|L')
 \ . \label{eq:ngng3A}
\ea

Returning now to the second terms in \Eqn{eq:ngng1} and following 
a similar derivation to the first term, we find 
\ba
\left<n_\g n_\g'\right>_{\rm B}
& = & 
\nbar(M)N^{(2)}_\g(M)\Phi(L|M) \Phi(L'|M)  
\Theta(\br|L)\Theta(\br'|L') U(\br-\bx|M)  U(\br'-\bx|M) \delta^{\rm D}(\bx-\bx')\delta^{\rm D}(M-M')
\nn \\
& & 
+\nbar(M) N_\g^{(1)}(M) \Phi(L|M) \Theta(\br|L) U(\br-\bx|M) \delta^{\rm D}(\bx-\bx')\delta^{\rm D}(M-M')
\dirac(L-L')\dirac(\br-\br') \ . \label{eq:ngng3B}
\ea

Following the derivation $\left<n_\g n_\g'\right>$ we may now
straightforwardly write down the results for the cases of the cross-
and auto-correlation of the synthetic galaxy-halo field with the real
one:
\ba
\left< n_\g n_s'\right>
& = & 
\alpha^{-1} \nbar(M)\nbar(M')N_\g(M)N_\g(M')
\Theta(\br|L)\Theta(\br'|L')U(\br-\bx|M)U(\br'-\bx'|M') \Phi(L|M) \Phi(L'|M') 
\label{eq:ngns} \ ; \\
\left< n_s n_s' \right>
& = & 
\alpha^{-2} \nbar(M)\nbar(M')N_\g(M) N_\g(M')
\Theta(\br|L)\Theta(\br'|L') U(\br-\bx|M)U(\br'-\bx'|M') \Phi(L|M) \Phi(L'|M') \nn \\
& & 
+\alpha^{-1} \nbar(M) N^{(2)}_\g(M) \Phi(L|M) \Phi(L'|M) \Theta(\br|L)\Theta(\br'|L') 
U(\br-\bx|M) U(\br'-\bx|M) 
\delta^{\rm D}(\bx-\bx')\delta^{\rm D}(M-M')
\nn \\
& &
+N_\g^{(1)}(M) \Phi(L|M) \Theta(\br|L)U(\br-\bx|M)\delta^{\rm D}(\bx-\bx')\delta^{\rm D}(M-M')
\dirac(L-L')\dirac(\br-\br') \ ,\label{eq:nsns}
\ea
where in the above we have made use of the following short-hand notation:
$\left< n_\g n_s'\right> \equiv \left<n_\g(\br,L,\bx,M) n_s(\br',L',\bx',M')\right>$ and
$\left< n_s n_s'\right> \equiv \left<n_s(\br,L,\bx,M) n_s(\br',L',\bx',M')\right>$.


\section{The galaxy covariance matrix}

\subsection{Comment on the covariance matrix of the matter power spectrum}\label{app:covm}

Starting with \Eqn{eq:covdef} and inserting \Eqn{eq:Pk} we find:
\ba 
{\rm Cov}\!\left[\hat{P}(\bk_1),\hat{P}(\bk_2)\right] =  
{\rm Cov}\!\left[|\Fgt(\bk_1)|^2,|\Fgt(\bk_2)|^2\right] 
-{\rm Cov}\!\left[|\Fgt(\bk_1)|^2,P_{\rm shot}\right] 
-{\rm Cov}\!\left[|\Fgt(\bk_2)|^2,P_{\rm shot}\right] 
+{\rm Var}\!\left[P_{\rm shot}\right] ,
\label{eq:def_cov}
\ea
where:
\ba 
{\rm Cov}\!\left[|\Fgt(\bk_1)|^2,|\Fgt(\bk_2)|^2\right] & \equiv & 
\left<|\Fgt(\bk_1)|^2|\Fgt(\bk_2)|^2\right>-
\left<|\Fgt(\bk_1)|^2\right> \left<|\Fgt(\bk_2)|^2 \right>   \ ;
\nn \\
{\rm Cov}\!\left[|\Fgt(\bk_i)|^2,P_{\rm shot}\right] & \equiv & 
\left<|\Fgt(\bk_i)|^2P_{\rm shot}\right>-
\left<|\Fgt(\bk_i)|^2\right> \left<P_{\rm shot}\right> \ \ \ ; \ \ i \in \{1,2\}  \ ;
\nn \\
{\rm Var}\!\left[P_{\rm shot}\right] & \equiv & 
\left<P_{\rm shot}^2\right>-\left<P_{\rm shot}\right>^2 \ .
\nn \ea

If we assume that the statistical uncertainties are dominated by
$|\Fgt(\bk_1)|^2$ and not $P_{\rm shot}$, then we may approximate the
covariance matrix as is written in \Eqn{eq:CovFF}.


\section{Derivation of the covariance matrix of the $\Fgt$ 
power spectrum}
\label{app:covGen}


To begin, we notice that we may rewrite the covariance matrix of
$\left|{\mathcal F}_{\g}(k)\right|^2$, which is given in
\Eqn{eq:CovFF}, as
\be
{\rm Cov}\!\left[|\Fgt(\bk_1)|^2,|\Fgt(\bk_2)|^2\right] 
=
\int \dk_3 \dk_4 \dirac(\bk_1+\bk_3) \dirac(\bk_2+\bk_4)
\left[\frac{}{}
\left<\Fg(\bk_1)\dots \Fg(\bk_4)\right> -
\left<\Fg(\bk_1)\Fg(\bk_3)\right>\left<\Fg(\bk_2)\Fg(\bk_4)\right> \right] \ .
\label{eq:covF0}
\ee
We see that in order to proceed we need the 4-point function of the
$\Fg(\bk)$ modes. On transforming to real space, this requirement is
transformed into the need to determine the four-point correlation:
\ba
\left<\Fg(\bk_1)\dots \Fg(\bk_4)\right> -
\left<\Fg(\bk_1)\Fg(\bk_3)\right>\left<\Fg(\bk_2)\Fg(\bk_4)\right> & & 
\nn \\
& &
\hspace{-6cm}= \int \dr_1 \dots \dr_4 
\left[\frac{}{}
\left<\Fg(\br_1)\dots \Fg(\br_4)\right>-
\left<\Fg(\br_1)\Fg(\br_3)\right>\left<\Fg(\br_2)\Fg(\br_4)\right>\right]
{\rm e}^{i\bk_1\cdot\br_1+\dots+\bk_4\cdot\br_4}
\label{eq:covF}
\ea
%


\subsection{Computing the 4-point correlation function of $\Fg(\br)$}
\label{app:B1}


Since the terms $\left<\Fg(\br_i)\Fg(\br_j)\right>$ are given by
\Eqn{eq:xihm2}, we are left with the task of computing the 4-point
correlation function of the field $\Fg(\br)$. Using our relation
\Eqn{eq:galden}, this is given by:
\ba \left<\Fg(\br_1)\dots \Fg(\br_4)\right> & = &
\frac{1}{A^2} \prod_{i=1}^4 \left\{\int dL_i \dx_i dM_i w(\br_i,L_i,\bx_i,M_i)\right\}
\nn \\
& & \times 
\left<\frac{}{}\left[n_\g(\br_1,L_1,\bx_1,M_1)-\alpha n_{\s}(\br_1,L_1\bx_1,M_1)\right]
\dots\left[n_\g(\br_4,L_4,\bx_4,M_4)-\alpha n_{\s}(\br_4,L_4,\bx_4,M_4)\right]\right> 
\nn \\
& = & 
\frac{1}{A^2} \prod_{i=1}^4 \left\{\int dL_i \dx_i dM_i w(\br_i,L_i,\bx_i,M_i)\right\}
\left\{\frac{}{}
\left<n_{\g,1}\dots n_{\g,4}\right> 
-\alpha \left[\frac{}{}\left<n_{\g,1}n_{\g,2}n_{\g,3} n_{\s,4}\right>
+{3\rm cyc}\right] \right. \nn \\
& & +\alpha^2 
\left[\frac{}{}
\left<n_{\g,1}n_{\g,2}n_{\s,3}n_{\s,4}\right>+{5\rm perm}\right]
-\alpha^3 \left[\frac{}{}
\left<n_{\g,1}n_{\s,2}n_{\s,3} n_{\s,4}\right>
+{3\rm cyc}\right]+\left. \frac{}{}\alpha^4\left<n_{\s,1}\dots n_{\s,4}\right>\right\}
\label{eq:FgFgFgFg}
\ea
with the short-hand notation identical to that used in
Appendix~\ref{app:corr}:  $n_{\g,i}\equiv n_{\g}(\br_i,L_i,\bx_i,M_i)$
and $n_{s,i}\equiv n_{\s}(\br_i,L_i,\bx_i,M_i)$.
Focusing on the first term in curly brackets on the right-hand-side,
and if we insert our galaxy-halo double delta expansion we find:
\ba 
\left<n_{\g,1}'\dots n_{\g,4}'\right>
& = & 
\left<
\sum_{i_1,i_2,i_3,i_4=1}^{\Nh}
\dirac(\bx_1'-\bx_{i_1})\dirac(M_1'-M_{i_1})
\dots
\dirac(\bx_4'-\bx_{i_4})\dirac(M_4'-M_{i_4})\left<\rm g\right>\right>_\h\ ,
\ea
where we have introduced the short-hand notation
\be \left<\rm g\right> \equiv \left<
\sum_{j_1,j_2,j_3,j_4=1}^{\Ng}
\dirac(\br_1-\br_{j_1})\dirac(L_1-L_{j_1})
\dots 
\dirac(\br_4-\br_{j_4})\dirac(L_4-L_{j_4})
\right>_{\g}\ .
\ee
As was done for the case of the two-point function, we may now split
the sum over haloes into five types of terms:
\ba 
\left<n_{\g,1}'\dots n_{\g,4}'\right> & = & 
\left<\Gamma_1\right>+\left<\Gamma_2\right>+\left<\Gamma_3\right>
+\left<\Gamma_4\right>+\left<\Gamma_5\right> \ ,
\ea
where the terms $\Gamma_i$ are defined:
\ba 
\Gamma_1 &= & 
\hspace{-0.5cm} \sum_{i_1\ne i_2\ne i_3\ne i_4}\hspace{-0.5cm} \dirac(\bx_1'-\bx_{i_1})\dots \dirac(\bx_4'-\bx_{i_4})
\dirac(M_1'-M_{i_1})\dots \dirac(M_4'-M_{i_4}) \left<\rm g\right> \ ; \nn \\
\Gamma_2 & = & 
\hspace{-0.5cm} \sum_{i_1\ne i_2\ne i_3=i_4}\hspace{-0.5cm} 
\dirac(\bx_1'\hspace{-0.1cm}-\hspace{-0.1cm}\bx_{i_1})\dirac(\bx_2'\hspace{-0.1cm}-\hspace{-0.1cm}\bx_{i_2})
\dirac(M_1'\hspace{-0.1cm}-\hspace{-0.1cm}M_{i_1})\dirac(M_2'\hspace{-0.1cm}-\hspace{-0.1cm}M_{i_2})
\prod_{p=3}^4\left\{
\dirac(\bx_p'\hspace{-0.1cm}-\hspace{-0.1cm}\bx_{i_p})\dirac(M_p'\hspace{-0.1cm}
-\hspace{-0.1cm}M_{i_p})\right\}\left<\rm g\right>+5 {\:\rm perms} \ ;\nn \\
\Gamma_3 & = & 
\hspace{-0.5cm} \sum_{i_1=i_2\ne i_3=i_4}
\prod_{p=1}^2\left\{\dirac(\bx_p'-\bx_{i_p})\dirac(M_p'-M_{i_p})\right\}
\prod_{q=3}^4\left\{\dirac(\bx_q'-\bx_{i_q})\dirac(M_q'-M_{i_q})\right\}\left<\rm g\right>
+2 {\,\rm perms} \ ; \nn \\
\Gamma_4 & = & 
\hspace{-0.5cm} \sum_{i_1\ne i_2=i_3=i_4}\hspace{-0.5cm} 
\dirac(\bx_1'-\bx_{i_1})\dirac(M_1'-M_{i_1})
\prod_{p=2}^{4}
\left\{\dirac(\bx_p'-\bx_{i_p})\dirac(M_p'-M_{i_p})\right\}\left<\rm g\right>
 +3 {\,\rm perms} \ ; \nn \\
\Gamma_5 & = & 
\hspace{-0.5cm} \sum_{i_1=i_2=i_3=i_4}
\prod_{p=1}^{4}\left\{\dirac(\bx_p'-\bx_{i_p})
\dirac(M_p'-M_{i_p})\right\}\left<\rm g\right> \ .
\ea

\noindent{\bf Computing $\left<\Gamma_1\right>$}: Integrating over the
Dirac delta functions and relabelling primed variables to unprimed, we
write:
\ba
\left<\Gamma_1\right>
& = & 
\hspace{-0.5cm}\sum_{i_1\ne i_2\ne i_3\ne i_4}^{\Nh} \hspace{-0.5cm}p(\bx_1,\dots,\bx_4,M_1,\dots,M_4) 
\left<\rm g\right> 
\nn\\
& = & \Nh(\Nh-1)(\Nh-2)(\Nh-3) p(\bx_1,M_1)\dots p(\bx_4,M_4) 
\left[\frac{}{}1+\left\{
\xi^\c_{12}+\xi^\c_{13}+\xi^\c_{14}+\xi^\c_{23}+\xi^\c_{24}+\xi^\c_{34} \right\}
\right. \nn \\
& & \left. \frac{}{}+\left\{\zeta^\c_{123}+\zeta^\c_{234}+\zeta^\c_{341}+\zeta^\c_{412}\right\} + 
\left\{\xi^\c_{12}\xi^\c_{34}+\xi^\c_{13}\xi^\c_{24}+\xi^\c_{14}\xi^\c_{23}\right\}+\eta^\c_{1234}
\right] \left<\rm g\right> \nn \\
& \approx & \nbar_1\dots \nbar_4
\left[\frac{}{}1
+\left\{\xi^\c_{12}+{\rm 5\, perms}\right\}
+\left\{\zeta^\c_{123}+{\rm 3\, perms}\right\}
+\left\{\xi^\c_{12}\xi^\c_{34}+{\rm 2\, perms}\right\}+ \eta^\c_{1234} 
\right]
\left<\rm g\right> \ ,
\nn\ea
where in the above we have decomposed the joint 4-point PDF into its
respective 1-point moments and set of correlation functions. In this
case $\zeta$ and $\eta$ denote the connected three- and four-point
correlation functions, respectively. Note also that we used the
following short-hand notation:
\be 
\xi^\c_{ij}\equiv\xi^\c(\bx_i,\bx_j,M_i,M_j) \hspace{0.3cm} ; \hspace{0.3cm}
\zeta^\c_{ijk}\equiv\zeta(\bx_i,\bx_j,\bx_k,M_i,M_j,M_k) \hspace{0.3cm} ; \hspace{0.3cm}
\eta^\c_{ijkl}\equiv \eta^\c(\bx_i,\bx_j,\bx_k,\bx_l,M_i,M_j,M_k,M_l).
\ee

\vspace{0.2cm}

\noindent{\bf Computing $\left<\Gamma_2\right>$}: We denote
$\dirac_{\h,ij}\equiv \dirac(M_i-M_j)\dirac(\bx_i-\bx_j)$ and
$\dirac_{\g, ij}\equiv \dirac(L_i-L_j) \dirac(\br_i-\br_j)$. Taking
the expectations and integrating over the delta functions we find:
\ba
\left<\Gamma_2\right>
& = & 
\hspace{-0.5cm} \sum_{i_1\ne i_2 \ne i_3=i_4}^{\Nh} \hspace{-0.5cm} p(\bx_1,\bx_2,\bx_3,M_1,M_2,M_3)\dirac_{\h,34}
\left<g\right>
+\hspace{-0.5cm} \sum_{i_1\ne i_2=i_3 \ne i_4}^{\Nh}\hspace{-0.5cm} p(\bx_1,\bx_2,\bx_4,M_1,M_2,M_4)\dirac_{\h,23}
\left<g\right> \nn \\
& + & 
\hspace{-0.5cm} \sum_{i_1=i_2 \ne i_3\ne i_4}^{\Nh} \hspace{-0.5cm} p(\bx_1,\bx_3,\bx_4,M_1,M_3,M_4)\dirac_{\h,12}
\left<g\right>
+ \hspace{-0.5cm} \sum_{i_1=i_3 \ne i_2\ne i_4}^{\Nh} \hspace{-0.5cm} p(\bx_1,\bx_2,\bx_4,M_1,M_2,M_4)\dirac_{\h,13}
\left<g\right> \nn \\
& +  & 
\hspace{-0.5cm} \sum_{i_1=i_4 \ne i_2\ne i_3}^{\Nh} \hspace{-0.5cm} p(\bx_1,\bx_2,\bx_3,M_1,M_2,M_3)\dirac_{\h,14}
\left<g\right>
+\hspace{-0.5cm} \sum_{i_2=i_4 \ne i_1\ne i_3}^{\Nh} \hspace{-0.5cm} p(\bx_1,\bx_2,\bx_3,M_1,M_2,M_3)\dirac_{\h,24}
\left<g\right>
\nn \\
& = & 
 \Nh(\Nh-1)(\Nh-2) \left[\frac{}{} p(\bx_1,\bx_2,\bx_3,M_1,M_2,M_3)
\left<g\right>\dirac_{\h,34}+p(\bx_1,\bx_2,\bx_4,M_1,M_2,M_4)\left<g\right> 
\dirac_{\h,23}\right. \nn \\ 
& + & p(\bx_1,\bx_3,\bx_4,M_1,M_3,M_4)\left<g\right>\dirac_{\h,12}
+p(\bx_1,\bx_2,\bx_4,M_1,M_2,M_4)\left<g\right>\dirac_{\h,13} + p(\bx_1,\bx_2,\bx_3,M_1,M_2,M_3)\left<g\right>
\dirac_{\h,14}
\nn \\
&  + &  \left. p(\bx_1,\bx_2,\bx_3,M_1,M_2,M_3)\left<g\right>\dirac_{\h,24} \frac{}{}\right] 
\nn \\
& \approx & \nbar_1\nbar_3\nbar_4 \left[1+\xi^\c_{13}+\xi^\c_{14}+\xi^\c_{34}+\zeta^\c_{134}\right]
\left<g\right>\dirac_{\h,12} 
+ \nbar_1\nbar_2\nbar_4
\left[1+\xi^\c_{12}+\xi^\c_{24}+\xi^\c_{24}+\zeta^\c_{124}\right]\left<g\right>\dirac_{\h,23}
\nn \\
& + & \nbar_1\nbar_2\nbar_3
\left[1+\xi^\c_{12}+\xi^\c_{13}+\xi^\c_{23}+\zeta^\c_{123}\right]\left<g\right>\dirac_{\h,14}
+ \nbar_1\nbar_2\nbar_4
\left[1+\xi^\c_{12}+\xi^\c_{14}+\xi^\c_{24}+\zeta^\c_{124}\right]\left<g\right> \dirac_{\h,23}
\nn \\
& + &  \nbar_1\nbar_2\nbar_3
\left[1+\xi^\c_{12}+\xi^\c_{23}+\xi^\c_{31}+\zeta^\c_{123}\right]\left<g\right> \dirac_{\h,24}
+ \nbar_1\nbar_2\nbar_3
\left[1+\xi^\c_{12}+\xi^\c_{23}+\xi^\c_{31}+\zeta^\c_{123}\right]\left<g\right>\dirac_{\h,34} \ .
\nn\ea
%

\vspace{0.2cm}

\noindent{\bf Computing $\left<\Gamma_3\right>$}: Again, on taking the
expectations and integrating over the delta functions we find:
\ba
\left<\Gamma_3\right>
& = & 
\hspace{-0.5cm}\sum_{i_1=i_2\ne i_3=i_4}\hspace{-0.5cm} p(\bx_1,\bx_3,M_1,M_3)\left<g\right>
\dirac_{\h,12}\,\dirac_{\h,34}
+ \hspace{-0.5cm}\sum_{i_1=i_3\ne i_2=i_4}\hspace{-0.5cm} p(\bx_1,\bx_2,M_1,M_2)\left<g\right>
\dirac_{\h,13}\,\dirac_{\h,24} + \hspace{-0.5cm}\sum_{i_1=i_4\ne i_2=i_3} \hspace{-0.5cm} 
p(\bx_1,\bx_2,M_1,M_2)\left<g\right>\dirac_{\h,14}\,\dirac_{\h,23}
\nn  \\
& = & 
\Nh(\Nh-1) \left[ p(\bx_1,\bx_3,M_1,M_3)\left<g\right>\dirac_{\h,12}\,\dirac_{\h,34}
+ p(\bx_1,\bx_2,M_1,M_2)\left<g\right>\dirac_{\h,13}\,\dirac_{\h,24}
+ p(\bx_1,\bx_2,M_1,M_2)\left<g\right>\dirac_{\h,14}\,\dirac_{\h,23}\right]
\nn  \\
& = & 
\frac{}{}\nbar_1\nbar_3\left[1+\xi^\g_{13}\right]\left<g\right>\dirac_{\h,12}\,\dirac_{\h,34}
+ \nbar_1\nbar_2\left[1+\xi^\g_{12}\right]\left<g\right>\left[\frac{}{}\dirac_{\h,13}\,\dirac_{\h,24}
+ \dirac_{\h,14}\,\dirac_{\h,23}\right]\ . 
\nn\ea


\vspace{0.2cm}

\noindent{\bf Computing $\left<\Gamma_4\right>$}: Again, on taking the
expectations and integrating over the delta functions we find:
\ba
\left<\Gamma_4\right>
& = & 
\hspace{-0.5cm}\sum_{i_1=i_2=i_3\ne i_4} \hspace{-0.5cm}p(\bx_1,\bx_4,M_1,M_4)\left<g\right>
\dirac_{\h,12}\,\dirac_{\h,13}
+\hspace{-0.5cm}\sum_{i_1=i_2=i_4\ne i_3}\hspace{-0.5cm}p(\bx_1,\bx_3,M_1,M_3)\left<g\right>\dirac_{\h,12}
\,\dirac_{\h,14}
\nn \\
& + & 
\hspace{-0.5cm} \sum_{i_1=i_3=i_4\ne i_2} \hspace{-0.5cm} p(\bx_1,\bx_2,M_1,M_2)\left<g\right>
\dirac_{\h,13}\,\dirac_{\h,14}
+ \hspace{-0.5cm} \sum_{i_2=i_3=i_4\ne i_1} \hspace{-0.5cm} p(\bx_1,\bx_2,M_1,M_2)\left<g\right>
\dirac_{\h,23}\,\dirac_{\h,24}
\nn \\
& = & 
N(N-1)\left[ \frac{}{} p(\bx_1,\bx_4,M_1,M_4)\left<g\right> \dirac_{\h,12}\,\dirac_{\h,13}
+ p(\bx_1,\bx_3,M_1,M_3)\left<g\right> \dirac_{\h,12}\,\dirac_{\h,14}\right. 
\nn \\
& + & \left. p(\bx_1,\bx_2,M_1,M_2)\left<g\right> \dirac_{\h,13}\,\dirac_{\h,14}
+ p(\bx_1,\bx_2,M_1,M_2)\left<g\right>\dirac_{\h,23}\,\dirac_{\h,24} \frac{}{}\right]
\nn \\
& = & 
\nbar_1\nbar_4\left[1+\xi^\g_{14}\right]\left<g\right>\dirac_{\h,12}\,\dirac_{\h,13}
+ \nbar_1\nbar_3\left[1+\xi^\g_{13}\right]\left<g\right>\dirac_{\h,12}\,\dirac_{\h,14}
+\nbar_1\nbar_2\left[1+\xi^\g_{12}\right]\left<g\right>\left[\dirac_{\h,13}\,\dirac_{\h,14}
+ \dirac_{\h,23}\,\dirac_{\h,24}\right] 
\nn\ea
%


\vspace{0.2cm}

\noindent{\bf Computing $\left<\Gamma_5\right>$}: Again, on taking the
expectations and integrating over the delta functions we find:
\ba
\left<\Gamma_5\right>
& = & 
\hspace{-0.5cm}\sum_{i_1=i_2=i_3=i_4}\hspace{-0.5cm} p(\bx_1,M_1)\left<g\right>
\dirac_{\h,12}\,\dirac_{\h,13}\,\dirac_{\h,14}=  
\frac{}{} \nbar_1\left<g\right>\dirac_{\h,12}\,\dirac_{\h,13}\,\dirac_{\h,14}
\nn\ea
%


\vspace{0.2cm}

\noindent
Collecting the terms $\left<\Gamma_1\right>$,
$\left<\Gamma_2\right>$, $\left<\Gamma_3\right>$,
$\left<\Gamma_4\right>$ and $\left<\Gamma_5\right>$, we write:
\ba 
\left<n_{\g,1} n_{\g,2} n_{\g,3} n_{\g,4}\right> & = &   
\nbar_1 \nbar_2 \nbar_3 \nbar_4 \left\{\frac{ }{ } \hspace{-0.2cm} 
\left[1+\xi^\c_{12}+\xi^\c_{13}+\xi^\c_{14}+\xi^\c_{23}+\xi^\c_{24}+\xi^\c_{34} 
+\zeta^\c_{123} + \zeta^\c_{124} + \zeta^\c_{134} + \zeta^\c_{234}
+ \xi^\c_{12}\xi^\c_{34}+\xi^\c_{13}\xi^\c_{24} 
\right.\right. \nn\\
& + & \left.\left.
\xi^\c_{14}\xi^\c_{23}+\eta^\c_{1234}\right]\left<g\right>\hspace{-0.1cm}
+\left[1+\xi^\c_{13}+\xi^\c_{14}+\xi^\c_{34}+\zeta^\c_{134}\right]
\hspace{-0.1cm}\frac{\dirac_{\h,12}}{\nbar_2}\left<g\right>\hspace{-0.1cm}
+\left[1+\xi^\c_{12}+\xi^\c_{24}+\xi^\c_{41}+\zeta^\c_{124}\right]
\hspace{-0.1cm}\frac{\dirac_{\h,23}}{\nbar_3}\left<g\right> \hspace{-0.1cm}
\right.\nn \\
& + & \left.
\left[1+\xi^\c_{12}+\xi^\c_{13}+\xi^\c_{23}+\zeta^\c_{123}\right]
\frac{\dirac_{\h,14}}{\nbar_4}\left<g\right>
+\left[1+\xi^\c_{12}+\xi^\c_{14}+\xi^\c_{24}+\zeta^\c_{124}\right]
\frac{\dirac_{\h,13}}{\nbar_3}\left<g\right>
\right.\nn \\
& + & \left.
\left[1+\xi^\c_{12}+\xi^\c_{23}+\xi^\c_{31}+\zeta^\c_{123}\right]
\frac{\dirac_{\h,24}}{\nbar_4}\left<g\right>
+\left[1+\xi^\c_{12}+\xi^\c_{23}+\xi^\c_{31}+\zeta^\c_{123}\right]
\frac{\dirac_{\h,34}}{\nbar_4}\left<g\right>
\right.\nn \\
& + & \left.
\left[1+\xi^\c_{13}\right]\hspace{-0.1cm}\frac{\dirac_{\h,12}\dirac_{\h,34}}{\nbar_2 \nbar_4}
\left<g\right>\hspace{-0.1cm} 
+\left[1+\xi^\c_{12}\right]\hspace{-0.1cm}\frac{\dirac_{\h,13}\dirac_{\h,24}}{\nbar_3 \nbar_4}
\left<g\right>\hspace{-0.1cm}
+\left[1+\xi^\c_{12}\right]\hspace{-0.1cm}\frac{\dirac_{\h,14}\dirac_{\h,23}}{\nbar_3 \nbar_4}
\left<g\right>\hspace{-0.1cm}
+\left[1+\xi^\c_{14}\right]\hspace{-0.1cm}\frac{\dirac_{\h,12}\dirac_{\h,13}}{\nbar_2 \nbar_3}
\left<g\right>\hspace{-0.1cm}
\right. \nn \\
& + & \left.
\left[1+\xi^\c_{13}\right]\hspace{-0.1cm}\frac{\dirac_{\h,12}\dirac_{\h,14}}{\nbar_2 \nbar_4}\left<g\right>
+\left[1+\xi^\c_{12}\right]\hspace{-0.1cm}\frac{\dirac_{\h,13}\dirac_{\h,14}}{\nbar_3 \nbar_4}\left<g\right>
+\left[1+\xi^\c_{12}\right]\hspace{-0.1cm}\frac{\dirac_{\h,23}\dirac_{\h,24}}{\nbar_3 \nbar_4}\left<g\right>
+\hspace{-0.1cm}\frac{\dirac_{\h,12}\dirac_{\h,13}\dirac_{\h,14}}{\nbar_2 \nbar_3 \nbar_4}\left<g\right> 
\hspace{-0.1cm}\right\}.
\nn \ea
Based on the above expression we are now in a position to immediately
write down the other 4-point function cases required to compute
\Eqn{eq:FgFgFgFg}:
\ba 
\left<n_{\g,1}n_{\g,2}n_{\g,3}n_{s,4}\right>
& = & \alpha^{-1}\nbar_1 \nbar_2 \nbar_3 \nbar_4 \left\{ \frac{}{} \hspace{-0.2cm}
\left(1+\xi^\c_{12}+\xi^\c_{13}+\xi^\c_{23}+\zeta^\c_{123}\right) \left<g\right>
+\left(1+\xi^\c_{13}\right)\frac{\dirac_{\h,12}}{\nbar_2}\left<g\right> \right.
\nn \\
& & \left.+\left(1+\xi^\c_{12}\right)\left[\frac{\dirac_{\h,23}}{\nbar_3} \left<g\right>
+\frac{\dirac_{\h,13}}{\nbar_3}\left<g\right>\right]
+\frac{\dirac_{\h,12}\dirac_{\h,13}}{\nbar_2 \nbar_3}\left<g\right> \right\}
\label{eq:ngngngns};
\\
\left<n_{\g,1}n_{\g,2}n_{s,3}n_{s,4}\right>
 &  = &
\alpha^{-2}\nbar_1 \nbar_2 \nbar_3 \nbar_4 \left\{ \frac{}{} \hspace{-0.2cm}
\left(1+\xi^\c_{12}\right)\left<g\right>+ \frac{\dirac_{\h,12}}{\nbar_2}\left<g\right>
+\alpha \left(1+\xi^\c_{12}\right)\frac{\dirac_{\h,34}}{\nbar_4}\left<g\right> 
+\alpha \frac{\dirac_{\h,12}\,\dirac_{\h,34}}{\nbar_2 \nbar_4}\left<g\right> \right\}
\label{eq:ngngnsns};
\\
\left<n_{\g,1}n_{s,2}n_{s,3}n_{s,4}\right>
&  = &
\alpha^{-3} \nbar_1 \nbar_2 \nbar_3 \nbar_4 \left\{ \frac{}{} \hspace{-0.2cm}
\left<g\right> + \alpha \left[\frac{\dirac_{\h,23}}{\nbar_3}+\frac{\dirac_{\h,24}}{\nbar_4}
+\frac{\dirac_{\h,34}}{\nbar_4}\right]\left<g\right> 
+\alpha \frac{\dirac_{\h,23}\dirac_{\h,24}}{\nbar_3 \nbar_4}\left<g\right> \right\}
\label{eq:ngnsnsns}\ ;
\\
\left<n_{s,1}n_{s,2}n_{s,3}n_{s,4}\right>
 &  = &
\alpha^{-4} \nbar_1\nbar_2 \nbar_3 \nbar_4 \left\{ \left<g\right> + 
\alpha \left[\frac{\dirac_{\h,12}}{\nbar_2} \left<g\right> + \frac{\dirac_{\h,23}}{\nbar_3}
\left<g\right> +\frac{\dirac_{\h,13}}{\nbar_3}\left<g\right> + \frac{\dirac_{\h,14}}{\nbar_4} 
\left<g\right> + \frac{\dirac_{\h,24}}{\nbar_4}\left<g\right> + \frac{\dirac_{\h,34}}{\nbar_4} 
\left<g\right> \right]
\right. \nn \\ 
& & + \left.
\alpha^2 \left[\frac{\dirac_{\h,12}\dirac_{\h,34}}{\nbar_2 \nbar_4}\left<g\right>
+ \frac{\dirac_{\h,13} \dirac_{\h,24}}{\nbar_3 \nbar_4}\left<g\right> 
+ \frac{\dirac_{\h,23} \dirac_{\h,14}}{\nbar_3\nbar_4}\left<g\right>
+ \frac{\dirac_{\h,12} \dirac_{\h,13}}{\nbar_2\nbar_3}\left<g\right> 
+ \frac{\dirac_{\h,12} \dirac_{\h,14}}{\nbar_2\nbar_4}\left<g\right> \right.\right.
\nn \\ 
& &  \left.\left.
+ \frac{\dirac_{\h,13} \dirac_{\h,14}}{\nbar_3\nbar_4}\left<g\right> 
+ \frac{\dirac_{\h,23} \dirac_{\h,24}}{\nbar_3\nbar_4}\left<g\right>\right] 
+ \alpha^3 \frac{\dirac_{\h,12} \dirac_{\h,13} \dirac_{\h,14}}
{\nbar_2 \nbar_3 \nbar_4} \left<g\right> \right\}.
\label{eq:nsnsnsns}
\ea

In order to compute the 4-point correlation function of $\Fg$, we
need to compute the sum of 4 terms of \Eqn{eq:ngnsnsns}, with
permuted location of g and s. This is given by:
\ba 
{L_1} & \equiv & 
\left<n_{\g,1}n_{s,2}n_{s,3}n_{s,4}\right> +
\left<n_{s,1}n_{\g,2}n_{s,3}n_{s,4}\right> +
\left<n_{s,1}n_{s,2}n_{\g,3}n_{s,4}\right> +
\left<n_{s,1}n_{s,2}n_{s,3}n_{\g,4}\right> 
\nn \\
& = & \nbar_1 \nbar_2 \nbar_3 \nbar_4\left\{ 4\alpha^{-3} \left<g\right>
+2\alpha^{-2}\left[\frac{\dirac_{\h,23}}{\nbar_3}\left<g\right>
+\frac{\dirac_{\h,24}}{\nbar_4}\left<g\right>+\frac{\dirac_{\h,34}}{\nbar_4}\left<g\right>
+ \frac{\dirac_{\h,13}}{\nbar_3} \left<g\right>+\frac{\dirac_{\h,14}}{\nbar_4}
\left<g\right>+ \frac{\dirac_{\h,12}}{\nbar_2}\left<g\right>\right] \right. \nn \\
& & + \left. \alpha^{-1}\left[\frac{\dirac_{\h,23}}{\nbar_3} \frac{\dirac_{\h,24}}{\nbar_4}\left<g\right>
+ \frac{\dirac_{\h,13}}{\nbar_3}\frac{\dirac_{\h,14}}{\nbar_4}\left<g\right>
+ \frac{\dirac_{\h,12}}{\nbar_2}\frac{\dirac_{\h,14}}{\nbar_4}\left<g\right>
+ \frac{\dirac_{\h,12}}{\nbar_2}\frac{\dirac_{\h,13}}{\nbar_3}\left<g\right>\right]\right\}.
\nn \ea
We also need to compute the sum of 6 terms of \Eqn{eq:ngngnsns}, with
permuted location of g and s. This is given by:
\ba 
L_2 & \equiv &  \left<n_{\g,1}n_{\g,2}n_{s,3}n_{s,4}\right> +{5\,\rm perms}\nn \\
& = & 
\nbar_1 \nbar_2 \nbar_3 \nbar_4 \alpha^{-2}\left\{
\left(6+\xi^\c_{12}+\xi^\c_{13}+\xi^\c_{14}+\xi^\c_{23}+\xi^\c_{24}+\xi^\c_{34}\right)
\left<g\right> +\frac{\dirac_{\h,12}}{\nbar_2}\left<g\right>
+\left[\frac{\dirac_{\h,13}}{\nbar_3}+\frac{\dirac_{\h,23}}{\nbar_3}\right]\left<g\right> \right.\nn\\
& + &\left. \left[\frac{\dirac_{\h,14}}{\nbar_4} +\frac{\dirac_{\h,24}}{\nbar_4} 
+ \frac{\dirac_{\h,34}}{\nbar_4}\right]\left<g\right> +\alpha \left(
\left[\left(1+\xi^\c_{12}\right)\frac{\dirac_{\h,34}}{\nbar_4}+\left(1+\xi^\c_{13}\right)
\frac{\dirac_{\h,24}}{\nbar_4}+\left(1+\xi^\c_{23}\right)\frac{\dirac_{\h,14}}{\nbar_4}\right]
\left<g\right> \right.\right.\nn \\
& + & 
\left.\left.\hspace{-0.2cm}\left[\left(1+\xi^\c_{14}\right)\frac{\dirac_{\h,23}}{\nbar_3}
\left<g\right>\hspace{-0.1cm}
+\left(1+\xi^\c_{24}\right)\frac{\dirac_{\h,13}}{\nbar_3}\right]\hspace{-0.1cm}
\left<g\right>\hspace{-0.1cm}
+\left(1+\xi^\c_{34}\right)\frac{\dirac_{\h,12}}{\nbar_2}\left<g\right>\hspace{-0.1cm} 
+2\frac{\dirac_{\h,12}}{\nbar_2}\frac{\dirac_{\h,34}}{\nbar_4} \left<g\right>\hspace{-0.1cm} 
+2\hspace{-0.1cm}\left[\frac{\dirac_{\h,13}}{\nbar_3}\frac{\dirac_{\h,24}}{\nbar_4} 
+\frac{\dirac_{\h,23}}{\nbar_3}\frac{\dirac_{\h,14}}{\nbar_4}\right]\hspace{-0.1cm}
\left<g\right>\hspace{-0.1cm} \right)\right\}.
\nn \ea
In addition we need to compute the sum of 4 terms of
\Eqn{eq:ngngngns}, again where the locations of g and s are
permuted. This sum can be written:
\ba 
L_3 & \equiv & 
\left<n_{\g,1}n_{\g,2}n_{\g,3}n_{s,4}\right>+
\left<n_{\g,1}n_{\g,2}n_{s,3}n_{\g,4}\right>+
\left<n_{\g,1}n_{s,2}n_{\g,3}n_{\g,4}\right>+
\left<n_{s,1}n_{\g,2}n_{\g,3}n_{\g,4}\right>\nn \\
& = & 
\alpha^{-1}\nbar_1\nbar_2\nbar_3\nbar_4\left\{\frac{}{}\hspace{-0.2cm}
\left(4+2\xi^\c_{12}+2\xi^\c_{13}+2\xi^\c_{23}
+2\xi^\c_{14}+2\xi^\c_{24}+2\xi^\c_{34}+
\zeta^\c_{123}+\zeta^\c_{124}+\zeta^\c_{134}+\zeta^\c_{234}
\right)\left<g\right>\right.
\nn \\
& + & 
\left.\left(1+\xi^\c_{13}\right)\frac{\dirac_{\h,12}}{\nbar_2}\left<g\right>
+\left(1+\xi^\c_{12}\right)\frac{\dirac_{\h,23}}{\nbar_3}\left<g\right>
+\left(1+\xi^\c_{12}\right)\frac{\dirac_{\h,31}}{\nbar_3}\left<g\right>\right.
+\left(1+\xi^\c_{14}\right)\frac{\dirac_{\h,12}}{\nbar_2}\left<g\right>
+\left(1+\xi^\c_{12}\right)\frac{\dirac_{\h,24}}{\nbar_4} \left<g\right>
\nn \\
& + & 
\left. \left(1+\xi^\c_{12}\right)\frac{\dirac_{\h,41}}{\nbar_4}\left<g\right> 
+\left(1+\xi^\c_{14}\right)\frac{\dirac_{\h,13}}{\nbar_3}\left<g\right>
+\left(1+\xi^\c_{13}\right)\frac{\dirac_{\h,14}}{\nbar_4}\left<g\right>
+\left(1+\xi^\c_{13}\right)\frac{\dirac_{\h,34}}{\nbar_4}\left<g\right>
+\left(1+\xi^\c_{24}\right)\frac{\dirac_{\h,23}}{\nbar_3}\left<g\right>
\right.
\nn \\
& + & 
\left. \left(1+\xi^\c_{23}\right)\frac{\dirac_{\h,24}}{\nbar_4}\left<g\right>
+\left(1+\xi^\c_{23}\right)\frac{\dirac_{\h,34}}{\nbar_4}\left<g\right>
+\frac{\dirac_{\h,12}}{\nbar_2}\frac{\dirac_{\h,13}}{\nbar_3}\left<g\right>
+\frac{\dirac_{\h,12}}{\nbar_2}\frac{\dirac_{\h,14}}{\nbar_4}\left<g\right> 
+\frac{\dirac_{\h,13}}{\nbar_3}\frac{\dirac_{\h,14}}{\nbar_4}\left<g\right>
+\frac{\dirac_{\h,23}}{\nbar_3}\frac{\dirac_{\h,24}}{\nbar_4}\left<g\right>\right\}.
\nn \ea
Collecting the terms $L_1$, $L_2$ and $L_3$ along with
$\left<n_{\g,1}n_{\g,2}n_{\g,3}n_{\g,4}\right>$ and
$\left<n_{s,1}n_{s,2}n_{s,3}n_{s,4}\right>$ and inserting them into
\Eqn{eq:FgFgFgFg}, and after some algebra we arrive at the following
arrangement:
\ba
\left<F_{\g,1}\dots F_{\g,4}\right> & = &
\frac{1}{A^2} \prod_{i=1}^4 \left\{\int dL_i \dx_i \,dM_i \,\nbar_i w_i\right\} 
\left\{\frac{}{} \eta^\c_{1234} \left<g\right>
+ \left[\xi^\c_{12}+\frac{(1+\alpha)}{\nbar_2}\dirac_{\h,12}\right]
 \left[\xi^\c_{34}+\frac{(1+\alpha)}{\nbar_4}\dirac_{\h,34}\right]\left<g\right>
\right.
\nn \\
&  & + \left.
\left[\xi^\c_{13}+\frac{(1+\alpha)}{\nbar_3}\dirac_{\h,13}\right]
\left[\xi^\c_{24}+\frac{(1+\alpha)}{\nbar_4}\dirac_{\h,24}\right]\left<g\right>
+\left[\xi^\c_{14}+\frac{(1+\alpha)}{\nbar_4}\dirac_{\h,14}\right]
\left[\xi^\c_{23}+\frac{(1+\alpha)}{\nbar_3}\dirac_{\h,23}\right]\left<g\right> \right. 
\nn \\
& & \left. 
+ \zeta^\c_{134}\frac{\dirac_{\h,12}}{\nbar_2}\left<g\right>
+\zeta^\c_{124}\frac{\dirac_{\h,23}}{\nbar_3} \left<g\right> 
+\zeta^\c_{123}\frac{\dirac_{\h,14}}{\nbar_4}\left<g\right>
+\zeta^\c_{124}\frac{\dirac_{\h,13}}{\nbar_3}\left<g\right>
+\zeta^\c_{123}\frac{\dirac_{\h,24}}{\nbar_4}\left<g\right>
+\zeta^\c_{123}\frac{\dirac_{\h,34}}{\nbar_4}\left<g\right> \right.
\nn \\
& & \left. 
+ \xi^\c_{13}\frac{\dirac_{\h,12}\,\dirac_{\h,34}}{\nbar_2\nbar_4}\left<g\right>
+\xi^\c_{12}\frac{\dirac_{\h,13}\,\dirac_{\h,24}}{\nbar_3\nbar_4}\left<g\right>
+\xi^\c_{12}\frac{\dirac_{\h,14}\,\dirac_{\h,23}}{\nbar_3\nbar_4}\left<g\right>
+\xi^\c_{14}\frac{\dirac_{\h,12}\dirac_{\h,13}}{\nbar_2\nbar_3}\left<g\right>
+\xi^\c_{13}\frac{\dirac_{\h,12}\dirac_{\h,14}}{\nbar_2\nbar_4}\left<g\right>\right.
\nn \\
& & \left.
+\xi^\c_{12}\frac{\dirac_{\h,13}\dirac_{\h,14}}{\nbar_3\nbar_4}\left<g\right>
+\xi^\c_{12}\frac{\dirac_{\h,23}\dirac_{\h,24}}{\nbar_3\nbar_4}\left<g\right>
+\frac{(1+\alpha^3)\dirac_{\h,12}\,\dirac_{\h,13}\,\dirac_{\h,14}}
{\nbar_2\nbar_3\nbar_4}\left<g\right>\right\},
\nn \ea
where in the above we used the short-hand notation $F_{\g,i}\equiv
\Fg(\br_i)$. In order to compute the covariance we also require the
second term in \Eqn{eq:covF}. On repeatedly using \Eqn{eq:xihm2} we
find that this can be written:
\be 
\left<F_{\g,1}F_{\g,3}\right> \left<F_{\g,2}F_{\g,4}\right>=  
\prod_{i=1}^4 \left\{\int dL_i \dx_i dM_i \,\nbar_i w_i\right\}
\left[\frac{}{}\xi^\c_{13}\left<g\right>+\frac{(1+\alpha)\dirac_{\h,13}}{\nbar_3}\left<g\right>
\right]
\left[\frac{}{}\xi^\c_{24}\left<g\right>+\frac{(1+\alpha)\dirac_{\h,24}}{\nbar_4}\left<g\right>
\right]. 
\nn\ee
Joining the last two equations, we write the covariance as:
\ba
& & \hspace{-1cm}
\left<\Fg(\br_1)\Fg(\br_2)\Fg(\br_3)\Fg(\br_4)\right> 
- \left<\Fg(\br_1)\Fg(\br_3)\right> \left<\Fg(\br_2)\Fg(\br_4)\right> = 
\frac{1}{A^2} \prod_{i=1}^4 \left\{\int dL_i \dx_i dM_i\, \nbar_i w_i\right\}
\left\{\frac{\hspace{-0.1cm}}{\hspace{-0.1cm}} \eta^\c_{1234} \left<g\right> + \right.
\nn \\
& & \left.\hspace{-0.2cm}
+ \left[\xi^\c_{12}+\frac{(1+\alpha)}{\nbar_2}\dirac_{\h,12}\right]
 \left[\xi^\c_{34}+\frac{(1+\alpha)}{\nbar_4}\dirac_{\h,34}\right]\left<g\right>
+\left[\xi^\c_{14}+\frac{(1+\alpha)}{\nbar_4}\dirac_{\h,14}\right]
 \left[\xi^\c_{23}+\frac{(1+\alpha)}{\nbar_3}\dirac_{\h,23}\right]\left<g\right>
+\zeta^\c_{134}\frac{\dirac_{\h,12}}{\nbar_2}\left<g\right>
\right.
\nn \\
& & \left.\hspace{-0.2cm}
+\zeta^\c_{124}\frac{\dirac_{\h,23}}{\nbar_3} \left<g\right>
+\zeta^\c_{123}\frac{\dirac_{\h,14}}{\nbar_4}\left<g\right>
+\zeta^\c_{124}\frac{\dirac_{\h,13}}{\nbar_3}\left<g\right>
+\zeta^\c_{123}\frac{\dirac_{\h,24}}{\nbar_4}\left<g\right>
+\zeta^\c_{123}\frac{\dirac_{\h,34}}{\nbar_4}\left<g\right>
+\xi^\c_{13}\frac{\dirac_{\h,12}\dirac_{\h,34}}{\nbar_2\nbar_4}\left<g\right>
+\xi^\c_{12}\frac{\dirac_{\h,13}\dirac_{\h,24}}{\nbar_3\nbar_4}\left<g\right> 
\right. \nn \\
& & \left. \hspace{-0.2cm}
+\xi^\c_{12}\frac{\dirac_{\h,14}\dirac_{\h,23}}{\nbar_3\nbar_4}\hspace{-0.1cm}\left<g\right>\hspace{-0.1cm}
+\xi^\c_{14}\frac{\dirac_{\h,12}\dirac_{\h,13}}{\nbar_2\nbar_3}\hspace{-0.1cm}\left<g\right>\hspace{-0.1cm}
+\xi^\c_{13}\frac{\dirac_{\h,12}\dirac_{\h,14}}{\nbar_2\nbar_4}\hspace{-0.1cm}\left<g\right>\hspace{-0.1cm}
+\xi^\c_{12}\frac{\dirac_{\h,13}\dirac_{\h,14}}{\nbar_3\nbar_4}\hspace{-0.1cm}\left<g\right>\hspace{-0.1cm}
+\xi^\c_{12}\frac{\dirac_{\h,23}\dirac_{\h,24}}{\nbar_3\nbar_4}\hspace{-0.1cm}\left<g\right>\hspace{-0.1cm}
+\frac{(1+\alpha^3)\dirac_{\h,12}\dirac_{\h,13}\dirac_{\h,14}}
{\nbar_2\nbar_3\nbar_4}\left<g\right>\right\}\nn\\
\label{eq:four1}
\ea

We now make the assumption that the fluctuations are close to
Gaussian, hence we take $\eta=\zeta=0$.

\ba
& & \hspace{-1.cm}
\left<\Fg(\br_1)\Fg(\br_2)\Fg(\br_3)\Fg(\br_4)\right> \hspace{-0.1cm}
-  \hspace{-0.1cm}\left<\Fg(\br_1)\Fg(\br_3)\right> \left<\Fg(\br_2)\Fg(\br_4)\right> = 
\frac{1}{A^2} \prod_{i=1}^4 \left\{\int dL_i \dx_i dM_i\, \nbar_i w_i\right\}
\left\{\frac{(1+\alpha^3)\dirac_{\h,12}\dirac_{\h,13}\dirac_{\h,14}}
{\nbar_2\nbar_3\nbar_4}\left<g\right> 
\right. \nn \\
& & 
\left. \frac{}{}\left[\xi^\c_{12}+\frac{(1+\alpha)}{\nbar_2}\dirac_{\h,12}\right]
 \left[\xi^\c_{34}+\frac{(1+\alpha)}{\nbar_4}\dirac_{\h,34}\right]\left<g\right>
+\left[\xi^\c_{14}+\frac{(1+\alpha)}{\nbar_4}\dirac_{\h,14}\right]
 \left[\xi^\c_{23}+\frac{(1+\alpha)}{\nbar_3}\dirac_{\h,23}\right]\left<g\right>
+\xi^\c_{14}\frac{\dirac_{\h,12}\dirac_{\h,13}}{\nbar_2\nbar_3}\left<g\right>
\right.\nn \\
& &  \left.
+\xi^\c_{13}\left[\frac{\dirac_{\h,12}\dirac_{\h,14}}{\nbar_2\nbar_4}\left<g\right>
+ \frac{\dirac_{\h,12}\dirac_{\h,34}}{\nbar_2\nbar_4}\left<g\right> \right]
+ \xi^\c_{12}\left[ \frac{\dirac_{\h,13}\dirac_{\h,14}}{\nbar_3\nbar_4}\left<g\right>
+ \frac{\dirac_{\h,13}\dirac_{\h,24}}{\nbar_3\nbar_4}\left<g\right>
+ \frac{\dirac_{\h,23}\dirac_{\h,14}}{\nbar_3\nbar_4}\left<g\right> 
+ \frac{\dirac_{\h,23}\dirac_{\h,24}}{\nbar_3\nbar_4}\left<g\right> \right]\right\}
\label{eq:four2}
\ea
On taking the limit that $\nbarg \Vu\gg 1$, the first and last three
terms will be sub-dominant \citep{Smith2009}. Using the linear bias
model from \Eqn{eq:linbias}, we write
\ba
& & \hspace{-1.cm}
\left<\Fg(\br_1)\Fg(\br_2)\Fg(\br_3)\Fg(\br_4)\right> 
- \left<\Fg(\br_1)\Fg(\br_3)\right> \left<\Fg(\br_2)\Fg(\br_4)\right> = 
\frac{1}{A^2} \prod_{i=1}^4 \left\{\int dL_i \dx_i dM_i\, \nbar_i w_i\right\}
\nn \\
& & 
\hspace{-0.6cm}\times \left\{\hspace{-0.1cm}
 \left[b_1b_2\xi_{12}+\hspace{-0.1cm}\frac{(1+\alpha)}{\nbar_2}\dirac_{\h,12}\right]\hspace{-0.1cm}
 \left[b_3b_4\xi_{34}+\hspace{-0.1cm}\frac{(1+\alpha)}{\nbar_4}\dirac_{\h,34}\right]\hspace{-0.1cm}\left<g\right>
+\hspace{-0.1cm}\left[b_1b_4\xi_{14}+\hspace{-0.1cm}\frac{(1+\alpha)}{\nbar_4}\dirac_{\h,14}\right]\hspace{-0.1cm}
 \left[b_2b_3\xi_{23}+\hspace{-0.1cm}\frac{(1+\alpha)}{\nbar_3}\dirac_{\h,23}\right]\hspace{-0.1cm}\left<g\right>
\hspace{-0.1cm}\right\}\hspace{-0.1cm}.
 \label{eq:four3}
\ea


\subsection{Averaging over the galaxy distributions}
\label{app:B2}


Let us now return to the evaluation of the expectation values for the
galaxy population. Consider again \Eqn{eq:four3} and let us look in
particular at the terms $\left<g\right>$ and any premultiplying Dirac
delta functions. For compactness, we shall use the following definition:
\be
f(p|q)\equiv \theta(\br_p|L_p) \Phi(L_p|M_q) U(\br_p-\bx_q|M_q)
\label{eq:fcomp}
\ee
We find that there are three types of terms forming the individual
$\left<g\right>$ factors:
\ba
\left<g\right> & \rightarrow & 
\prod_{p=1}^{4} \left\{ N^{(1)}_{\g,p} f(p|p) \right\} \ ;
\nn \\
\dirac_{\h,14}\left<g\right> & \rightarrow & 
\prod_{p=1}^3\left\{ N^{(1)}_{\g,p} f(p|p)\right\} \left[\frac{N^{(2)}_{\g,1}}{N^{(1)}_{\g,1}}
f(4|1) + \dirac_{\g,14}\right] \dirac_{\h,14} \ ;
\nn \\
\dirac_{\h,12}\dirac_{\h,34} \left<g\right>
& \rightarrow & \hspace{-0.2cm}\prod_{p\in\{1,3\}} \hspace{-0.1cm}\left\{ N^{(1)}_{\g,p} f(p|p)\right\} 
\left[\frac{N^{(2)}_{\g,1}}{N^{(1)}_{\g,1}}
f(2|1) +  \dirac_{\g,12} \right]\left[\frac{N^{(2)}_{\g,3}}{N^{(1)}_{\g,3}}f(4|3) + \dirac_{\g,34}\right]
\dirac_{\h,12}\dirac_{\h,34} \ .
\ea
The rest of the $\left<g\right>$ factors can be worked out in a
similar way. Embedding them into \Eqn{eq:four3}, and using
\Eqns{eq:covF0}{eq:covF}, we arrive at the expression for the
covariance of the power spectrum estimator in the Gaussian
approximation:
\ba
{\rm Cov}\!\left[|\Fg(\bk_1)|^2,|\Fg(\bk_2)|^2\right] &\hspace{-0.2cm} =\hspace{-0.2cm} &
\left|\int \frac{\dq}{(2\pi)^3} P(\bq) G_{(1,1)}(\bk_1, \bq) G_{(1,1)}(\bk_2, -\bq) + (1+\alpha)
\left[G_{(2,0)}(\bk_1+\bk_2, {\bf 0}) + G(\bk_1, \bk_2)\right] \right|^2 \nn \\
&  \hspace{-2.8cm} + & \hspace{-1.4cm}
\left|\int \frac{\dq}{(2\pi)^3} P(\bq) G_{(1,1)}(\bk_1, \bq) G_{(1,1)}(-\bk_2, -\bq) + (1+\alpha)
\left[G_{(2,0)}(\bk_1-\bk_2, {\bf 0}) + G(\bk_1, -\bk_2)\right] \right|^2,
\label{eq:GaussCov}
\ea
where we have defined two more functions:
\ba
G_{(l,m)}(\bk, \bq) & \equiv &\int dM \nbar(M) b^m(M) N^{(1)}_{\g}(M) 
\tilde{\mathcal W}^U_{(l)}(\bk, \bq, M) \ ;\nn \\
G(\bk_1, \bk_2) & \equiv &  \int dM \nbar(M) N^{(2)}_{\g}(M)\int\frac{\dq}{(2\pi)^3}
\tilde{\mathcal W}^U_{(1)}(\bk_1, \bq, M) \tilde{\mathcal W}^U_{(1)}(\bk_2, -\bq, M) \ .
\ea


\subsection{The covariance matrix in the large-scale limit}
\label{app:B3}


In the large-scale limit, the profiles of the galaxies behave like
Dirac delta functions, e.g. $U(\br-\bx|M)\rightarrow \dirac(\br-\bx)$.
It is straightforward to show that in this limit the above-defined
functions become:
\[
G_{(l,m)}(\bk, \bq)  \rightarrow  \tilde{\mathcal G}^{(1)}_{(l, m)}(\bk+\bq) \ \ \ ; \ \ \
G(\bk_1, \bk_2)  \rightarrow  \tilde{\mathcal G}^{(2)}_{(1, 0)}(\bk_1+\bk_2),
\]
where $\tilde{\mathcal G}^{(n)}_{(l,m)}$ are the Fourier transforms of
the functions defined by CITE later. With these changes,
\Eqn{eq:GaussCov} can be expressed as:
\ba
{\rm Cov}\!\left[|\Fg(\bk_1)|^2,|\Fg(\bk_2)|^2\right] &\hspace{-0.2cm} =\hspace{-0.2cm} &
\left|\int \frac{\dq}{(2\pi)^3} P(\bq) \tilde{\mathcal G}^{(1)}_{(1,1)}(\bk_1+\bq) 
\tilde{\mathcal G}^{(1)}_{(1,1)}(\bk_2-\bq) + (1+\alpha)
\left[\tilde{\mathcal G}^{(1)}_{(2,0)}(\bk_1+\bk_2) + \tilde{\mathcal G}^{(2)}_{(1,0)}(\bk_1+\bk_2)\right] 
\right|^2 \nn \\
&  \hspace{-2.8cm} + & \hspace{-1.4cm}
\left|\int \frac{\dq}{(2\pi)^3} P(\bq) \tilde{\mathcal G}^{(1)}_{(1,1)}(\bk_1+\bq) 
\tilde{\mathcal G}^{(1)}_{(1,1)}(-\bk_2-\bq) + (1+\alpha)
\left[\tilde{\mathcal G}^{(1)}_{(2,0)}(\bk_1-\bk_2) + \tilde{\mathcal G}^{(2)}_{(1,0)}(\bk_1-\bk_2)
\right] \right|^2,
\nn \ea
which is exactly \Eqn{eq:CovLS}. This concludes our proof of it.


\section{Shell averaging the covariance matrix of the $\Fgt$ 
power spectrum}
\label{app:qresult}


A reasonable approximation when computing the shell-averaged power, is
that if the shells are narrow compared to the scale over which the
power spectrum varies, one can factor the latter out of the integrals
in \Eqn{eq:CovLS3}, writing:
\ba
{\rm Cov}\!\left[|\Fgt(k_i)|^2,|\Fgt(k_j)|^2\right] 
& = &
2{\overline P}^2(k_i)
\int_{V_{i}} \frac{\dk_1}{V_i}
\int_{V_{j}} \frac{\dk_2}{V_j} 
 \tilde{\mathcal Q}^{(1,1)}_{(1,1|1,1)}(\bk_1+\bk_2)\tilde{\mathcal Q}^{(1,1)}_{(1,1|1,1)}(-\bk_1-\bk_2)
\nn \\
& & 
+
4(1+\alpha)
\overline{P}(k_i) 
\int_{V_{i}} \frac{\dk_1}{V_i}
\int_{V_{j}} \frac{\dk_2}{V_j} 
\tilde{\mathcal Q}^{(1,1)}_{(1,1|1,1)}(\bk_1+\bk_2)\tilde{\mathcal Q}^{(2)}_{(1|0)}(-\bk_1-\bk_2)
\nn \\
& & 
+
4(1+\alpha)
\overline{P}(k_i) 
\int_{V_{i}} \frac{\dk_1}{V_i}
\int_{V_{j}} \frac{\dk_2}{V_j} 
\tilde{\mathcal Q}^{(1,1)}_{(1,1|1,1)}(\bk_1+\bk_2)\tilde{\mathcal Q}^{(1)}_{(2|0)}(-\bk_1-\bk_2)
\nn \\
& & 
+
4(1+\alpha)^2
\int_{V_{i}} \frac{\dk_1}{V_i}
\int_{V_{j}} \frac{\dk_2}{V_j} 
\tilde{\mathcal Q}^{(2)}_{(1|0)}(\bk_1+\bk_2)\tilde{\mathcal Q}^{(1)}_{(2|0)}(-\bk_1-\bk_2)
\nn \\
& & 
+
2(1+\alpha)^2\int_{V_{i}} \frac{\dk_1}{V_i}
\int_{V_{j}} \frac{\dk_2}{V_j} 
\tilde{\mathcal Q}^{(2)}_{(1|0)}(\bk_1+\bk_2)\tilde{\mathcal Q}^{(2)}_{(1|0)}(-\bk_1-\bk_2)
\nn \\
& & 
+
2(1+\alpha)^2\int_{V_{i}} \frac{\dk_1}{V_i}
\int_{V_{j}} \frac{\dk_2}{V_j} 
\tilde{\mathcal Q}^{(1)}_{(2|0)}(\bk_1+\bk_2)\tilde{\mathcal Q}^{(1)}_{(2|0)}(-\bk_1-\bk_2)
\label{eq:CovLS4} \ .
\ea
Our task now is to solve the integrals forming the terms of
\Eqn{eq:CovLS4}. In general, this integrals have the form:
\ba
\int_{V_{i}} \frac{\dk_1}{V_i} \int_{V_{j}} \frac{\dk_2}{V_j} 
\tilde{\mathcal Q}_{(l_1,l_2|m_1,m_2)}^{(n_1,n_2)}(\bk_1+\bk_2)
\tilde{\mathcal Q}_{(l'_1,l'_2|m'_1,m'_2)}^{(n'_1,n'_2)}(-\bk_1-\bk_2) \nn \\
& & \hspace{-7cm} = 
\int_{V_{i}} \frac{\dk_1}{V_i} \int_{V_{j}} \frac{\dk_2}{V_j} \int \dr_1 \dr_2
{\mathcal Q}_{(l_1,l_2|m_1,m_2)}^{(n_1,n_2)}(\br_1) 
{\mathcal Q}_{(l'_1,l'_2|m'_1,m'_2)}^{(n'_1,n'_2)}(\br_2)
{\rm e}^{i(\bk_1+\bk_2)\cdot(\br_1-\br_2)} \nn \\
& &  \hspace{-7cm} = 
\int \dr_1 \dr_2 {\mathcal Q}_{(l_1,l_2|m_1,m_2)}^{(n_1,n_2)}(\br_1) 
{\mathcal Q}_{(l'_1,l'_2|m'_1,m'_2)}^{(n'_1,n'_2)}(\br_2)
\int_{V_{i}} \frac{\dk_1}{V_i} {\rm e}^{i \bk_1\cdot(\br_1-\br_2)}
\int_{V_{j}} \frac{\dk_2}{V_j} {\rm e}^{i \bk_2\cdot(\br_1-\br_2)} \nn \\
& &  \hspace{-7cm} = 
\int \dr_1 \dr_2 {\mathcal Q}_{(l_1,l_2|m_1,m_2)}^{(n_1,n_2)}(\br_1) 
{\mathcal Q}_{(l'_1,l'_2|m'_1,m'_2)}^{(n'_1,n'_2)}(\br_2)
\overline{j_0}(k_i|\br_1-\br_2|) \overline{j_0}(k_j|\br_1-\br_2|) \nn \\
& &  \hspace{-7cm} = 
\int \dr_{21} \overline{j_0}(k_ir_{21}) \overline{j_0}(k_jr_{21}) 
\Sigma_{(l_1,l_2|m_1,m_2)(l'_1,l'_2|m'_1,m'_2)}^{(n_1,n_2)(n'_1,n'_2)}(r_{21}) 
\label{eq:INT}
\ea
In the above, we have defined the shell-average of the spherical
Bessel function as
\be \overline{j_0}(k_i r)\equiv \frac{1}{V_i}\int_{k_i-\Delta k/2}^{k_i+\Delta k/2} 
dk_1 k_1^2 4\pi j_0(k_1 r) \ .
\ee
To obtain the last line of \Eqn{eq:INT}, we made a change of variables
$\br_{21}=\br_2-\br_1$, and defined the correlation function of the
weighted survey window function to be,
\be 
\Sigma_{(l_1,l_2|m_1,m_2)(l'_1,l'_2|m'_1,m'_2)}^{(n_1,n_2)(n'_1,n'_2)}(r_{21}) 
\equiv \int \frac{{\rm d}^2\hat{\br}_{21}}{4\pi}\int \dr_1 
{\mathcal Q}_{(l_1,l_2|m_1,m_2)}^{(n_1,n_2)}(\br_1)
{\mathcal Q}_{(l'_1,l'_2|m'_1,m'_2)}^{(n'_1,n'_2)}(\br_{21}+\br_1)\ . 
\label{eq:Xi:APP}
\ee
In the limit that the survey volume is large, the weighted survey
window correlation function is very slowly varying over nearly all
length scales of interest, and so can be approximated by its value
at zero-lag. Using the orthogonality relation of the Bessel functions,
%
$\int_0^{\infty} dr r^2 j_\alpha(ur)j_\alpha(vr) = 
( \pi/2u^2)\dirac(u-v)$
%
we write:
\ba
\int_{V_{i}} \frac{\dk_1}{V_i} \int_{V_{j}} \frac{\dk_2}{V_j} 
\tilde{\mathcal Q}_{(l_1,l_2|m_1,m_2)}^{(n_1,n_2)}(\bk_1+\bk_2)
\tilde{\mathcal Q}_{(l'_1,l'_2|m'_1,m'_2)}^{(n'_1,n'_2)}(-\bk_1-\bk_2) 
\approx
\frac{(2\pi)^3}{V_i} \Sigma_{(l_1,l_2|m_1,m_2)(l'_1,l'_2|m'_1,m'_2)}^{(n_1,n_2)(n'_1,n'_2)}(0) 
\delta^{K}_{i,j} \ . 
\label{eq:Qintegrals}
\ea
We shall now apply this result to the six terms of \Eqn{eq:CovLS4} and
write for each of them:
\ba 
\Sigma^{(1,1)(1,1)}_{(1,1|1,1)(1,1|1,1)}(0) &=& \int\dr \left[{\mathcal
    Q}^{(1,1)}_{(1,1|1,1)}(\br)\right]^2 = \int\dr \left[{\mathcal
    G}^{(1)}_{(1,1)}(\br)\right]^4
 \ ;\\
\Sigma_{(1,1|1,1)(1|0)}^{(1,1)(2)}(0) 
&=& \int\dr {\mathcal Q}^{(1,1)}_{(1,1|1,1)}(\br){\mathcal Q}^{(2)}_{(1|0)}(\br)
= \int\dr \left[{\mathcal G}^{(1)}_{(1,1)}(\br)\right]^2{\mathcal G}^{(2)}_{(1,0)}(\br)
\ ;\\
\Sigma_{(1,1|1,1)(2|0)}^{(1,1)(1)}(0) 
&=& \int\dr {\mathcal Q}^{(1,1)}_{(1,1|1,1)}(\br){\mathcal Q}^{(1)}_{(2|0)}(\br)
= \int\dr \left[{\mathcal G}^{(1)}_{(1,1)}(\br)\right]^2{\mathcal G}^{(1)}_{(2,0)}(\br)
\ ;\\
\Sigma^{(2)(1)}_{(1|0)(2|0)}(0) 
&=& \int\dr {\mathcal Q}^{(2)}_{(1|0)}(\br){\mathcal Q}^{(1)}_{(2|0)}(\br)
= \int\dr {\mathcal G}^{(2)}_{(1,0)}(\br){\mathcal G}^{(1)}_{(2,0)}(\br)
\ ;\\
\Sigma^{(2)(2)}_{(1|0)(1|0)}(0) 
&=& \int\dr \left[{\mathcal Q}^{(2)}_{(1|0)}(\br)\right]^2
= \int\dr \left[{\mathcal G}^{(2)}_{(1,0)}(\br)\right]^2
\ ;\\
\Sigma^{(1)(1)}_{(2|0)(2|0)}(0) 
&=& \int\dr \left[{\mathcal Q}^{(1)}_{(2|0)}(\br)\right]^2
= \int\dr \left[{\mathcal G}^{(1)}_{(2,0)}(\br)\right]^2
\ea
Finally, putting together all these terms we write our final
expression for the shell-averaged covariance as:
\ba
{\rm Cov}\!\left[|\Fgt(k_i)|^2,|\Fgt(k_j)|^2\right] = 
\frac{2(2\pi)^3}{V_{i}}{\overline P}^2(k_i)\delta^{K}_{i,j}
\int \dr \left\{ \left[{\mathcal G}^{(1)}_{(1,1)}(\br)\right]^2 + 
\frac{(1+\alpha)}{\overline{P}(k_i)} \left[ 
{\mathcal G}^{(2)}_{(1,0)}(\br) + {\mathcal G}^{(1)}_{(2,0)}(\br)\right]
\right\}^2 \ .
\nn\ea
which is in fact \Eqn{eq:cov_final} from the main text.


\section{Functional derivatives}
\label{app:functionals}


In order to compute the functional derivatives of $\mathcal{N}$ and
$\mathcal{D}$ making up $F[w]$, we must first work out the functional
derivatives of the functions $\Gb$ and the normalisation $A$.


\subsection{Functional derivatives of the $\Gb$ functions and normalisation $A$}

For small variations in the path of $w$ we find that the functional
derivative of $\Gb$ can be written:
\ba
\Gb^{(1)}_{(1,1)}[w+\delta w] & = & \hspace{-0.1cm} 
\int\hspace{-0.1cm} dM \nbar(M) b(M) N^{(1)}_{\g}(M)\hspace{-0.1cm} 
\int \hspace{-0.1cm} dL \Phi(L|M)\Theta(\br|L)\left[w(\br, L, M)+\delta w(\br, L, M)\right] 
= \Gb^{(1)}_{(1,1)}[w] + \delta\Gb^{(1)}_{(1,1)}[w]; \nn \\
\delta\Gb^{(1)}_{(1,1)}[w] & \equiv & \int\hspace{-0.1cm} dM \nbar(M) b(M) 
N^{(1)}_{\g}(M)\hspace{-0.1cm} \int \hspace{-0.1cm} dL \Phi(L|M)\Theta(\br|L)
\delta w(\br, L, M) \ ; 
\label{eq:devG111} \\
\Gb^{(2)}_{(1,0)}[w+\delta w] & = & \int\hspace{-0.1cm} dM \nbar(M) N^{(2)}_{\g}(M)\hspace{-0.1cm}
\left\{\int dL \Phi(L|M)\Theta(\br|L)\left[w(\br, L, M)+\delta w(\br, L, M)\right]\right\}^2
=\Gb^{(2)}_{(1,0)}[w] + \delta\Gb^{(2)}_{(1,0)}[w]; \nn \\
\delta\Gb^{(2)}_{(1,0)}[w] & \equiv & 2\int\hspace{-0.1cm} dM \nbar(M) N^{(2)}_{\g}(M) 
\overline{\mathcal W}_1(\br, M) \int dL \Phi(L|M)\Theta(\br|L) 
\delta w(\br, L, M) \ ; 
\label{eq:devG210} \\
\Gb^{(1)}_{(2,0)}[w+\delta w] & = & \hspace{-0.1cm} \int\hspace{-0.1cm} dM 
\nbar(M) N^{(1)}_{\g}(M)\hspace{-0.1cm} \int dL \Phi(L|M)\Theta(\br|L)
\left[w(\br, L, M)+\delta w(\br, L, M)\right]^2 = \Gb^{(1)}_{(2,0)}[w]
+ \delta  \Gb^{(1)}_{(2,0)}[w] \ ; \nn \\
\delta \Gb^{(1)}_{(2,0)}[w] & \equiv & 2 \int\hspace{-0.1cm} dM \nbar(M) 
N^{(1)}_{\g}(M)\hspace{-0.1cm} \int \hspace{-0.1cm} dL \Phi(L|M)\Theta(\br|L)
w(\br, L, M) \delta w(\br, L, M). 
\label{eq:devG120}
\ea
In the above we have neglected the terms containing $[\delta w]^n$
with $n\geq 2$, and we have used a similar definition to
\Eqn{eq:Gscaled} and defined:
\be
\overline{\mathcal W}_l(\br, M) = A^{l/2}{\mathcal W}_l(\br, M) \ .
\label{eq:Wbar}
\ee
Again, for small variations in the value of $w$, the functional
derivative of the normalization constant $A$ can be written:
\ba
A[w+\delta w] &=& \hspace{-0.1cm} \int \hspace{-0.1cm} \dr \left(\Gb^{(1)}_{(1,1)}
[w+\delta w]\right)^2 = \hspace{-0.1cm} \int \hspace{-0.1cm} \dr
\left( \Gb^{(1)}_{(1,1)}[w] + \delta \Gb^{(1)}_{(1,1)}[w]\right)^2
= \hspace{-0.1cm} \int \hspace{-0.1cm} \dr\left[ \Gb^{(1)}_{(1,1)}(\br) \right]^2
+ 2 \hspace{-0.1cm} \int\hspace{-0.1cm}  \dr\, \Gb^{(1)}_{(1,1)}(\br) 
\delta \Gb^{(1)}_{(1,1)}[w] \nn \\ 
&=& A[w] + \delta A[w] \ , \nn \\
\delta A[w] & \equiv & 2 \int \dr \, \Gb^{(1)}_{(1,1)}(\br)  \int dM \, \nbar(M) b(M) 
N^{(1)}_{\g}(M)\hspace{-0.1cm} \int \hspace{-0.1cm} dL \Phi(L|M)\Theta(\br|L)
\delta w(\br, L, M) \ .
\label{eq:devA}
\ea
%
\vspace{-0.3cm}
\subsection{Functional derivative of $\mathcal{N}[w(\br,L,M)]$}
\vspace{-0.3cm}

Consider \Eqn{eq:N_def}, we may write the functional derivative as:
\ba
\delta \mathcal{N}[w] & = & 2\int \hspace{-0.1cm}\dr \left\{
\left[{\mathcal G}^{(1)}_{(1,1)}(\br)\right]^2 + c \left[ 
{\mathcal G}^{(2)}_{(1,0)}(\br) + {\mathcal G}^{(1)}_{(2,0)}(\br)\right]\right\}
\hspace{-0.1cm}\left\{ 2 {\mathcal G}^{(1)}_{(1,1)}(\br) \delta {\mathcal G}^{(1)}_{(1,1)}[w]
+ c \left[ \delta{\mathcal G}^{(2)}_{(1,0)}[w] 
+ \delta{\mathcal G}^{(1)}_{(2,0)}[w]\right]\hspace{-0.1cm}\right\} \ .
\nn\ea
Using the functional derivatives of
Eqs.~(\ref{eq:devG111}),~(\ref{eq:devG210}),~(\ref{eq:devG120}) to
calculate the terms in the parenthesis on the right-hand side, we
obtain the functional derivative of the numerator $\mathcal{N}$:
\ba 
\delta \mathcal{N}[w] &=& 
4\int \dr\, dM dL \left\{\left(\left[{\mathcal G}^{(1)}_{(1,1)}(\br)\right]^2
+ c \left[{\mathcal G}^{(2)}_{(1,0)}(\br) + {\mathcal
    G}^{(1)}_{(2,0)}(\br) \right]\right) \nbar(M) N^{(1)}_{\g}(M)
\Phi(L|M)\Theta(\br|L) \right. \nn \\ 
 & \times & \left.\left[ {\mathcal
    G}^{(1)}_{(1,1)}(\br) b(M) + \overline{\mathcal W}_1(\br, M)
  N^{(2)}_{\g}(M)/N^{(1)}_{\g}(M) + w(\br, L, M) \right]\right\} \delta
w(\br, L, M).
\label{eq:deriv_N}
\ea
%
\vspace{-0.3cm}
\subsection{Functional derivative of $\mathcal{D}[w(\br,L,M)]$}
\vspace{-0.3cm}
Since $\mathcal{D}=A^2$, we have $\delta\mathcal{D}[w]=2 A[w] \delta
A[w]$.  Using the functional derivative in \Eqn{eq:devA}, the
functional derivative of $\mathcal{D}[w]$ is given by
\ba
\delta \mathcal{D}[w] & = & 4 A[w] \int \dr \, dM dL \left\{
{\mathcal G}^{(1)}_{(1,1)}(\br)\,\nbar(M) b(M) N^{(1)}_{\g}(M) 
\Phi(L|M)\Theta(\br|L)\right\}\delta w(\br, L, M) \ .
\label{eq:deriv_D}
\ea  

\end{document}